\documentclass[twocolumn,appendixfloats]{emulateapj}\usepackage{amsmath}
\slugcomment{Accepted for publication in The Astrophysical Journal}
\usepackage{graphicx}
\usepackage{rotating}
\usepackage{longtable}

\usepackage{natbib}
\def\micron{{$\, {\mu}$m}}

\def\eqw{{EQW$_{6.2\mu m}$}}
\def\sil{{$s_{9.7\mu m}$}}

\shorttitle{MIR Properties of Nearby LIRGs}
\shortauthors{Stierwalt et al.}

\begin{document}
\title{Mid-Infrared Properties of Luminous Infrared Galaxies II: Probing the Dust and Gas Physics of the GOALS Sample}
\author{S. Stierwalt\altaffilmark{1,2}, L. Armus\altaffilmark{1}, V. Charmandaris\altaffilmark{3,4,5}, T. Diaz-Santos\altaffilmark{1}, J. Marshall\altaffilmark{1}, A. Evans\altaffilmark{2,6}, S. Haan\altaffilmark{1,7}, J. Howell\altaffilmark{1}, K. Iwasawa\altaffilmark{8}, D.C. Kim\altaffilmark{6}, E.J. Murphy\altaffilmark{1}, J.A. Rich\altaffilmark{9}, H.W.W. Spoon\altaffilmark{10}, H. Inami\altaffilmark{1,11}, A.O. Petric\altaffilmark{1,12}, V. U\altaffilmark{13}}

\altaffiltext{1}{Spitzer Science Center, California Institute of Technology, 1200 E. California Blvd., Pasadena, CA 91125. {\textit{e-mail:}} sabrinas@virginia.edu}
\altaffiltext{2}{Department of Astronomy, University of Virginia, P.O. Box 400325, Charlottesville, VA 22904.}
\altaffiltext{3}{Department of Physics, University of Crete, GR-71003, Heraklion, Greece}
\altaffiltext{4}{Institute for Astronomy, Astrophysics, Space Applications \& Remote Sensing, National Observatory of Athens, GR-15236, Penteli, Greece}
\altaffiltext{5}{Chercheur Associ\'e, Observatoire de Paris, F-75014,  Paris, France}
\altaffiltext{6}{National Radio Astronomy Observatory, 520 Edgemont Road, Charlottesville, VA 22903.}
\altaffiltext{7}{CSIRO Astronomy \& Space Science, Marsfield NSW 2122, Australia.}
\altaffiltext{8}{INAF-Observatorio Astronomico di Bologna, Via Ranzani 1, Bologna, Italy.}
\altaffiltext{9}{The Observatories, Carnegie Institute of Washington, 813 Santa Barbara Street, Pasadena, CA 91101.}
\altaffiltext{10}{Department of Astronomy, Cornell University, Ithaca, NY, 14853.}
\altaffiltext{11}{National Optical Astronomy Observatory, 950 N. Cherry Ave, Tucson, AZ 85719.}
\altaffiltext{12}{Department of Astronomy, California Institute of Technology, 1200 E. California Blvd., Pasadena, CA 91125.}
\altaffiltext{13}{Department of Physics and Astronomy, University of California, Riverside, CA 92507.}
\begin{abstract}
The Great Observatories All-Sky LIRG Survey (GOALS) is a comprehensive, multiwavelength study of luminous infrared galaxies (LIRGs) in the local universe. Here we present the results of a multi-component, spectral decomposition analysis of the low resolution mid-IR {\it{Spitzer}} IRS spectra from 5-38\micron\ of 244 LIRG nuclei. The detailed fits and high quality spectra allow for characterization of the individual PAH features, warm molecular hydrogen emission, and optical depths for both silicate dust grains and water ices. We find that starbursting LIRGs, which make up the majority of the GOALS sample, are very consistent in their MIR properties (i.e. $\tau_{9.7\mu m}$, $\tau_{ice}$, neon line ratios, and PAH feature ratios). However, as their \eqw~decreases, usually an indicator of an increasingly dominant AGN, LIRGs cover a larger spread in these MIR parameters. The contribution from PAH emission to the total IR luminosity (L(PAH)/L(IR)) in LIRGs varies from 2-29\% and LIRGs prior to their first encounter show significantly higher L(PAH)/L(IR) ratios on average. We observe a correlation between the strength of the starburst (represented by IR8 $=$L$_{IR}$/L$_{8\mu m}$) and the PAH fraction at 8\micron\ but no obvious link between IR8 and the 7.7 to 11.3 PAH ratio, suggesting that the fractional PDR emission, and not the overall grain properties are associated with the rise in IR8 for galaxies off the starburst main sequence. We detect crystalline silicate features in $\sim$6\% of the sample but only in the most obscured sources (\sil\ $<$ -1.24). Ice absorption features are observed in $\sim$11\% (56\%) of GOALS LIRGs (ULIRGs) in sources with a range of silicate depths. Most GOALS LIRGs have L(H$_2$)/L(PAH) ratios elevated above those observed for normal star-forming galaxies and exhibit a trend for increasing L(H$_2$)/L(PAH) ratio with increasing L(H$_2$). While star formation appears to be the dominant process responsible for exciting the H$_2$ in most of the GOALS galaxies, a subset of LIRGs ($\sim$10\%) show excess H$_2$ emission that is inconsistent with PDR models and may be excited by shocks or AGN-induced outflows. 
\end{abstract}

\section{Introduction}

Luminous Infrared Galaxies (LIRGs; L$_{IR} > 10^{11}$L$_{\odot}$) are relatively rare in the local
universe but by z$\sim$1 they dominate the co-moving infrared (IR) energy density
\citep{lefloch, magnelli}. Thus not only do LIRGs represent a
galaxy population that dominates the energy budget in the IR at an
epoch when star formation was at its peak, but they also reflect a set
of galaxies that has undergone rapid evolution since z$\sim$1-2. 

Given their high IR luminosities, LIRGs, especially those at z$\sim$0, offer an ideal extragalactic laboratory for studying the features that dominate in the mid-infrared (MIR), particularly the emission from polycyclic aromatic hydrocarbons (PAHs). PAH emission is believed to originate from the transitive heating of small carbonaceous grains by UV photons in photodissociation regions (PDRs) and thus is a good tracer of star formation, even when that star formation is heavily dust obscured. The various possible rotational and vibrational modes of the carbon and hydrogen atoms that form PAH molecules have been well-studied in the laboratory and so clues as to the ionization state and size distribution of the extragalactic dust grains can be determined from measurements of various PAH ratio strengths. More than 85\% of the MIR emission from LIRGs is produced by star formation \citep[and not by an AGN;][]{petric}, and so LIRGs provide a variety of conditions (weak AGN, a range of merger stages, varying amounts of warm H$_2$, deep silicate obscuration) within which to study PAH emission. In addition to grain size distribution and ionization state, PAHs, when linked to other galaxy properties, can reveal whether or not warm H$_2$ emission originates from within PDRs (i.e. is traced by PAH emission) versus being triggered by shocks or X-ray heating \citep[see for example][]{roussel, zakamska}, as well as the environments in which IR and UV emission decouple (through comparisons with the infrared excess, IRX). 

A major effort to study LIRGs in the local universe is the Great Observatories All-sky LIRG Survey \citep{GOALS}. The GOALS sample of 180 LIRGs and 22 ULIRGs represents a complete subset of the IRAS Revised Bright Galaxy Catalog \citep[RBGS;][]{rbgs} covering a range of dust obscuration and merger stages. In our first paper, we presented the low resolution IRS spectra for 244 nuclei within the 202 GOALS LIRG and ULIRG systems and some of the basic related measurements \citep[\eqw, \sil, MIR slope, as well as merger stage;][]{paperI}. Here we present the results of a detailed spectral decomposition of these low resolution MIR LIRG and ULIRG spectra using the method described in \cite{cafe}. 

In Section 2, we briefly discuss the sample selection, as well as the data reduction and spectral fitting method. In Section 3, we present the results of the spectral decomposition, specifically how PAH emission relates to other PAH features (via PAH feature ratios), neon fine structure lines, the total L(IR), merger stage, obscuration (due to silicate dust grains and water ices), and UV properties (i.e. IRX). In Section 4, we discuss warm molecular hydrogen in LIRGs, including whether it is confined to PDRs, and some of the LIRGs with excess H$_2$ emission in the GOALS sample. We also explore the effects weak AGN may have on the MIR properties of LIRGs. Finally, we summarize our conclusions in Section 5.

\section{Observations and Data Analysis \label{obs}}

\subsection{The Sample}
Of the 180 luminous ($10^{11} < L_{IR}/L_{\odot} < 10^{12}$) and 22 ultraluminous ($L_{IR} > 10^{12} L_{\odot}$) IR systems
comprising the GOALS sample, several are multiple interacting galaxies that are resolved in the MIR. As a result, 244 individual galaxy nuclei were
targeted for MIR spectroscopy using the low resolution Short-Low (SL: 5.5-14.5 \micron) and
Long-Low (LL: 14-35 \micron) modules of the {\it{Spitzer}} Infrared
Spectrograph \citep[IRS;][]{IRS}. New staring mode observations were obtained for 157 of
the LIRG systems (PID 30323; PI L. Armus), and archival low resolution
staring or
mapping mode data were acquired for the remainder of the sample. 

All 202 systems are nearby but cover a range of distances (15 Mpc $<$
D $<$ 400 Mpc) and so the resulting projected IRS slit width varies
from source to source. At the median galaxy distance of 100 Mpc, the
nuclear spectrum covers the central 1.8 kpc in SL and the central 5.2
kpc in LL.  Since NGC1068 (the nearest GOALS galaxy at D = 15.9
Mpc) saturates the spectrograph, the closest
sources for which results are presented here are NGC2146 (D = 17.5
Mpc) and NGC1365 (D = 17.9 Mpc) which are probed on sub-kpc scales by both the SL and LL
slits. For the most distant galaxy in the sample, the ULIRG
IRAS07251-0248 (D = 400 Mpc), the SL and LL slit widths translate to
$\sim$7 kpc and 21 kpc respectively.

\subsection{Data Reduction}
The data reduction is described in detail and all 244 reduced GOALS spectra are presented in \cite{paperI} along with tabulated values for 6.2-\micron~PAH equivalent
widths, apparent silicate absorption feature depths, and MIR slopes for the entire sample. Briefly, staring mode spectroscopic data were reduced using the S17 and S18.7 IRS pipelines from the
Spitzer Science
Center\footnote{For more information see: http://ssc.spitzer.caltech.edu/irs/features.html}. One dimensional spectra were
extracted using the standard extraction aperture and point source
calibration modes in
SPICE\footnote{For more information see: http://ssc.spitzer.caltech.edu/postbcd/doc/spice.pdf} which
employs a tapered extraction aperture that averages roughly to a size
of 10$\farcs6 \times 36\farcs$6 in LL and $3\farcs7 \times 9\farcs5$ in SL. 
For the 18 systems with archival mapping mode data, spectra
 were extracted using CUBISM
 \citep{cubism}. 

The three most prominent absorption features in the MIR are those due
to silicate dust grains at 9.7 and 18.5\micron~and ices
at 6.0\micron. In all three cases, the extent of the extinction can be
quantified in two different ways: the strength of the absorption feature, $s_{\lambda}$ and the optical depth, 
$\tau_{\lambda}$. The strength $s_{\lambda}$ is measured directly from
the data (and thus does not rely on the result of the spectral
decomposition) via: $s_{\lambda} = log(f_{\lambda}/C_{\lambda})$ where
$f_{\lambda}$ is the measured flux at the central wavelength of the
absorption feature and $C_{\lambda}$ is the estimated level of the
continuum flux in the absence of the absorption feature,
as derived from an extrapolation to the surrounding continuum. Thus,
$s_{\lambda} > 0$ indicates emission at that
wavelength. This method for calculating $s_{9.7\mu m}$ follows that of
\cite{spoon} and the resulting values for the GOALS galaxies are
presented in \cite{paperI}.

We also use our spectral
decomposition method to derive optical depths corrected for extinction
for the ices at 6.0\micron~and the silicates at 9.7\micron,
$\tau_{ice}$ and $\tau_{9.7\mu m}$.
The level of dust obscuration in each source is determined by assuming
the PAH-emitting dust grains are intermixed with a colder dust
component that is positioned between the observer and a warmer dust
component \citep{cafe}. 
The warm dust component is thus assumed to be extinguished according to $f_{warm} =
f_{warm,0}~e^{-\tau_{\lambda}}$ (screen geometry), while the extinction law applied to
the PAHs follows $f_{PAH} =
f_{PAH,0}~\frac{1-e^{-\tau_\lambda}}{\tau_\lambda}$ (mixed geometry), where
$f_{warm,0}$ and $f_{PAH,0}$ are the emitted fluxes before they have
been extinguished. 

\subsection{Spectral Fitting}\label{recap}
To measure line and PAH feature fluxes, silicate optical depths, water ice
absorption, and dust temperatures, each low resolution spectrum was
fit using the Continuum And Feature Extraction (CAFE) spectral decomposition method developed by \cite{cafe}. 
CAFE requires no prior knowledge of galaxy type or geometry and so is well suited for fitting the SEDs for sources that include
{\it{both}} a starburst and an AGN and for sources where
we do not have prior knowledge of the viewing angle geometry. An observed spectrum is decomposed into emission from old stellar
populations (the interstellar radiation field), PAHs, atomic and molecular
lines, and thermally heated
graphite and silicate grains at several characteristic
temperatures. We
assume that warm dust grains are located behind a screen of obscuring
graphite and silicate grains that are thoroughly mixed with PAHs, and that
the amount of extinction to the warm dust and PAH components is a free
parameter of the fitting procedure.
To make spectral fitting possible, each spectrum from SL is scaled up
by a constant multiplicative factor to match the larger LL slit, which
covers nine times the area covered by the SL slit (i.e. slit widths of $10\farcs7$
versus $3\farcs6$). The scale factors are calculated from the wavelength coverage overlap
in the SL1 and LL2 modules near 14\micron~and are presented for individual sources in
\cite{paperI}. Excluding the 17 systems for which an additional nucleus falls within the LL slit compared to the SL slit, the scale factors range from 0.91 to 2 for 93\% of the nuclei. The remaining fifteen sources with larger scale factors (i.e. scale factors between 2 and 5) are due to extended structure outside of the SL slit. Whenever features from SL and LL are compared in the following sections, the possible consequences of this scaling are discussed.

Of the 244 nuclear spectra in the GOALS sample, 10 are excluded from the analysis that follows because they lack complete IRS observations (NGC4922, IRASF08339+6517, ESO550-IG025 and IC4518), they saturate the spectrograph (NGC1068)\footnote{For a detailed discussion of the IRS spectra of NGC1068, see \cite{howell1068}.}, or the archival SL staring mode observations were not centered on the galaxy nucleus (IIIZw035, IRASF03359+1523, MCG+08-18-013,
IRASF17132+5313, and MCG-01-60-022).
For the remaining 234 sources, an individual PAH or line feature is considered detected
if the signal-to-noise ratio is greater than 2.5.
Among the fitted sources are two galaxies
for which only Short-Low spectra are available (NGC2388 and VV705) so
these are excluded from any plots involving features at $\lambda_{obs}
> 14$\micron~(i.e. the PAH emission complex or the H$_2$S(1) line at
$\sim$17\micron). Also excluded from any analysis at $\lambda_{obs} >
14$\micron~are 12 sources for which multiple nuclei fell in the Long-Low
slit but not in the Short-Low aperture (CGCG448-020, ESO077-IG014,
ESO173-G015, ESO255-IG007, ESO343-IG013, ESO440-IG058 (northern nucleus
only), IRAS03582+6012,
IRASF06076-2139, NGC5653, NGC6090, NGC3690 (western nucleus only), and NGC5256).
Only 3 sources do not have
enough silicate absorption, PAH, or emission line features to produce a
reliable fit, IRAS05223+1908, Mrk 231 (or UGC08058), and NGC1275, and so are
considered PAH and spectral line nondetections.  
Six sources are affected
by very deep silicate absorption (ESO203-IG001;
IRAS03582+6012\_E; IRAS08572+3915, IRAS15250+3609; ESO374-IG032;
and NGC4418) and the detailed structure between 8\micron-20\micron~
is not well-reproduced by the fit. These galaxies are
discussed in more detail in Sections \ref{bfit} \& \ref{crystals}. 

We compare the results from CAFE to those derived from the PAH
fitting program PAHFIT \citep{jdsings} for the same set of GOALS
sources and find the results to be largely consistent. For the PAH
feature ratio of L(6.2\micron)/L(7.7\micron), the results for 90\% of the
GOALS sample agree
within 10\%. The biggest difference between the two fitting techniques
results from different treatments of extinction due to both silicates
\citep[since CAFE includes silicate feature emission in unextinguished dust
components, see][]{cafe} and ices (which PAHFIT does not model). 
The difficulty of applying other MIR dust models, normally used for starburst galaxies, to the heavily obscured GOALS LIRGs was also noted by \cite{dopitaSED} who derived silicate absorption parameters using an empirical fitting technique instead. 
These different approaches used by CAFE and PAHFIT affect most noticeably the flux
derived for the 11.3-\micron~PAH complex which is nearest to the center
of the 9.7-\micron~silicate absorption feature. For the dustiest
sources, the resulting PAH fluxes derived 
from the two methods
differ by as much as 40\%. However, the derived
L(11.3\micron)/L(7.7\micron) flux ratio as measured by CAFE and PAHFIT
agrees within 10\% for 70\% of the measured sources. 

The results of the spectral decomposition for the GOALS sample as obtained by CAFE are presented in Tables \ref{pahtable} (PAH strengths, ice \& silicate optical depths), and \ref{H2table} (H$_2$ emission line strengths). No extinction correction has been applied to the H$_2$ line fluxes.

\section{Results}\label{results} 

The basic MIR parameters measured directly from the spectra without requiring the detailed spectral decomposition (i.e. EQW$_{6.2\mu m}$, $s_{9.7\mu m}$, and F$_{\nu}$[30\micron]/F$_{\nu}$[15\micron]) were presented in \cite{paperI}. We found that although local LIRGs cover a large range of MIR properties and any single LIRG cannot represent the class as a whole, the majority (63\%) of LIRGs have EQW$_{6.2\mu m} >$ 0.4\micron, $s_{9.7\mu m}$ $>$ -1.0, and MIR slopes in the range of 4 $<$ F$_{\nu}$[30\micron]/F$_{\nu}$[15\micron] $<$ 10. Although less numerous (only 18\% of the sample), LIRGs with the largest contributions from AGN (those with EQW$_{6.2\mu m} <$ 0.27\micron) cover a wider range of MIR slopes and silicate strengths than those sources of higher equivalent width that make up the majority of the sample. The sources with extremely low PAH equivalent widths (\eqw$<$0.1\micron) separate into two distinct types: relatively unobscured sources with a very hot dust component (and thus very shallow MIR slopes) and heavily dust obscured nuclei with a steep temperature gradient.

\begin{figure*}[htp]
\begin{center}
\includegraphics[height=1.7in,width=2.1in]{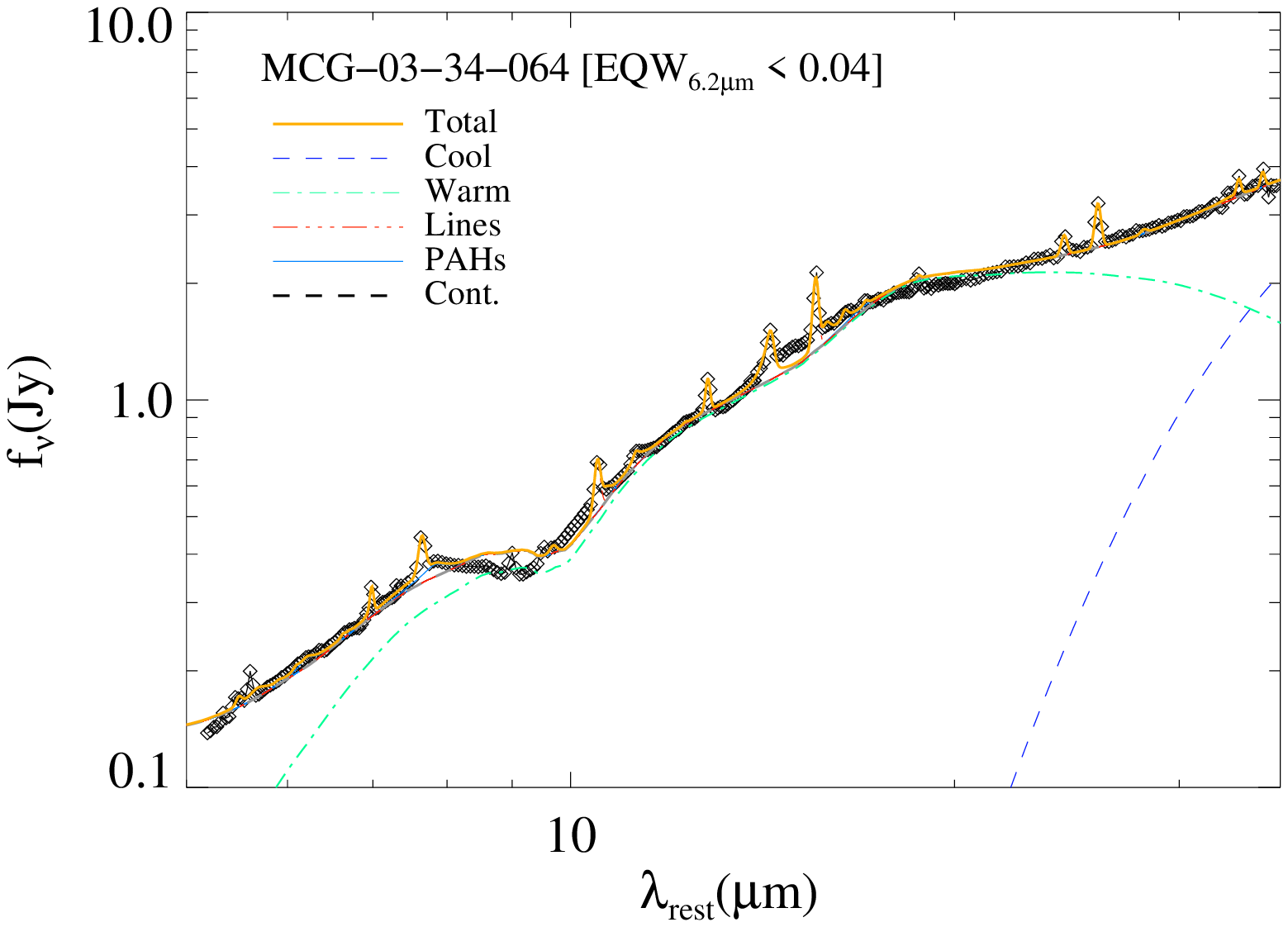}
\includegraphics[height=1.7in,width=2.1in]{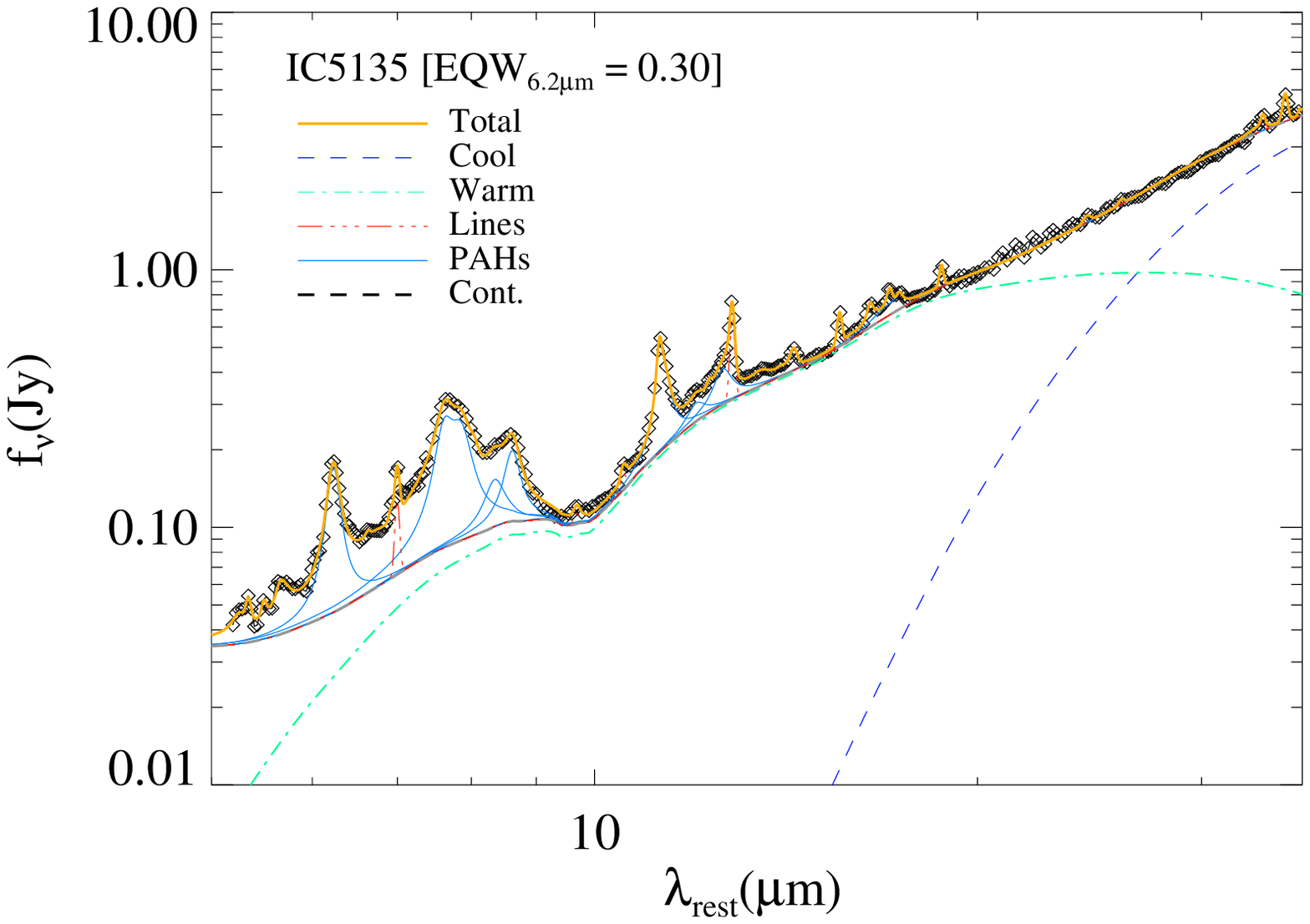}
\includegraphics[height=1.7in,width=2.1in]{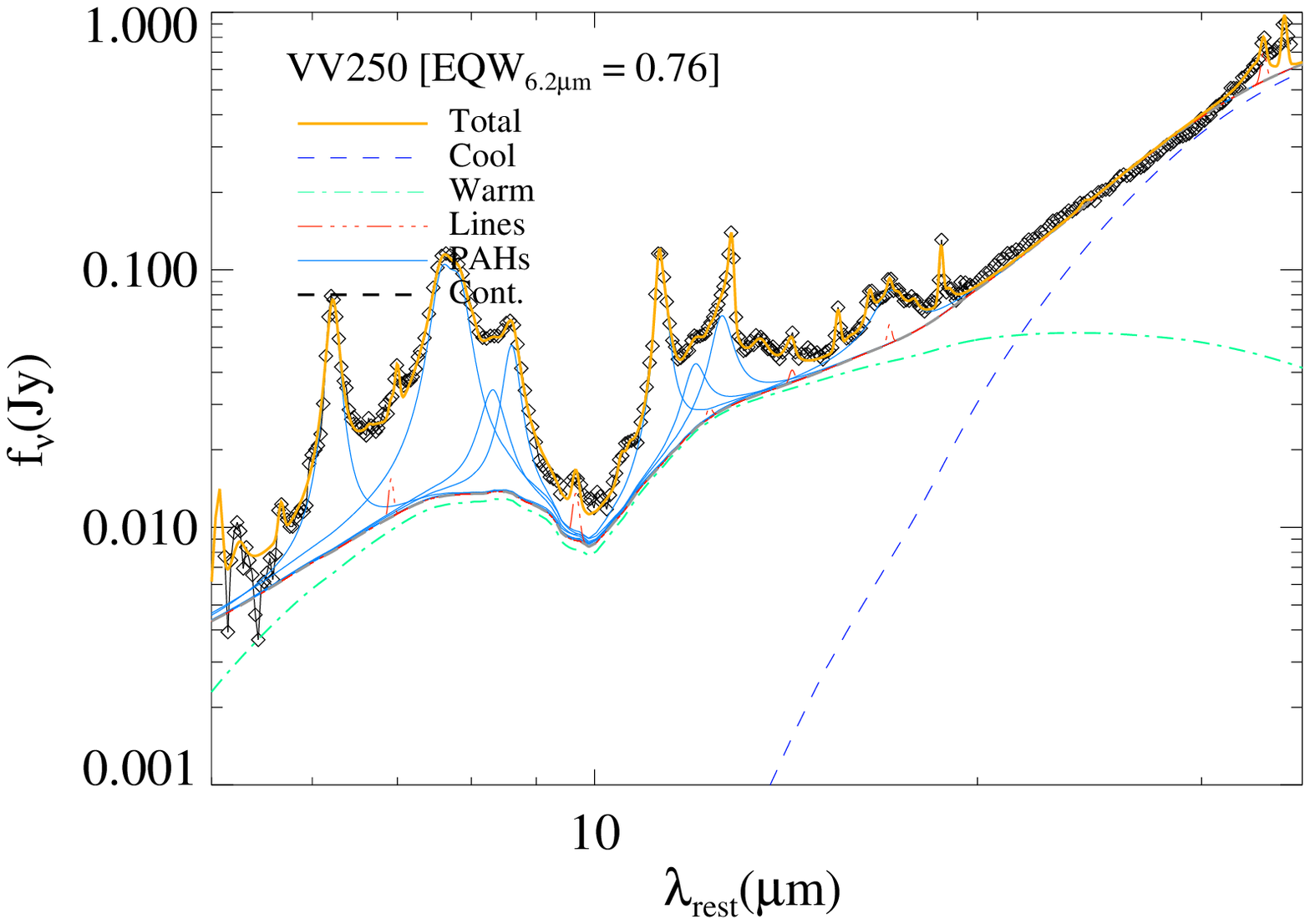}
\includegraphics[height=1.7in,width=2.1in]{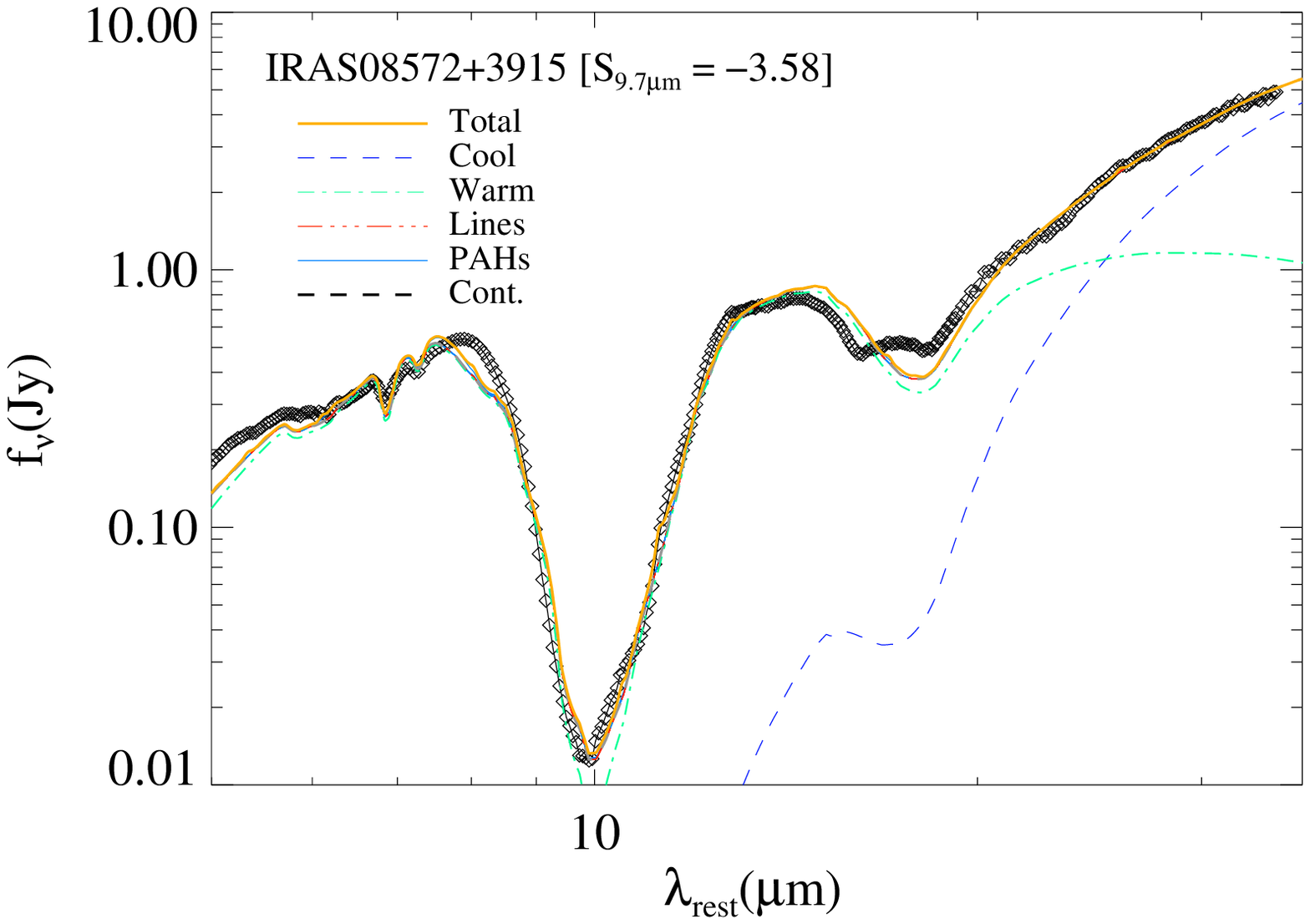}
\includegraphics[height=1.7in,width=2.1in]{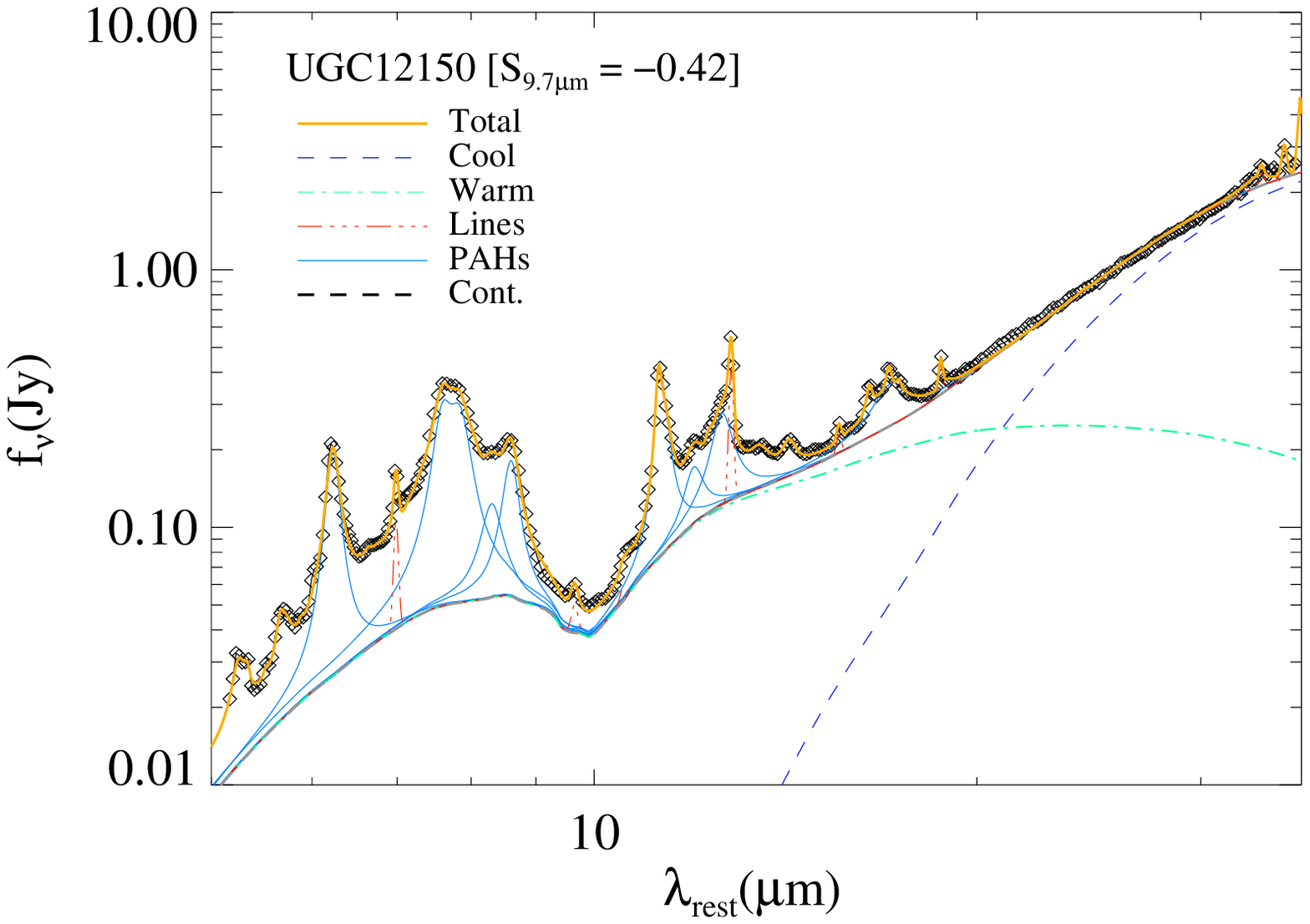}
\includegraphics[height=1.7in,width=2.1in]{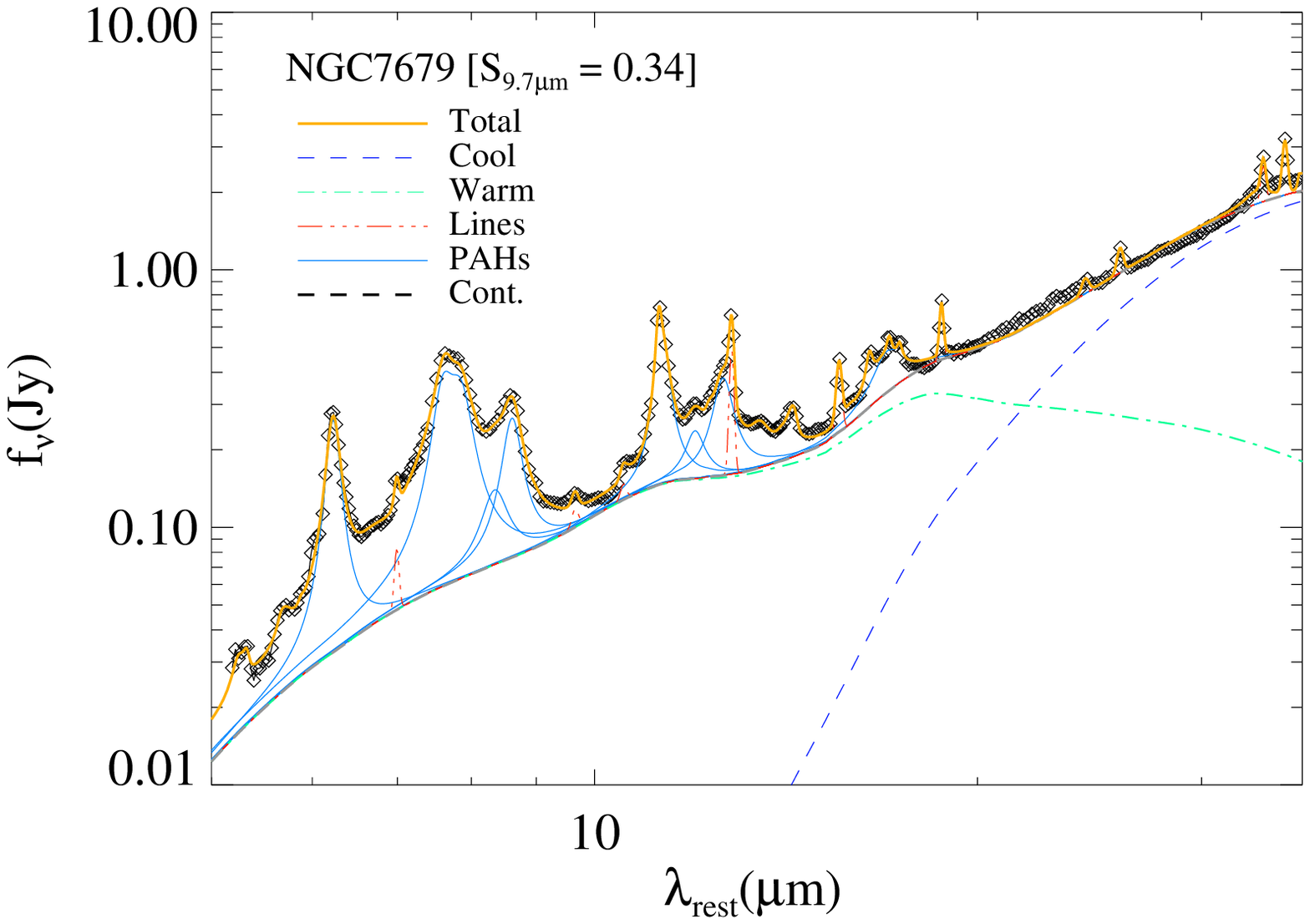}
\includegraphics[height=1.7in,width=2.1in]{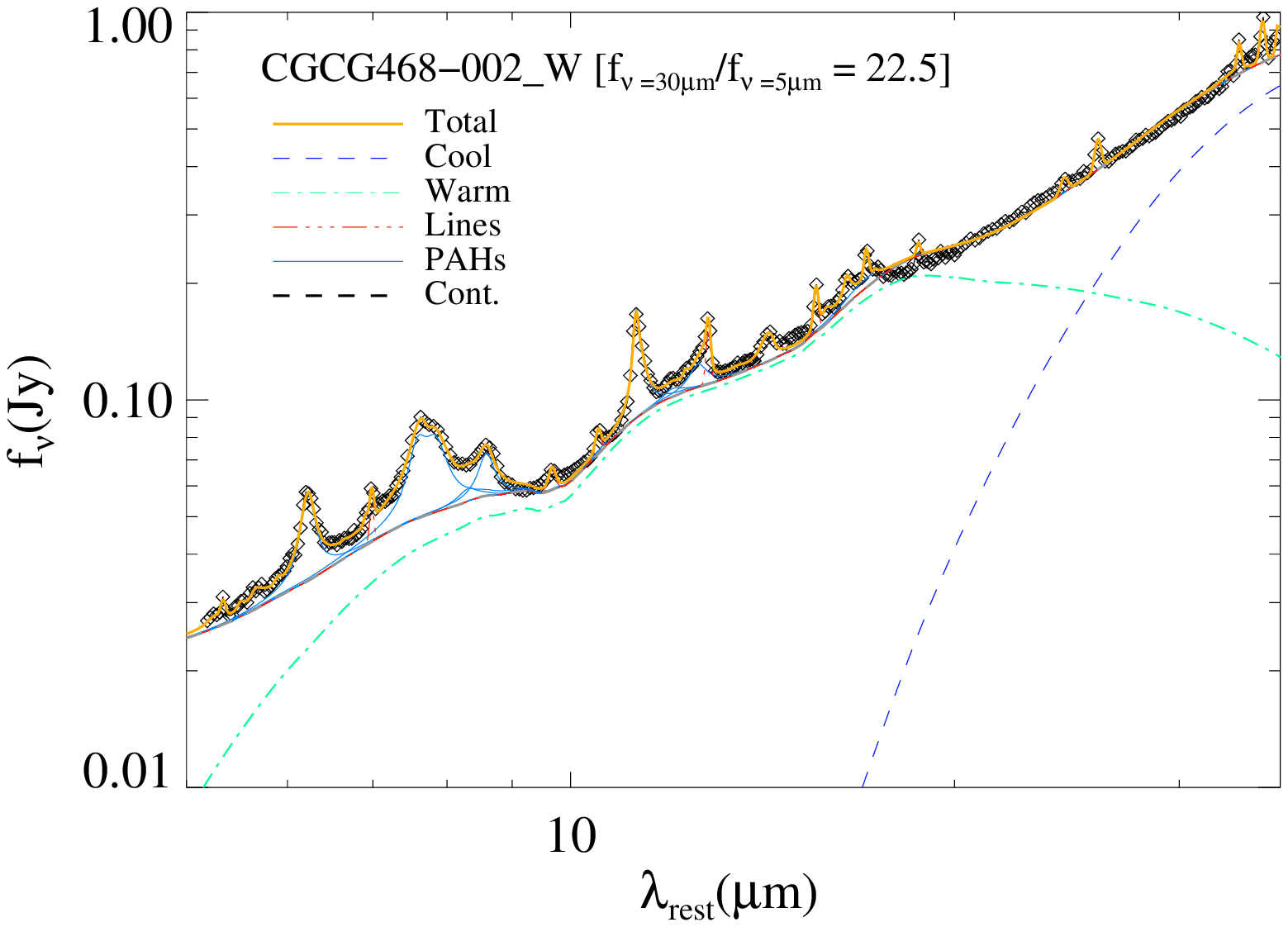}
\includegraphics[height=1.7in,width=2.1in]{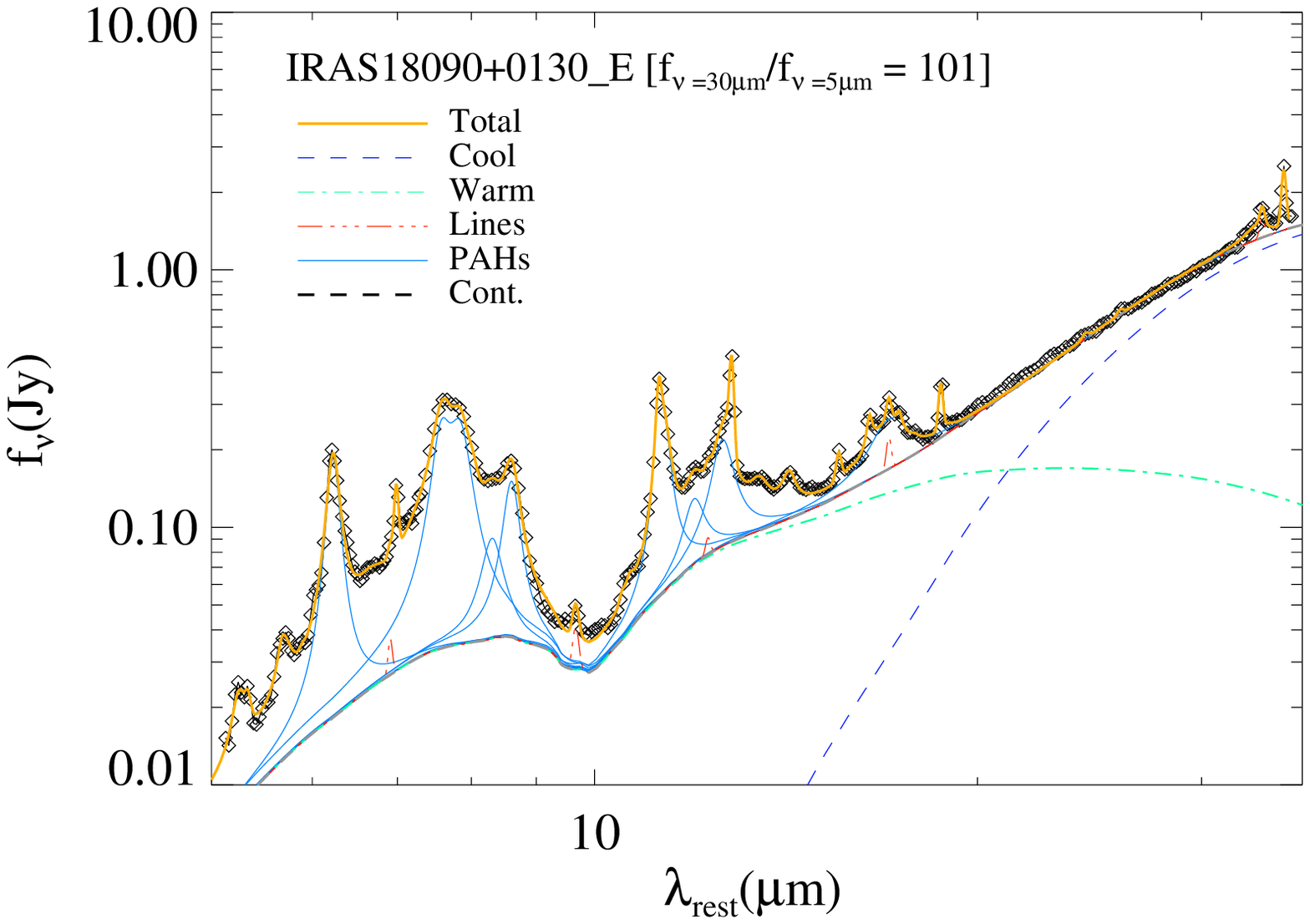}
\includegraphics[height=1.7in,width=2.1in]{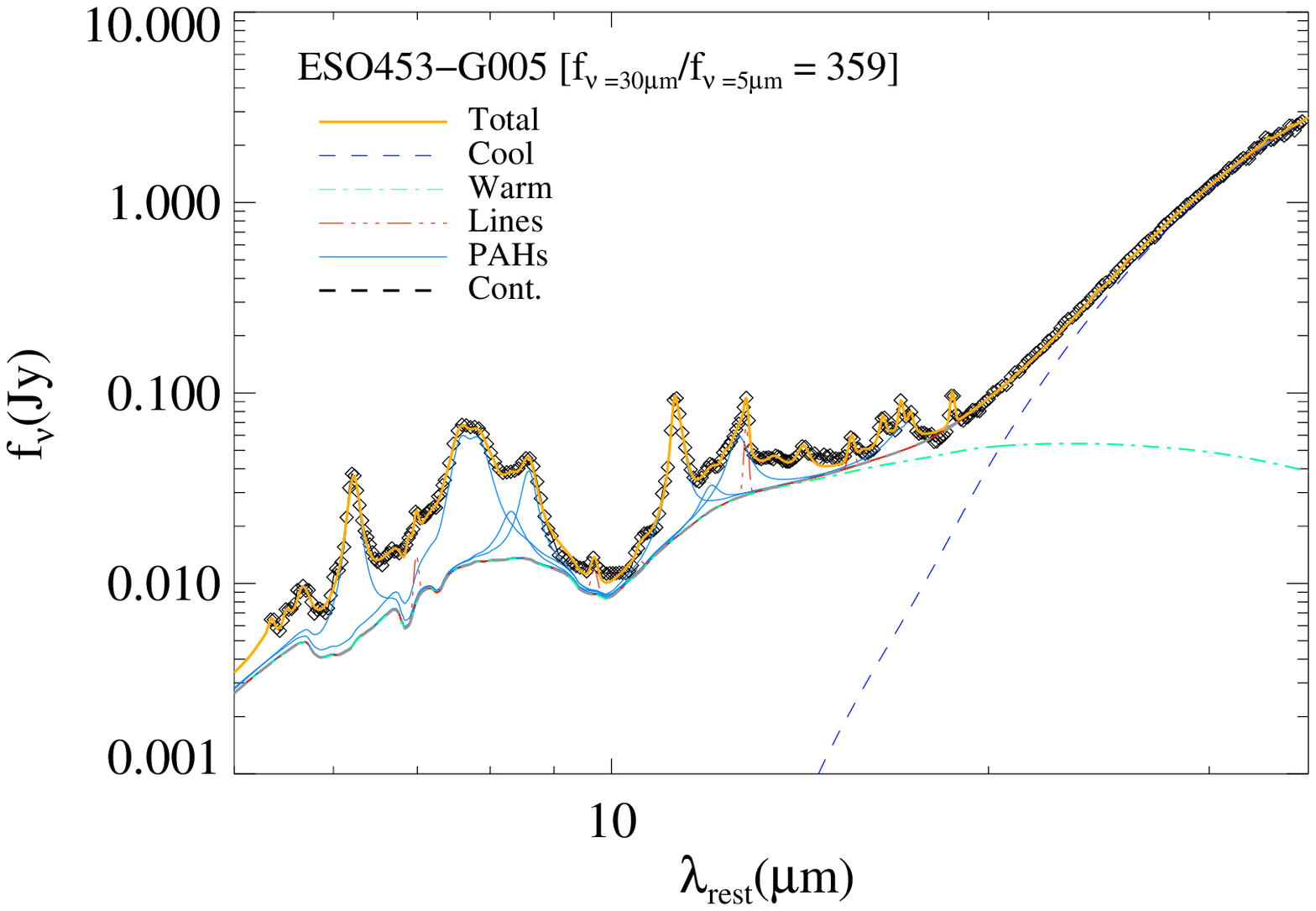}
\caption{Low Resolution IRS Spectra with CAFE spectral decomposition results for
  9 sources representing the ranges in 6.2-\micron~PAH equivalent
  width (top row), silicate depth (middle row), and MIR slope
  (F$_{\nu}$[30\micron]/F$_{\nu}$[15\micron]; bottom row) 
  covered by the GOALS sample. The overall fitted model is shown in
  yellow. Also shown are the warm and cool dust components (green and
  blue dashed lines), the overall continuum fit (gray dashed line),
  selected individual PAH emission features (blue solid lines), and
  unresolved atomic and molecular spectral line features (red dashed
  lines). 
\label{repspecs}}
\end{center}
\end{figure*}

Figure \ref{repspecs} shows nine example spectra that represent the ranges of 6.2-\micron~PAH equivalent width, silicate depth and MIR slope covered by the sample together with the fits produced by CAFE. 

\subsection{A Uniform PAH Spectral Signature in Starburst LIRGs}\label{PAHratios}

\begin{figure*}[htp]
\begin{center}
\includegraphics[height=4.8in,width=6.4in]{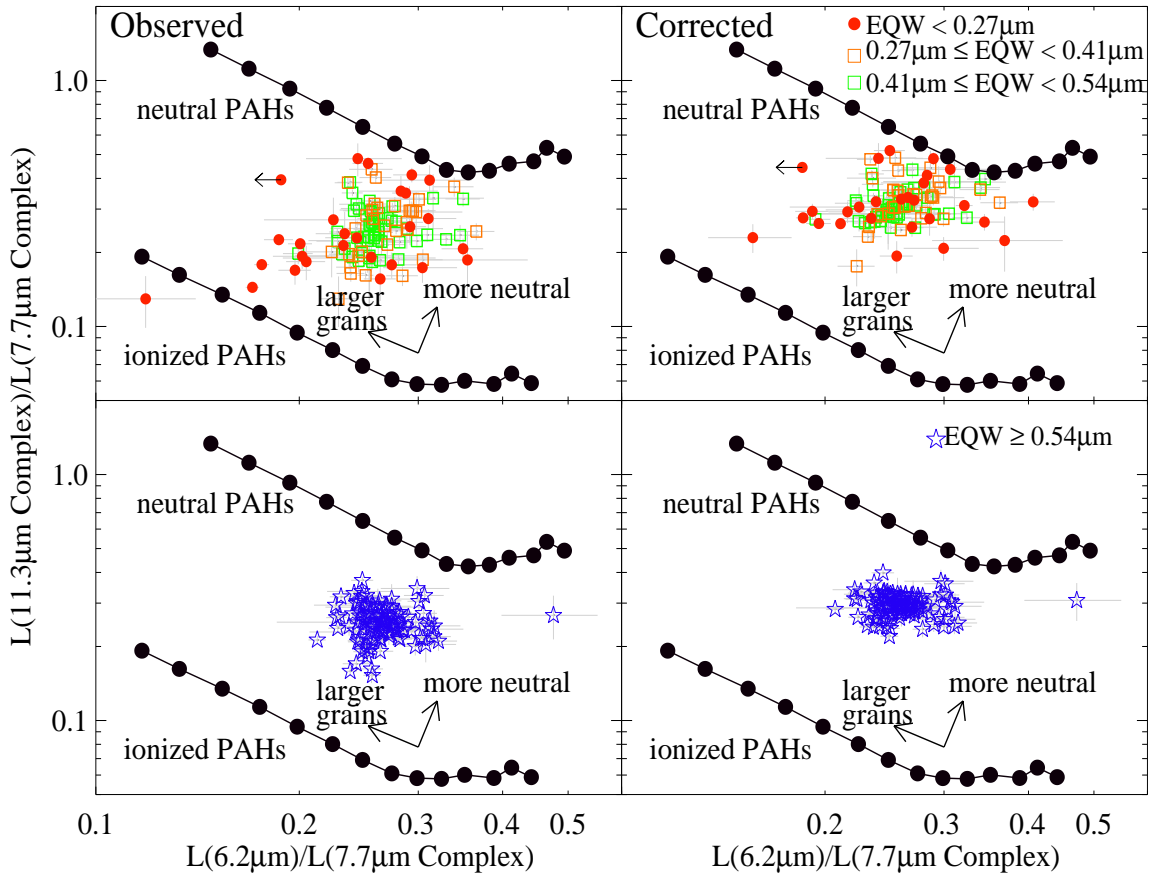}
\caption{PAH Diagnostics: L($6.2\mu
  m$)/L($7.7\mu m$ Complex) vs L($11.3\mu m$ Complex)/L($7.7\mu m$
  Complex) for 226 GOALS sources color coded by the 6.2-\micron~PAH
  equivalent width (EQW$_{6.2\mu m}$) as observed (left panels) and corrected for extinction (right panels). The highest equivalent width sources (\eqw$\geq$0.54\micron) are plotted separately in the lower panels for clarity. Less variation is seen in the PAH
  ratios once we account for the obscuring dust. While the mean locus remains the same, there is a marked increase in the range of PAH ratios seen in the AGN-dominated sources compared to those that are starburst-dominated, suggesting an influence of the AGN on the PAH grains. The expected ratios for purely neutral and
  purely ionized PAHs derived from the dust models of \citep{draineli} are shown by a solid black lines
  with PAH grain size increasing toward the upper left. 
\label{ODfig5}}
\end{center}
\end{figure*}

Dust grain geometry, size distribution, and ionization state can all affect the PAH feature ratios observed in the nuclear spectra. Our spectral decomposition method allows for the separation of line and PAH emission from the dust continuum and from the effects of silicate absorption and thus enables  full modeling of the contribution from each component. The three most dominant PAH features (at 6.2, 7.7, and 11.3\micron) are plotted as ratios for GOALS galaxies both with and without an extinction correction in Figure \ref{ODfig5}. Note that all three PAH features are observed in the SL portion of the spectrum and so their ratios are unaffected by the SL-to-LL multiplicative scale factors discussed in Section \ref{recap}. 

The two tracks in each panel of Figure \ref{ODfig5} represent the theoretical models of \cite{draineli} for purely ionized (lower track) and purely neutral (upper track) dust grains with grain size increasing toward the upper left. The 11.3-\micron~PAH is thought to originate from carbon-hydrogen bending modes of neutral grains, while the PAH emission at 6.2 and 7.7\micron~is produced by carbon-carbon stretching modes of cations. Thus higher L(11.3\micron~Complex)/L(7.7\micron~Complex) ratios may indicate a lower fraction of ionized grains \citep{li2001, allamandola}. Changes in PAH luminosity ratios may also reflect differing dust grain sizes as smaller dust grains are more likely to have higher-frequency vibrational modes and thus radiate at shorter wavelengths.

When extinction corrected (right panels), the LIRGs are highly concentrated around the average values $\langle$L(11.3\micron~Complex)/L(7.7\micron~Complex)$\rangle = 0.31\pm0.10$ and $\langle$L(6.2\micron)/L(7.7\micron~Complex)$\rangle = 0.26\pm0.06$. Such a tight clustering is not observed, however, for the measured values without correction for dust extinction (left panels) where the dispersion in the PAH ratios is twice as high.
Since the PAH feature at 11.3\micron~is most affected by the nearby silicate absorption feature at 9.7\micron, the L(11.3\micron~Complex)/L(7.7\micron~Complex) ratio is more strongly affected by the extinction correction than the L(6.2\micron)/L(7.7\micron~Complex) ratio. Correcting the PAH ratios for extinction both reduces the dispersion in the distribution shown in Figure \ref{ODfig5} and shifts the entire distribution up, favoring a more neutral grain population.

In LIRGs where starbursts dominate the MIR emission 
 \citep[i.e. \eqw$\geq$0.54\micron, the average \eqw\ observed for the starbursts of][lower panels]{brandlEW}, very little variation is seen in the PAH feature ratios (and thus the ionization states and grain size distributions) after correcting for extinction. However, the scatter increases with decreasing PAH equivalent width.
The lowest equivalent width sources (EQW$_{6.2\mu m} < $0.27\micron; red circles) are not offset from the higher EQW sources but instead have an increased dispersion that is $\sim$35\% larger than that observed for the highest EQW sources. A low PAH equivalent width requires a larger contribution from the hot dust continuum relative to PAH emission which typically indicates the presence of an AGN. Therefore, LIRGs with a larger fraction of their MIR emission produced by an AGN also have a much larger spread in their PAH ratios, an effect that is explored in Section \ref{agndisc}.

\subsection{The 17-\micron~PAH Complex}
\begin{figure}[htp]
\begin{center}
\includegraphics[height=2.4in,width=3.5in, viewport=20 0 500 345,clip]{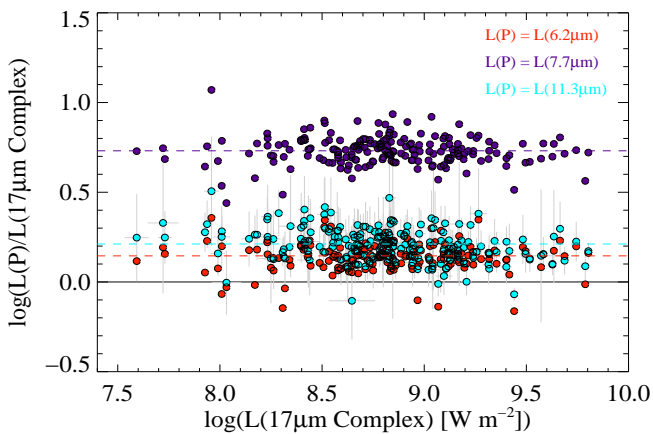}
\caption{PAH Diagnostics: Luminosities of the three strongest PAH features, L($6.2\mu m$) in red, L($7.7\mu m$ Complex) in purple, and L($11.3\mu m$ Complex) in blue, for 196 GOALS sources, weighted by and correlated with the luminosity of the 17-\micron~PAH complex. The red, purple, and blue dashed lines show the median values for each of the three PAH fractions. Only the uncertainties on the 11.3-\micron~PAH fraction are shown for clarity.
\label{pah17}}
\end{center}
\end{figure}

In addition to the more dominant PAH emission features at 6.2, 7.7, and 11.3\micron, a strong emission feature at 17\micron~is also thought to
originate from PAH bending modes \citep{pah17smith, pah17werner, pah17peeters}. Together these four PAH features make up on average 73\% of the total PAH emission in each GOALS LIRG. As shown in Figure \ref{pah17}, the luminosities of the 6.2, 7.7, and 11.3-\micron~PAH features (red, purple, and blue circles) are 1.4-5.5 times stronger than that of the 17-\micron\ PAH. 

\begin{figure}[htp]
\begin{center}
\includegraphics[height=2.3in,width=3.5in]{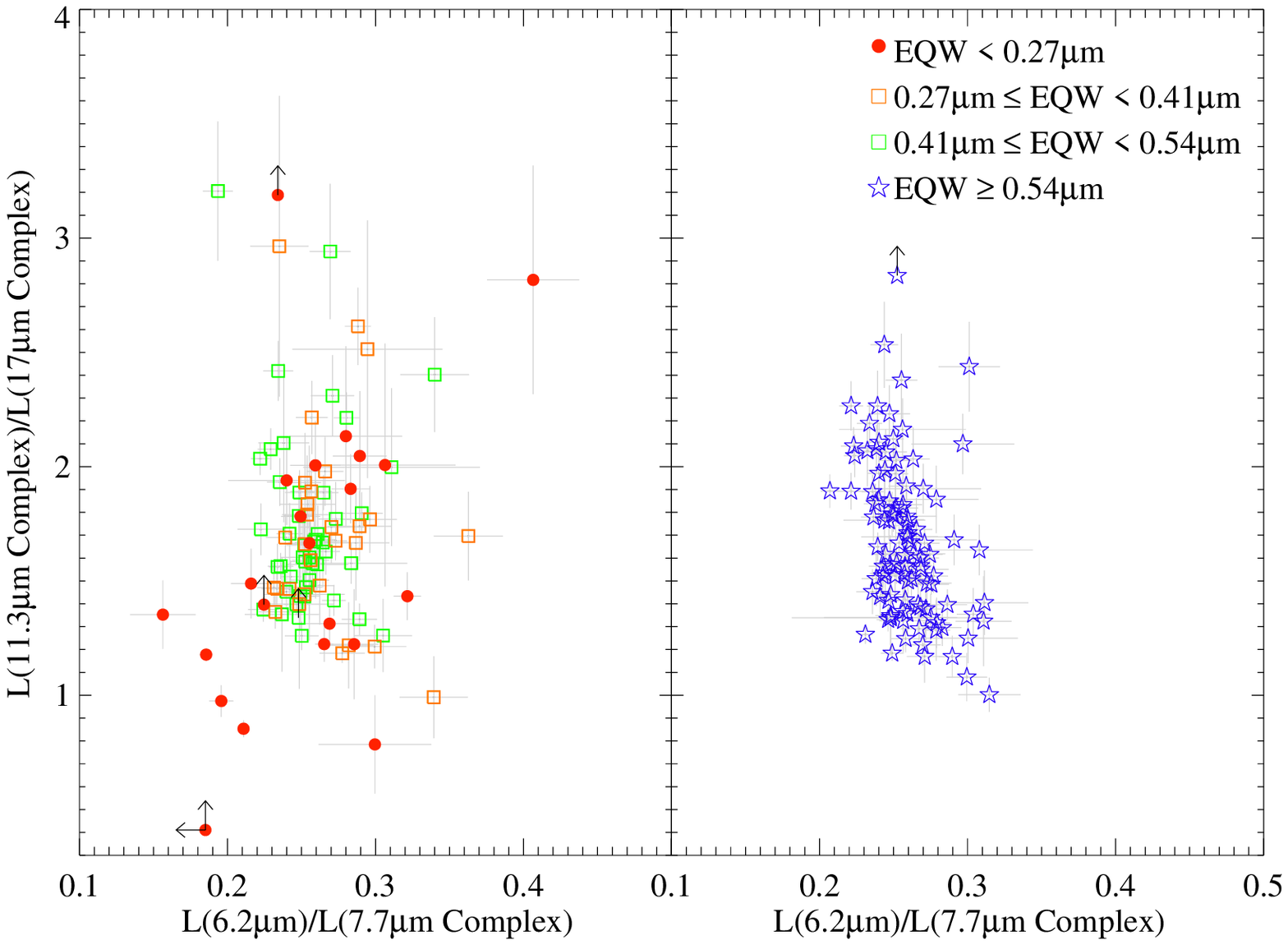}
\caption{PAH Diagnostics: PAH luminosity ratios of the four strongest PAH features: L($6.2\mu m$)/L($7.7\mu m$ Complex) vs L($11.3\mu m$
  Complex)/L($17\mu m$ Complex) for 203 GOALS sources color-coded by the
  equivalent width of the 6.2$\mu m$ PAH (EQW$_{6.2\mu
    m}$) with starburst dominated sources (i.e. \eqw $\geq$0.54) plotted separately in the right panel. Among the four strongest PAH features, the L($11.3\mu m$
  Complex)/L($17\mu m$ Complex) ratio exhibits the largest spread, varying by a factor of $\sim$5.
\label{JDfig13}}
\end{center}
\end{figure}

Although the correlations in Figure \ref{pah17} support the association of the 17-\micron~feature with PAH emission, the inclusion of the 17-\micron~feature causes the largest spread in PAH ratios observed for the GOALS sample. In Figure \ref{JDfig13}, the L($11.3\mu m$ Complex)/L($17\mu m$ Complex) ratio is plotted against the L($6.2\mu m$)/L($7.7\mu m$ Complex) ratio and found to vary by a factor of five. The sources with the highest EQW$_{6.2\mu m}$ (starburst-dominated galaxies; blue stars) are not as highly clustered in ($11.3\mu m$ Complex)/L($17\mu m$ Complex) as they are in the L($11.3\mu m$ Complex)/L($7.7\mu m$ Complex) ratio (seen in Figure \ref{ODfig5}). 

The 17-\micron~PAH feature is observed in the LL module and so some of the scatter in Figure \ref{JDfig13} may be due to the difference in SL and LL aperture sizes. However, the increased dispersion in L($11.3\mu m$ Complex)/L($17\mu m$ Complex) cannot be the result of aperture effects alone. If the scaling factors used to boost the SL-derived flux to match that observed within the LL aperture are removed from the L(11.3\micron~Complex) measurements (i.e. we divide the 11.3-\micron~PAH flux by the same scale factor applied to the SL spectrum to allow the fitting), the spread in the L($11.3\mu m$ Complex)/L($17\mu m$ Complex) remains the same and the sources pushing the upper and lower range of the ratio show no significant changes. Only the mean ratio for the sample as a whole decreases slightly to L($11.3\mu m$ Complex)/L($17\mu m$ Complex)$\sim$1.2. There is also no clear correlation observed between the PAH ratio and the distance to the galaxy, suggesting that while changing the fraction of the galaxy intercepted by the IRS slit may cause some points to move around in Figure \ref{JDfig13}, the overall distribution remains the same.

The ranges of PAH luminosity ratios shown in Figures \ref{ODfig5} and \ref{JDfig13} for the GOALS sources are similar to those covered by the nearby, normal star-forming galaxies from the {\it{Spitzer}} Infrared Nearby Galaxies Survey \citep[SINGS;][]{jdsings}, the 24-\micron~selected systems of the 5mJy Unbiased {\it{Spitzer}} Extragalactic Survey \citep[5MUSES;][]{5muses}, the UV-selected nearby galaxies in the {\it{Spitzer}} SDSS Galaxy Spectroscopic Survey \citep[SSGSS;][]{SSGSS}, and the slightly higher redshift LIRG sample of \cite{shipley}.  A similar clustering in PAH ratios was also observed for galactic photodissociation regions (PDRs) and Magellanic HII regions \citep{galliano}. These similarities suggest that the PDRs don't have fundamentally different properties in these different galaxy samples. One exception to this homogeneity across galaxy type and luminosity, the L($7.7\mu m$)/L($11.3\mu m$) ratio, is discussed in the next section.

\subsection{PAHs and Neon Fine Structure Lines}
\begin{figure}[htp]
\begin{center}
\includegraphics[height=2.4in,width=3.5in, viewport=20 0 500 345,clip]{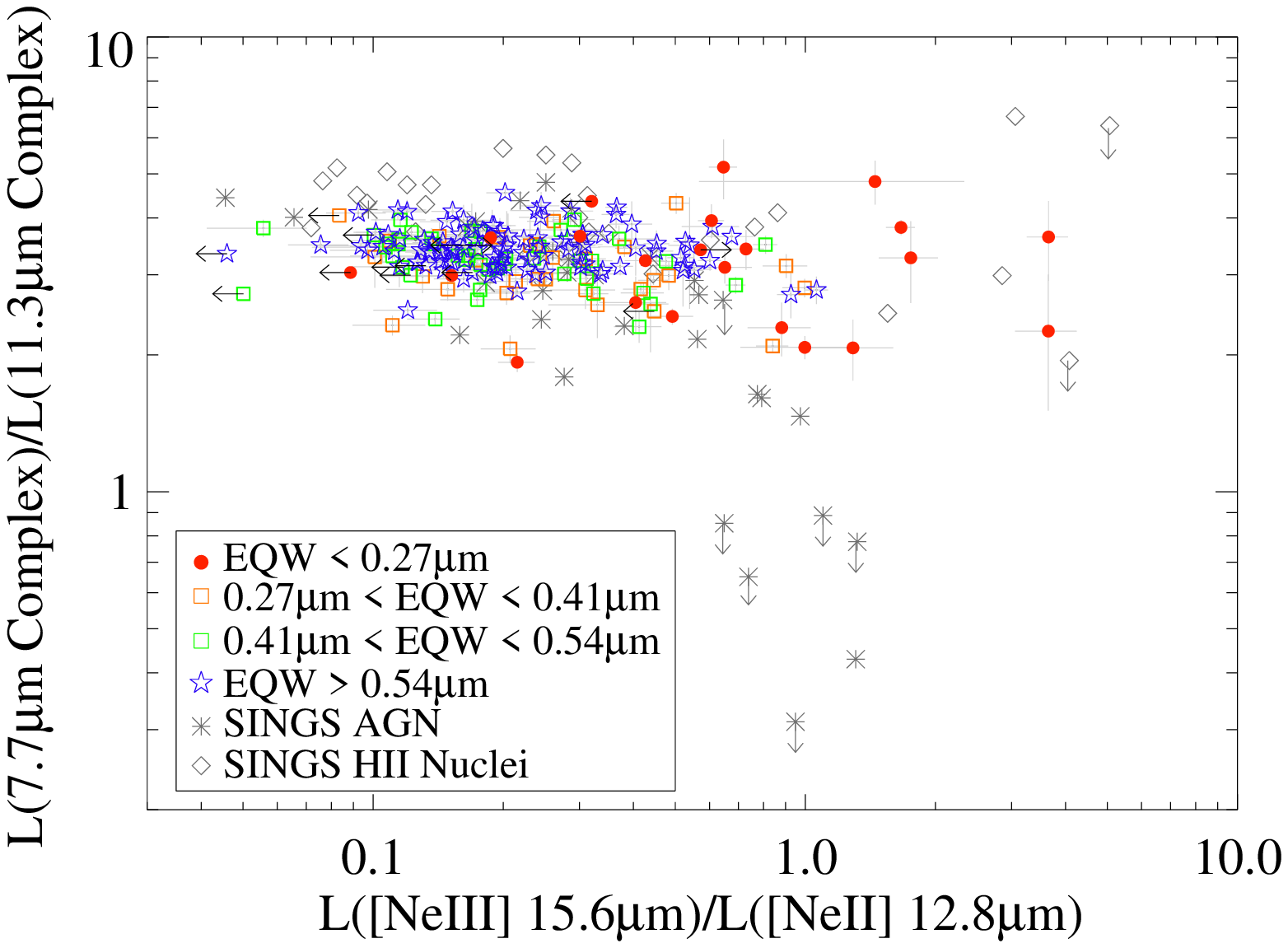}
\caption{PAH Diagnostics: Fine structure line ratio
  L([NeIII] 15.6$\mu m$)/L([NeII] 12.8$\mu m$) vs the PAH feature ratio
  L($7.7\mu m$ Complex)/L($11.3\mu m$ Complex) for 203 GOALS sources color-coded by the
  equivalent width of the 6.2$\mu m$ PAH. The starburst LIRGs (blue stars) cover the same range of ratios as the SINGS lower luminosity starbursts (black diamonds) but
  the AGN-dominated LIRGs (red circles) do not reach the low PAH ratios observed for the SINGS AGN (black asterisks). 
\label{JDfig14}}
\end{center}
\end{figure}

Two of the most prominent emission lines in the MIR, [NeIII] at
15.6\micron~and [NeII] at 12.8\micron, provide a direct diagnostic for
determining the ionization state of the gas or the hardness of
the radiation field \citep{dopita, groves}. In Figure \ref{JDfig14}, we plot the L([NeIII]
15.6\micron)/L([NeII] 12.8\micron) ratio\footnote{The neon ratios plotted here are derived from the CAFE fits to the low resolution spectra which follow the same distributions as the emission line fluxes derived from the high resolution spectra. The emission line fluxes are roughly 40\% higher than those measured in the short-high module in part due to the larger slit size of the long-low module. We refer the reader to \cite{inamihires} for a detailed discussion and comparison to models of the fine structure lines.} 
which does not have a strong dependence on dust
grain size, against the L($7.7\mu m$ Complex)/L($11.3\mu m$ Complex)
ratio which varies both with ionization state and grain size\footnote{Although the inverse PAH ratio was plotted in Figure \ref{ODfig5} and discussed in Section \ref{PAHratios}, the L($7.7\mu m$ Complex)/L($11.3\mu m$ Complex) ratio is used as a tracer of grain size and ionization state in \cite{jdsings}, so we quote it here for ease of comparison.}. The PAH ratio varies by less than a factor of three across
nearly two orders of magnitude of neon ratios, and thus over a large range
of ionization states.

Despite covering the same range of ionization states, the SINGS nearby star-forming galaxies (shown by the black symbols in Figure \ref{JDfig14}) show a variation in the L($7.7\mu m$ Complex)/L($11.3\mu m$ Complex) ratio of up to a factor of 10 \citep{jdsings}. Specifically, six of the SINGS sources for which $>$50\% of the MIR emission is derived from an AGN (black asterisks) are found at low PAH luminosity ratios (L($7.7\mu m$)/L($11.3\mu m$) $<$ 2) that are not reached by the GOALS sample. GOALS galaxies with low EQW$_{6.2\mu m}$, and thus a likely significant contribution from an AGN, clearly favor higher neon ratios as is observed for AGN-dominated SINGS sources, but these AGN only increase the
scatter of the PAH feature ratio to both higher and lower values rather than just favoring low L($7.7\mu m$)/L($11.3\mu m$). Other PAH studies have claimed to see a preference for sources hosting an AGN to be found in the lower right corner of Figure \ref{JDfig14} \citep{5muses, SSGSS}, but all are based on only a few data points and, like GOALS, do not reproduce the clear tail observed for SINGS. The dispersion in this PAH ratio (and lack thereof) is discussed in more detail in Section \ref{agndisc}.

The extinction corrections applied by CAFE (and not applied by PAHFIT used to produce the SINGS results) are not responsible for the lack of GOALS galaxies with L($7.7\mu m$)/L($11.3\mu m$) $<$ 2. As shown in Figure \ref{ODfig5}, removing the extinction correction would shift the GOALS galaxies to even higher L($7.7\mu m$)/L($11.3\mu m$) ratios. The applied SL-to-LL scaling factors also do not affect the L($7.7\mu m$)/L($11.3\mu m$) ratio, as both features are measured from the SL spectrum. 

\subsection{PAH Contribution to IR Luminosity}\label{pahfrac}

\subsubsection{Tracing L$_{PAH}$ with \eqw}
In the absence of more detailed MIR spectral decomposition, the equivalent widths of the strongest PAH features have been used as proxies for the relative strengths of dust continuum versus PAH feature emission. In particular, the EQW of the 6.2 or 7.7\micron~features, which are both prominent and not substantially affected by extinction or blending with other MIR features, have been studied extensively as function of various physical parameters such as the metallicity (i.e. \cite{wuBCDs}) or AGN strength \cite[i.e.][]{genzelULIRGs, laurent, armusULIRGs, quest, petric}. Our detailed, multicomponent spectral decomposition allows for the separation of not only the main PAH features discussed so far at 6.2, 7.7, 11.3, and 17\micron, but also the weaker PAH emission features (i.e. at 8.6, 12, 12.6, 13.6, 14.2, and 16.4\micron) and thus provides an accurate measure of the total PAH contribution to the infrared luminosity (L(PAH)/L(IR)).

\begin{figure}[ht]
\begin{center}
\includegraphics[height=2.25in,width=3in]{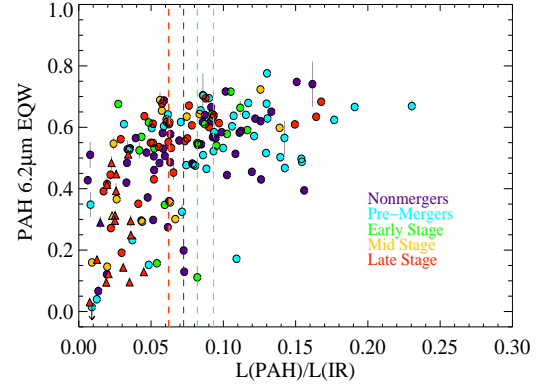}
\includegraphics[height=2.25in,width=3in]{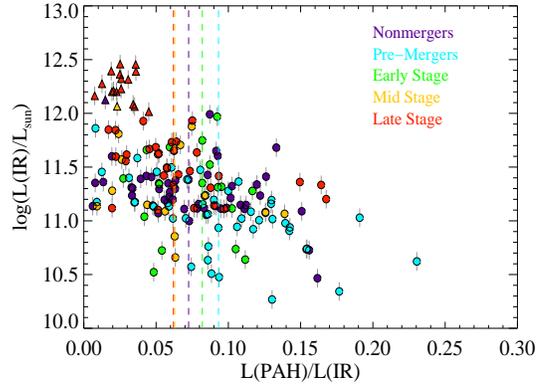}
\caption{Total PAH Luminosity: The contribution of the total PAH luminosity L(PAH) to the total
  IR luminosity L(IR) in 174 GOALS LIRGs (circles) and ULIRGs (triangles) is plotted against the
  equivalent width of the 6.2$\mu m$ PAH (left panel) and log(L(IR)/L$_{\odot}$) (right panel). L(PAH) is derived from all measurable PAH features (i.e. at 6.2, 7.7, 8.6, 11.3 12, 12.6, 13.6, 14.2, 16.4, and 17 \micron) and then scaled up to account for emission outside the slit. Both plots are color-coded by merger stage as classified from HST and IRAC 3.6-\micron~imaging \citep[see ][]{paperI}. The 3-$\sigma$
  clipped average of L(PAH)/L(IR) for each merger class is shown by a
  vertical dashed line.
\label{lpahmerge}}
\end{center}
\end{figure}

The L(PAH)/L(IR) fraction is shown in Figure \ref{lpahmerge} as it relates to EQW$_{6.2\mu m}$, L(IR), and merger stage for the GOALS sample. L(PAH) is derived from the CAFE fit to each nuclear spectrum and then scaled up to account for emission outside the slit. This multiplicative factor, which ranges from 0.8 to 8.8 and is given for all sources in Table \ref{pahtable}, represents the fraction of the emission at 8\micron~intercepted by the IRS slit: $F_{tot}^{IRAC}[8\mu m] / F_{slit}^{IRS}[8\mu m]$ where $F_{tot}^{IRAC}[8\mu m]$ is the total flux of a source as measured from its IRAC 8-\micron~image and $F_{slit}^{IRS}[8\mu m]$ is the flux within the IRS slit derived by convolving the IRAC 8-\micron~filter with the low resolution IRS spectrum. Eight sources with large factors (i.e. F$_{tot}^{IRAC}$[8\micron]/F$_{slit}^{IRAC}$[8\micron] $>$ 10) --- usually due to detailed, extended structure outside of the slit --- were not included (see \cite{paperI} for IRS slit projections shown on the IRAC 8-\micron~images). The L(PAH)/L(IR) for these eight sources ranged from 0.04 to 0.19. The total IR luminosities for all 202 U/LIRG systems were presented in \cite{GOALS} and derived using the definitions of \cite{sandersLIRGs}\footnote{$L_{IR}/L_{\odot} = 4\pi(D_L[m])^2 (F_{IR}[W m^{-2}])/3.826\times10^{26}[W m^{-2}]$ and $F_{IR} = 1.8\times 10^{-14}(13.48f_{12\mu m} + 5.16f_{25\mu m} + 2.58f_{60\mu m} + f_{100\mu m}[W m^{-2}])$}. In cases of multiple nuclei, the total L(IR) for the system is divided according to the ratio of the fluxes at 70 \micron\ for each nuclei. In a small number of cases, 70-\micron\ images are not available and so 24-\micron\ flux ratios are used instead. 

For the LIRGs in Figure \ref{lpahmerge}, the fraction of PAH-to-IR luminosity ranges from 0 to 23\% with a mean value of 8.3$\pm$5.8\%. These are similar to the mean value and range covered by the lower luminosity SINGS galaxies \citep{jdsings} and by LIRGs at slightly higher redshifts \citep[0.02 $< z <$ 0.6;][]{lagache, shipley}. With a standard deviation of nearly 6\%, however, the GOALS LIRGS are nearly evenly distributed across a wide range of PAH fractions. Even the starburst-dominated LIRGs (\eqw$\geq$0.54\micron) are found to have a standard deviation of 4.7\% about a mean PAH-to-IR luminosity of 9.5\%.

There is no linear correlation in either panel of Figure \ref{lpahmerge}, but clear trends are observed. LIRGs with larger luminosity contributions from PAHs (L(PAH)/L(IR) $>$ 10\%) cover a narrow range of IR properties (i.e. \eqw~$>$ 0.4\micron~and log(L(IR)/L$_{\odot}$) $<$ 11.5). However, LIRGs with lower luminosity contributions from PAHs (L(PAH)/L(IR) $<$ 5\%) are observed over nearly the full range of L(IR) and \eqw. This suggests that sources can have a high \eqw~and still have low L(PAH)/L(IR), but cannot have a high L(PAH)/L(IR) without a high \eqw.

The GOALS U/LIRGs in Figure \ref{lpahmerge} are also color coded by merger stage as presented in Table 1 of \cite{paperI}. Each galaxy is classified in one of five categories: nonmergers (no sign of merger activity or massive neighbors; shown in purple), pre-mergers (galaxy pairs prior to a first encounter; light blue), early-stage mergers (post-first encounter with galaxy disks still symmetric and in tact but with signs of tidal tails; green), mid-stage mergers (showing amorphous disks, tidal tails, and other signs of merger activity; yellow), or late-stage mergers (two nuclei in a common envelope; red). The 3-$\sigma$ clipped average of L(PAH)/L(IR) for each merger class is shown by a vertical dashed line. 

As shown in Figure \ref{lpahmerge}, LIRGs in an early-merger state have a slightly higher median PAH contribution (8.4\%; green line) than nonmergers (7.3\%; purple line), mid-stage mergers (6.3\%; yellow line) and late-stage mergers (6.2\%; red line) and are not found with ratios $<$3\%. Pre-mergers, or double galaxy systems prior to a first encounter (i.e. no signs of tidal tails) have the highest median PAH contribution to the IR luminosity at 12.8\% (light blue line). These galaxies may more closely resemble lower luminosity starbursting galaxies. The apparent decrease in L(PAH)/L(IR) for late stage mergers compared to pre-mergers is consistent with a higher fraction of low \eqw\ sources and an excess of IR emission not associated with star formation. As shown in Figure 8 of \cite{paperI}, the fraction of starburst-dominated LIRGs declines significantly for later merger stages while the fraction of composite sources (those with a weak AGN that does not yet dominate over star formation in the MIR) increases. Despite these slight differences in merger stage medians, galaxies at each merger stage are observed over nearly the full range of L(PAH)/L(IR).

\begin{figure}[htp]
\begin{center}
\includegraphics[height=2.4in,width=3.5in,viewport=20 0 500 345,clip]{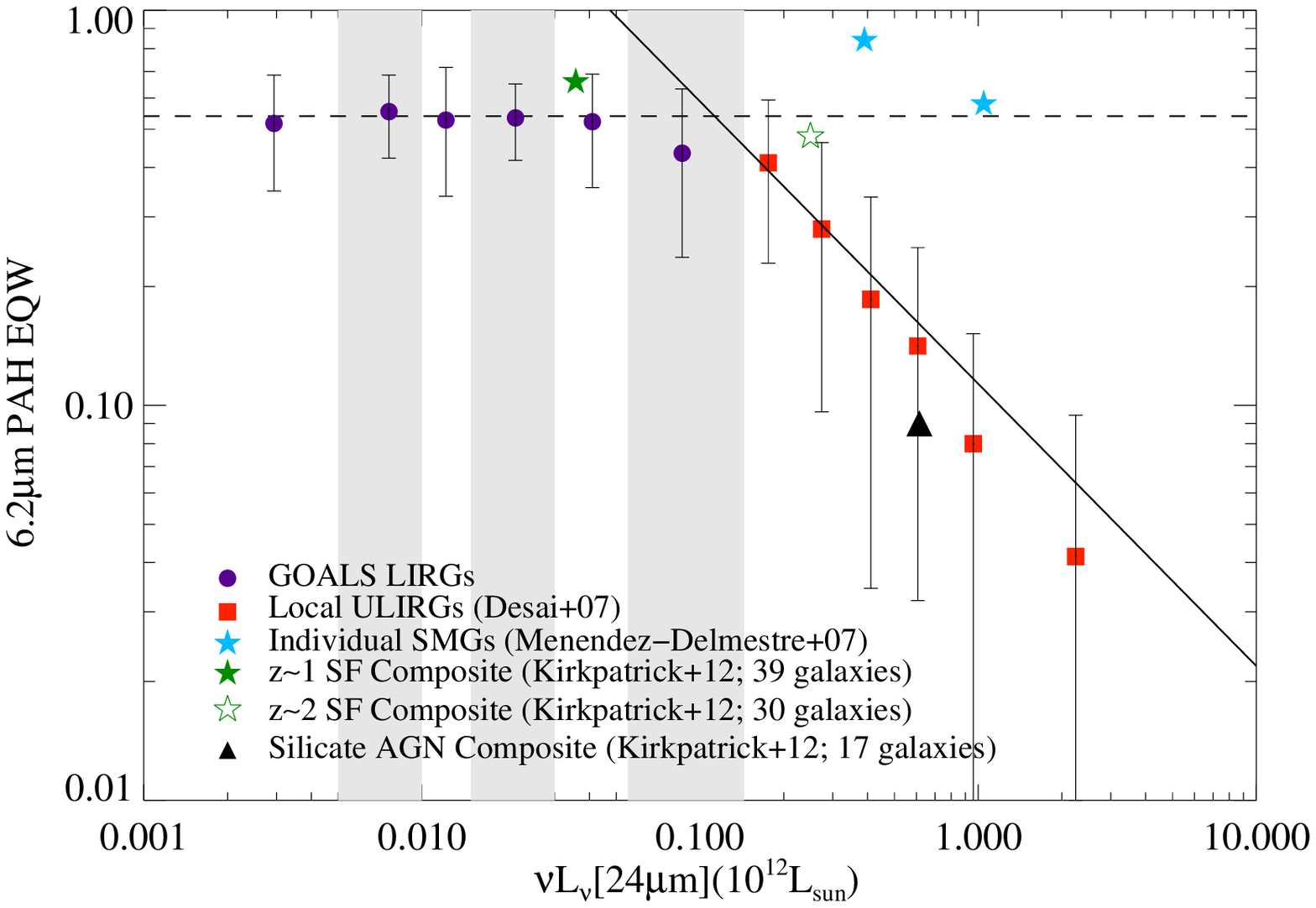}
\caption{The 6.2\micron~EQW as a function of 24-\micron~luminosity for the GOALS LIRGs (purple circles) and the ULIRGs of \cite{vandanaULIRGs} (red squares) both binned according to $\nu$L$_{\nu}$[24\micron] with a roughly equal number of galaxies per bin. The LIRG luminosity bins are indicated by the shaded regions and the ULIRG luminosity bins are shown in Figure 9 of \cite{vandanaULIRGs}. Two individual high-z SMGs from \cite{preSMGkarin} are shown as blue stars and the composite star-forming galaxies of \cite{kirk} are shown as closed (z$\sim$1) and open (z$\sim$2) green stars. The black triangle represents the composite spectrum of \cite{kirk} for AGN showing 9.7\micron\ silicate absorption. The horizontal dashed line shows the average \eqw\ observed for the lower luminosity starburst galaxies of \cite{brandlEW} and the solid line is the linear fit to the ULIRG sample from \cite{vandanaULIRGs}. 
\label{desai}}
\end{center}
\end{figure}

The GOALS ULIRGs (triangles) cover a much narrower range of L(PAH)/L(IR) (between 2 and 7\%) with a median value of 2.5$\pm$1.4\%, suggesting that PAH emission becomes substantially weaker in the ULIRGs. To illustrate this point more clearly, in Figure \ref{desai}, we plot the \eqw\ as a function of $\nu$L$_{\nu}$[24\micron] for 190 GOALS LIRGs (purple circles) and a larger sample of 107 local ULIRGs from \cite{vandanaULIRGs} (red squares; see their Fig. 9) both binned by $\nu$L$_{\nu}$[24\micron] (as indicated by the shaded regions) and then averaged in \eqw. 

From Figure \ref{desai}, we observe that over two orders of magnitude in $\nu$L$_{\nu}$(24\micron), the local LIRGs display an \eqw~of 0.5-0.7\micron, very close to the average for local starbursts (0.54\micron; dashed line) measured by \cite{brandlEW}. This suggests that for the given luminosity range, the integrated star forming properties of the galaxies as well as the conditions of the ISM and the PDRs surrounding the massive star forming regions are all very similar. Only when we reach 24-\micron\ luminosities close to 10$^{11}$L$_{\odot}$ is a decrease observed and the \eqw~of the GOALS LIRGs sample drops to a value of $\sim$0.4\micron, approaching the trend indicated by the ULIRGs of \cite{vandanaULIRGs}. 

Interestingly, higher redshift submillimeter galaxies \citep[][blue stars]{preSMGkarin} and composite star forming galaxies at z of 1 and 2 \cite[][green stars]{kirk}, do not follow this downturn. Instead we see that despite having higher rest-frame 24\micron\ luminosities than the LIRGs, their \eqw~remains high (\eqw$>$0.48\micron). This is consistent with the larger sizes and gas fractions observed for SMGs compared to local ULIRGs \citep[i.e.][]{daddiGasFrac, genzelGasFrac, elbaz}. At \eqw$=$0.09\micron, the composite spectrum from the \cite{kirk} sample representing AGN showing signs of silicate absorption at 9.7\micron\ (black triangle) is consistent with these objects having a very compact, highly obscured emitting region with a strong temperature gradient similar to the local ULIRGs, but the composite spectrum only represents a small fraction of the higher redshift galaxies in their sample.

\subsubsection{PAHs and the Starburst Main Sequence}
\begin{figure*}[htp]
\begin{center}
\includegraphics[height=3in,width=7in,viewport= 20 0 500 200,clip]{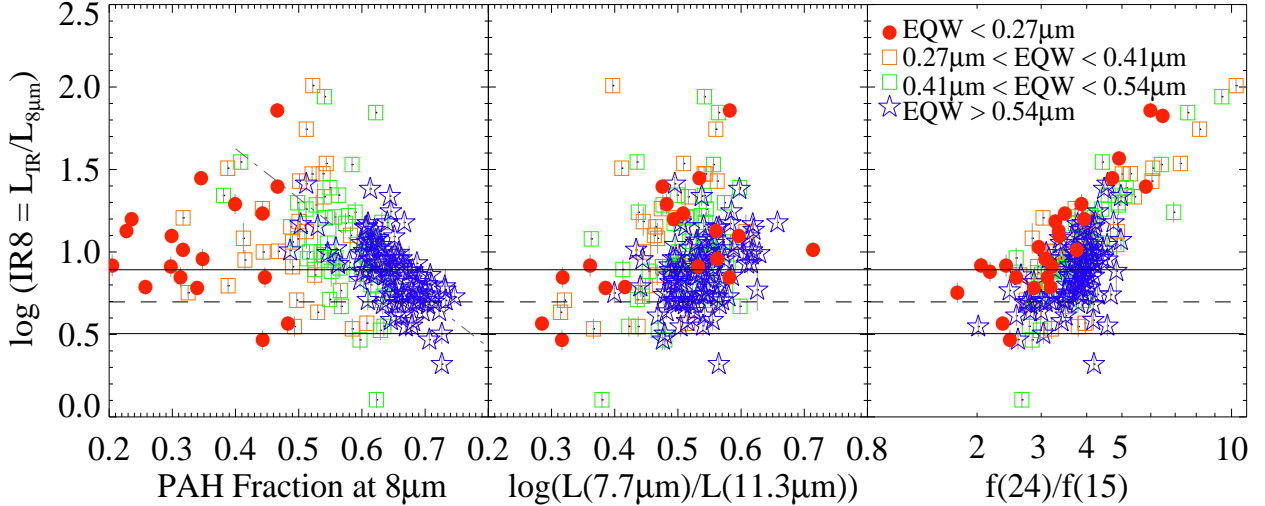}
\caption{IR8 $=$ L$_{IR}$/L$_{8\mu m}$ (where L$_{8\mu m}$ is the broadband 8\micron\ luminosity) as a function of $left$: PAH contribution to the IR luminosity at 8\micron, $middle$: the ratio of the 7.7\micron\ and 11.3\micron\ PAH features and $right$: the continuum flux ratio f(24\micron)/f(15\micron) for 203 GOALS galaxies. The main sequence followed by normal, star-forming galaxies ($\pm$1$\sigma$) is shown as dashed (solid) lines. A rough correlation for sources with \eqw$\geq$0.41\micron~(shown by the dashed-dotted line) is observed between the PAH contribution to the flux at 8\micron\ and IR8 (left), although IR8 increases by nearly an order of magnitude while the PAH fraction decreases by only a factor of $\sim$1.4 (left panel). There is no link between IR8 and the grain size distribution or ionization state (middle panel), while IR8 increases with increasing f(24\micron)/f(15\micron) at all equivalent widths. 
\label{ir8}}
\end{center}
\end{figure*}

Recent investigations have revealed that galaxies with high specific star formation rates (i.e. starbursts) separate themselves from a so-called main sequence of normal, star-forming galaxies in the ratio of total IR luminosity to rest-frame 8\micron\ luminosity, IR8 $=$L$_{IR}$/L$_{8\mu m}$ \citep{daddiMainSeq, elbaz}. In particular, those galaxies with the most intense, compact starbursts have higher values of IR8 \citep[][see also \cite{tanio1} for a discussion of how although the range of distances, and thus spatial extents probed by the slit, may contribute in a few specific cases, it is clearly not the dominant effect for the GOALS sample]{elbaz}. As shown in Figure \ref{ir8}, most GOALS galaxies are found above the star-forming main sequence (dashed line at log(IR8) $\sim$ 0.7). There is an inverse correlation (dashed-dotted line in the left panel) between IR8 and the PAH fraction at 8\micron\footnote{We calculate the PAH fraction at 8\micron~as: FRAC$_{PAH}$[8\micron] $=$ (f$_{\nu}^{obs}$(7.7\micron\ PAH) $+$ f$_{\nu}^{obs}$(8\micron\ PAH))/f$_{\nu}^{IRAC}$(8\micron) where the PAH fluxes are not corrected for extinction and f$_{\nu}^{IRAC}$(8\micron) is the total 8\micron~flux within the IRS slit derived from convolving the IRAC 8-\micron~filter with the low resolution IRS spectrum.} for sources with \eqw$>$0.41\micron~which suggests that compact, star-bursting LIRGs with high specific star formation rates have less PAH emission with respect to their IR emission than do star-forming galaxies on the main sequence. This relation is parameterized by $log(IR8) = 2.8 (\pm .3) - 3 (\pm .5) \times FRAC_{PAH}[8\mu m]$. LIRGs likely to harbor an AGN (i.e. those with \eqw$<$0.41\micron; red circles \& orange squares) show no correlation between IR8 and the contribution from PAH emission to the total flux. There is no obvious link between IR8 and the 7.7-\micron\ to 11.3\micron\ PAH ratio (middle panel), but IR8 increases with the continuum flux ratio f(24\micron)/f(15\micron) (right panel). 

Taken together, the three panels in Figure \ref{ir8} suggest that destruction of smaller dust grains is not the cause of the higher IR8 observed in more compact starburst systems. Instead, since the grain size or ionization distribution is not changing but the overall flux contribution from the PAHs is decreasing for the higher \eqw\ sources, there instead is likely less PDR emission relative to the emission in the IR. The increasing L(IR) shifts the peak emission to warmer dust temperatures which results in an increase in the emission at 24\micron~relative to that at 15\micron. A relative decrease in PDR emission is consistent with the results of \cite{inamihires} who saw no correlation between IR8 and the hardness of the radiation field as measured via emission lines from the ionized gas and of \cite{tanioCII} who observed a correlation between [CII]/FIR, the average dust temperature, and the luminosity surface density of the associated MIR emitting region. 

A recent analysis \citep{magdis} of the Herschel and {\it{Spitzer}} data for the 5MUSES, 24-\micron~selected sample of galaxies \citep{5muses} finds an anti-correlation between IR8 and the 6.2-\micron~PAH equivalent width among star-forming galaxies and ascribes the change to variations in the PAH emission.  We see a similar anti-correlation of IR8 with the fraction of PAH emission at 8\micron. However, we see no correlation of IR8 with the PAH band ratios, and the relatively small change in the PAH fraction at 8\micron\ (a factor of 1.4) cannot account for the rise of 5-10 times seen in IR8 among starburst-dominated LIRGs in Figure \ref{ir8}.  This suggests that while the PAH emission decreases relative to the MIR continuum (less relative PDR emission), the change in IR8 is dominated by a rise in the IR emission, driven mostly by the FIR (leading to increased f(24)/f(15) as shown in the right panel of Figure \ref{ir8}), and not an overall change in the  properties of the small grains.

\subsection{Sources of PAH Obscuration}\label{obscured}

\subsubsection{Silicate Absorption at 9.7 \& 18.5\micron} \label{bfit}
\begin{figure*}[htp]
\begin{center}
\includegraphics[height=5.7in,width=3.2in]{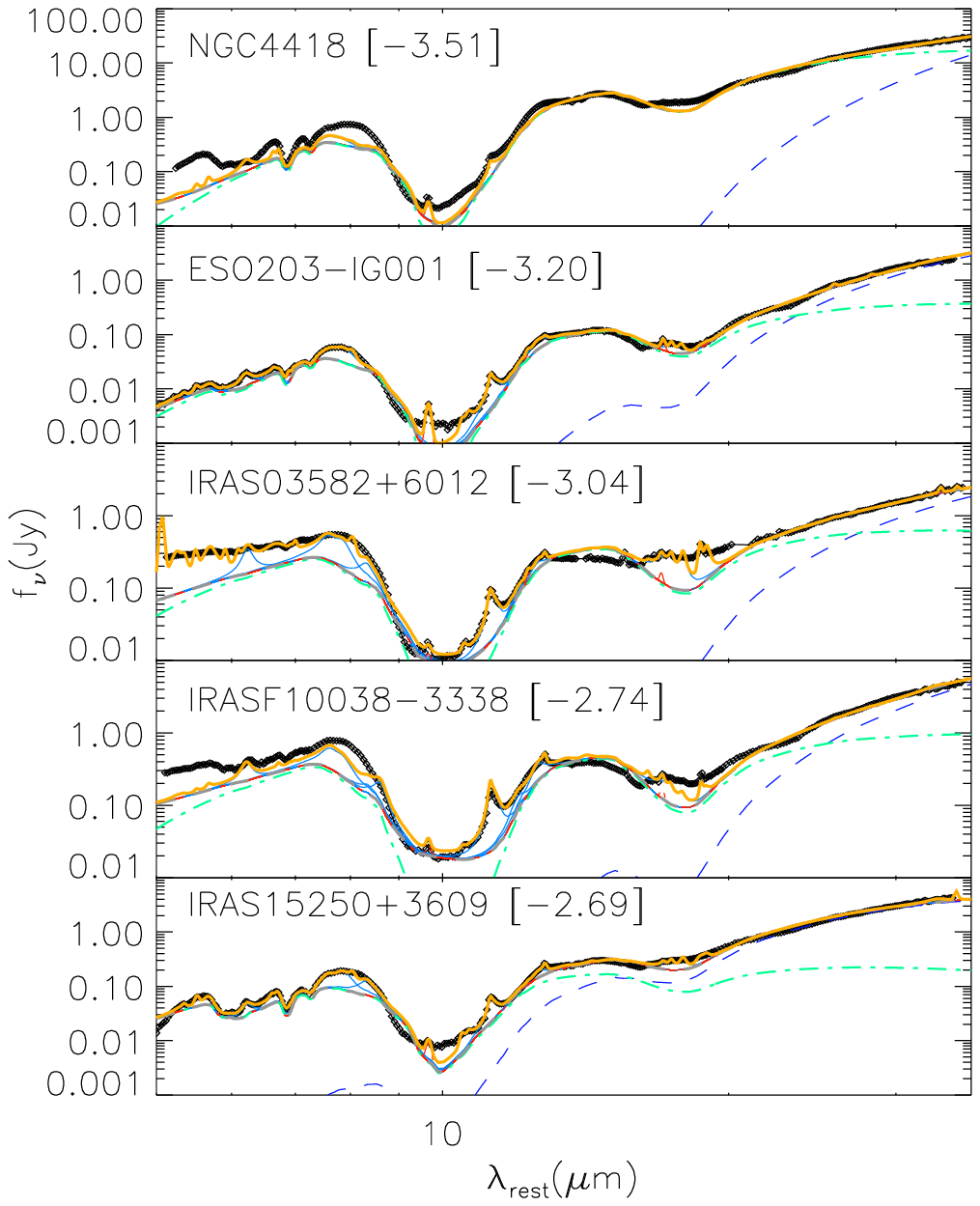}
\hspace{-.7cm}
\includegraphics[height=5.7in,width=3.2in]{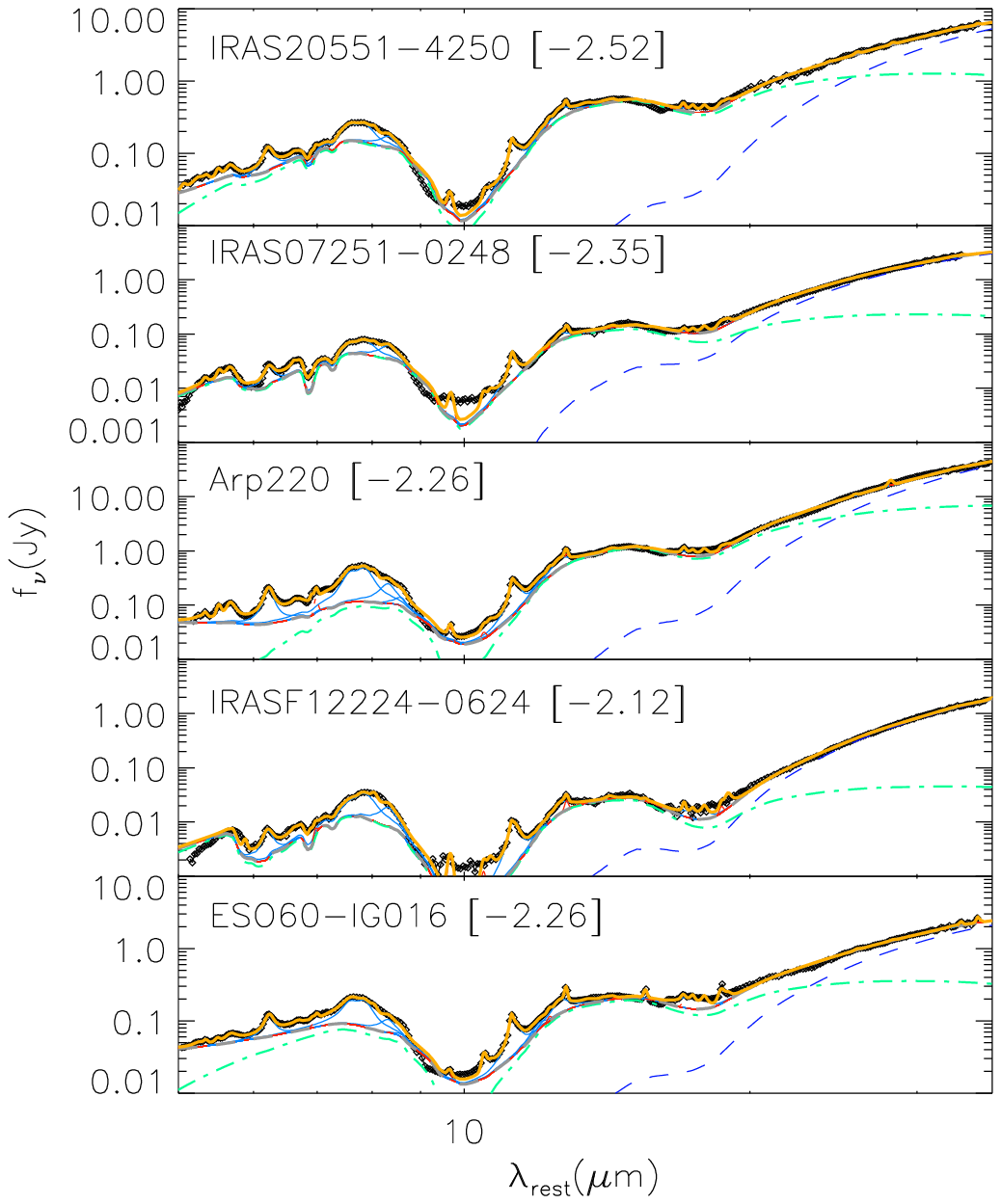}
\caption{Low Resolution IRS Spectra with CAFE spectral decomposition results for
  10 GOALS galaxies with the deepest 9.7-\micron~silicate absorption. Together
  with IRAS08572+3915 (shown in Figure \ref{repspecs}), these 11 sources make up
  the entirety of the GOALS sample with $s_{9.7\mu m} <$ -1.75. 
  The overall fitted model is shown in yellow and model components are
  color-coded as in Figure \ref{repspecs}. The $s_{9.7\mu m}$ value is given after each source name in brackets.
\label{deep}}
\end{center}
\end{figure*}

The Si-O stretching and O-Si-O bending resonances of silicate dust
grains produce broad absorption features in the MIR at 9.7\micron~and
18.5\micron. Due to the large widths of these features (often
spanning $>$1\micron), the silicate grains producing them are likely amorphous.
The distribution of silicate
strengths at 9.7\micron~($s_{9.7\mu m}$) for the entire GOALS sample
and the relation of $s_{9.7\mu m}$ to other
MIR properties were presented in \cite{paperI}. We summarize those results here by noting that  -3.58 $< s_{9.7\mu m} < $0.52 for the GOALS sample and that the GOALS ULIRGs
are on average more highly obscured than the LIRGs with $\langle{s_{9.7\mu m}}\rangle_{ULIRGs}$ = -1.28 and
$\langle{s_{9.7\mu m}}\rangle_{LIRGs}$ = -0.34. 
The MIR spectra for the most heavily obscured GOALS galaxies are given
in Figure \ref{deep}. Together with IRAS08572+3915 (shown in Figure
\ref{repspecs}), these 11 sources make up the the entirety of the GOALS
sample with $s_{9.7\mu m} <$ -1.75. 

For two of these LIRGs, IRAS03582+6012\_E and
ESO374-IG032, only a poor fit can be made at $\lambda
\lesssim$ 10\micron~and near 10\micron~and 18\micron~where the deep
silicate absorption features are present. In each case, the fit produced from the dust model assumed by CAFE falls below the observed spectrum in these regions suggesting these sources have an excess of hot dust emission contributing 
below 10\micron. These sources also show the shallowest MIR slopes of
the entire sample at F$_{\nu}$[30\micron]/F$_{\nu}$[5\micron] = 9.55
and F$_{\nu}$[30\micron]/F$_{\nu}$[5\micron] = 5.14, respectively. 
ESO374-IG032 hosts an hydroxyl (OH) megamaser
\citep{OHMkazes, OHMdarling} which
are often indicators of major mergers and large amounts of very dense
molecular gas in LIRGs \citep{OHMbaan}.

\subsubsection{Crystalline Silicates at 23\micron}\label{crystals}

In a few of the heavily obscured sources (\sil\ $<$ -1.75), CAFE produces a good quality fit to most of the MIR spectrum, but the detailed structure between 8-10\micron~or between 18-20\micron~is not well-represented. A different dust model, specifically one that includes the presence of crystalline silicates, might be necessary to explain the detailed spectral structure in these obscured sources. Crystalline silicates appear as absorption features at 11, 16, 19, 23, and 28\micron. The overlap that occurs with the stronger, broader absorption features at 9.7 and 18.5\micron~due to amorphous silicate grains affects the shape and depth of the broad absorption features which then cannot be reproduced well by CAFE. Crystalline silicates, specifically the Mg$_2$SiO$_4$ molecule, have been shown to be important in the heavily obscured, late merger stage of ULIRGs \citep{spoonULIRGs}. The presence of crystalline dust grains indicates ongoing star formation, specifically in the cooler outer regions since they are observed in absorption, and, given the strength of the features relative to those produced by amorphous silicates, \cite{spoonULIRGs} suggest the crystalline silicates observed in the heavily obscured ULIRGs are produced by massive stars on short timescales. 

\begin{figure*}[htp]
\begin{center}
\includegraphics[height=5.7in,width=3.2in]{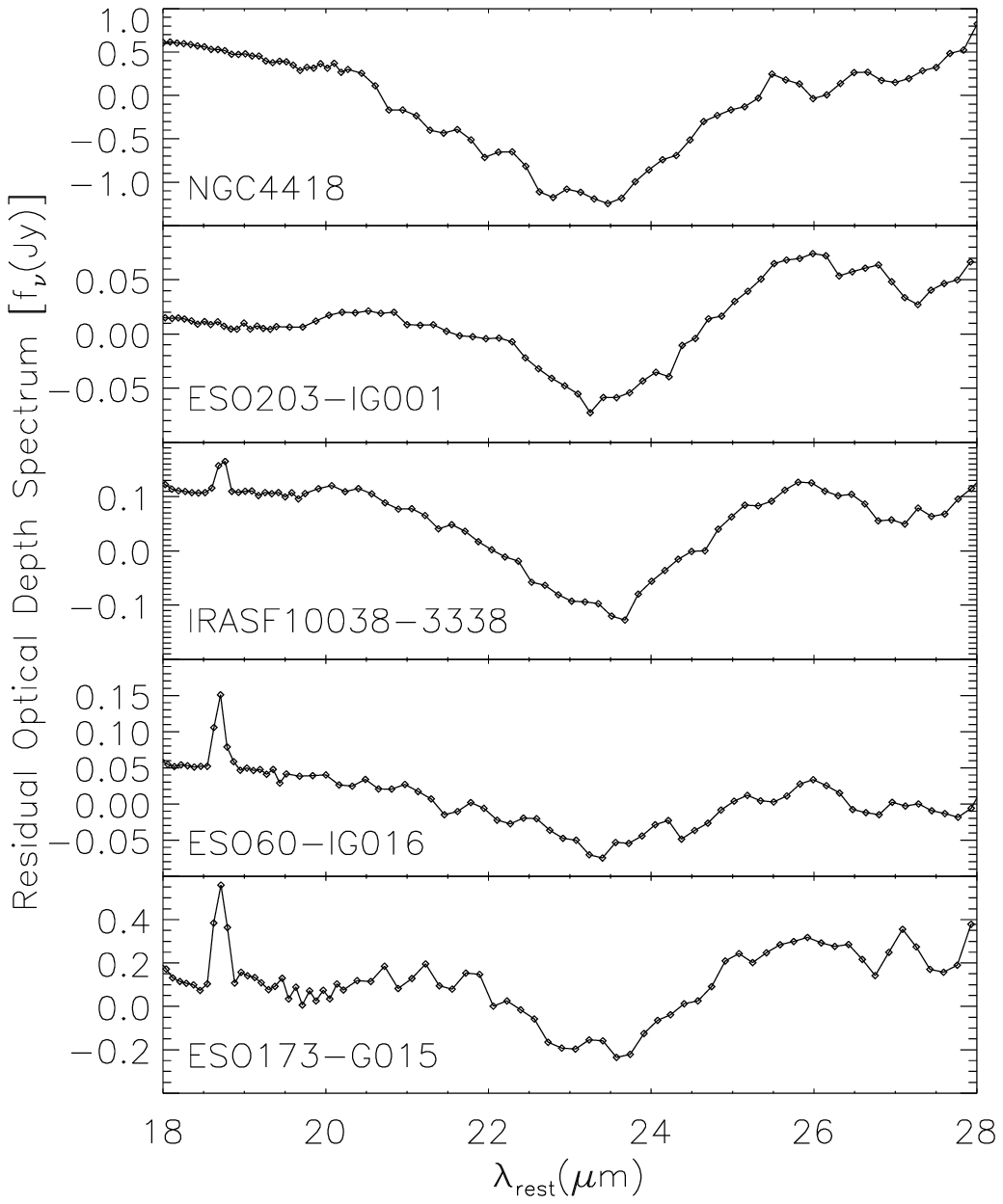}
\hspace{-.6cm}
\includegraphics[height=5.7in,width=3.2in]{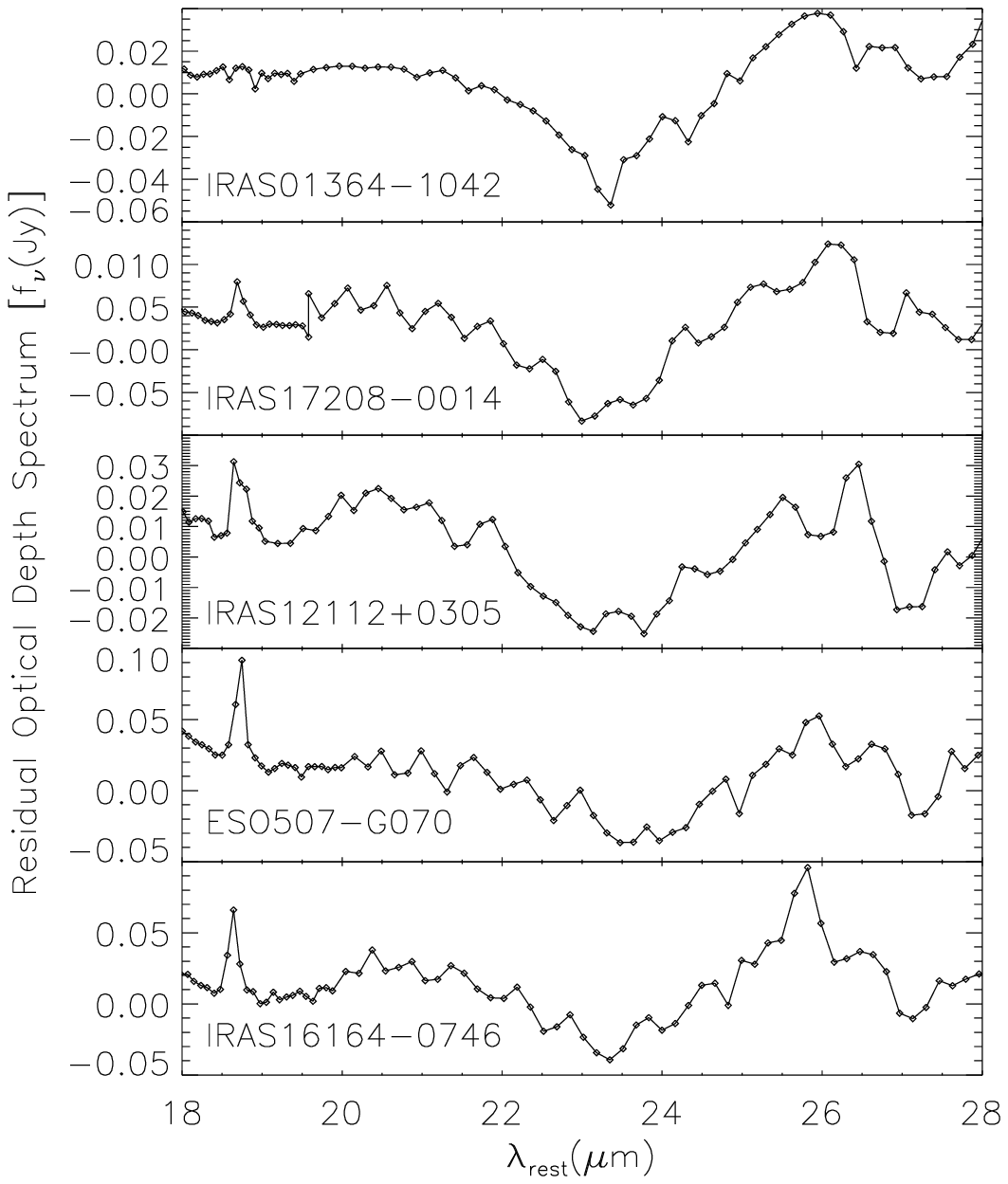}
\caption{Residual Optical Depth spectra from 18-28\micron\ for the 8 LIRGs and 2 of the 6 ULIRGs for which a crystalline silicate feature at 23\micron\ is observed in the low resolution spectra in order of decreasing silicate strength, \sil. (The remaining 4 ULIRGs for which the feature is also detected, Arp220, IRAS08572+3915, IRAS15250+3609, and IRAS20551-4250, were included in the \cite{spoonULIRGs} high resolution crystalline silicate study and so are not reproduced here.) 
\label{crystspecs}}
\end{center}
\end{figure*}

\begin{figure}[htp]
\begin{center}
\includegraphics[height=2.4in,width=3.5in,viewport=20 0 500 345,clip]{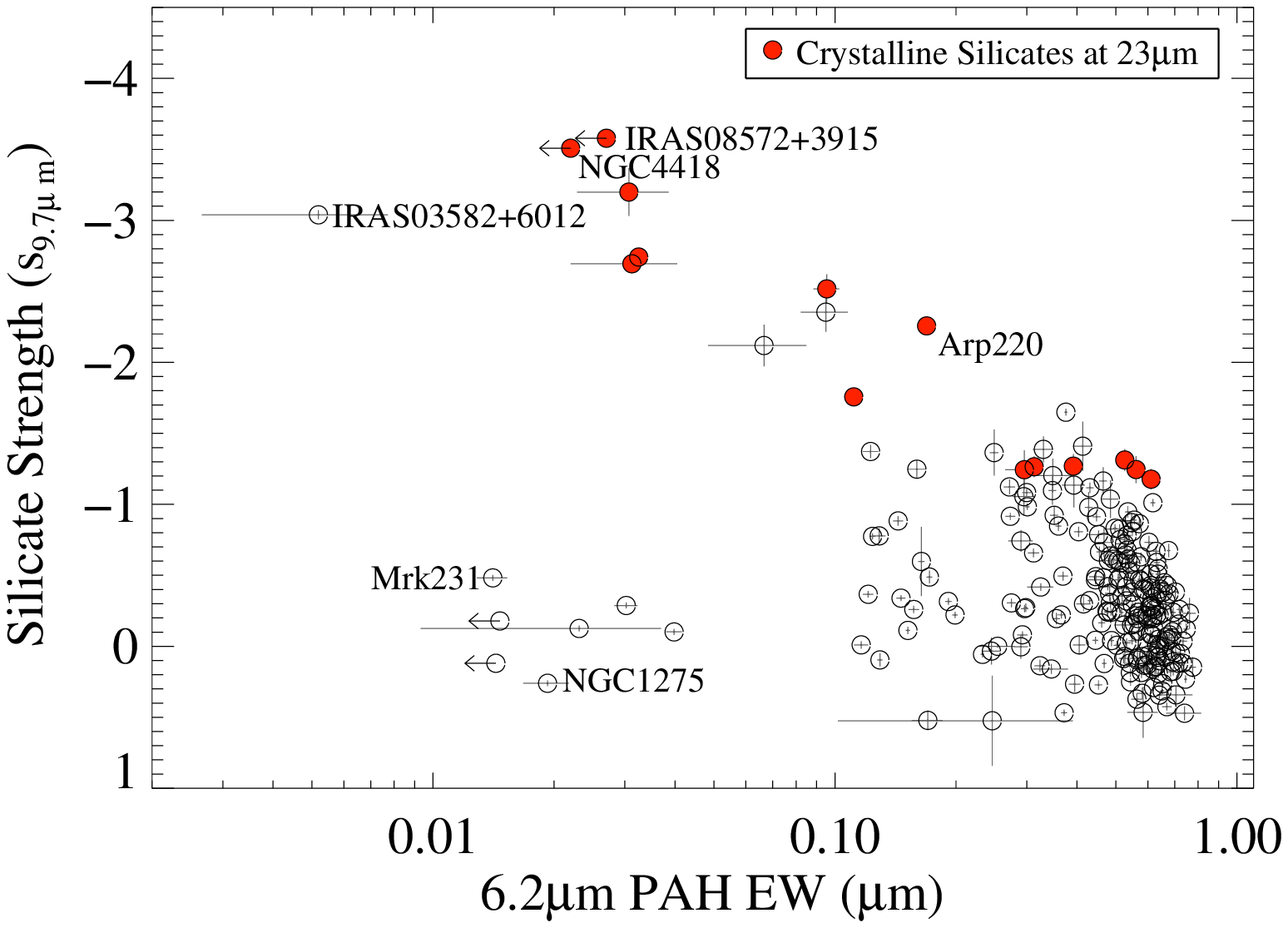}
\caption{Equivalent width of the 6.2\micron\ PAH versus silicate strength for 239 GOALS U/LIRGs. Crystalline silicates are rare among the entire GOALS sample but comprise a large fraction (70\%) of those sources that are heavily obscured.
\label{crystspoon}}
\end{center}
\end{figure}

The crystalline silicate feature at 23\micron\ is the feature most clearly seen at the resolutions of the SL and LL IRS modules and is detected in 8 LIRGs and 6 ULIRGs of the GOALS sample. Four of these six ULIRGs (Arp220, IRAS08572+3915, IRAS15250+3609, and IRAS20551-4250) were included in the higher resolution \cite{spoonULIRGs} study of 12 ULIRGs for which crystalline silicate features at 23\micron\ as well as at 11, 16, 19, and 28\micron\ were reported. In Figure \ref{crystspecs}, we plot the residual optical depth spectra for the remaining 2 ULIRGs and 8 additional LIRGs for which we detect the 23\micron\ crystalline silicate feature. We consider the crystalline silicates at 23\micron\ detected if the feature is $\gtrsim$3$\sigma$ in the residual optical depth spectrum (i.e. the original source spectrum with the CAFE fits to the amorphous cool and warm dust components removed). In Figure \ref{crystspoon}, we show that while $\sim$6\% of the overall GOALS sample is detected in crystalline silicates at 23\micron, all of these detections have \sil\ $<$ -1.24, and thus $\sim$70\% of these more obscured LIRGs and ULIRGs are detected in crystalline silicates over a range in \eqw.

\subsubsection{Ice Absorption at 6.0\micron}\label{ice}

Another form of opacity arises near 6.0\micron~in the form of water
ice absorption. The first extragalactic detections of this absorption feature were in the LIRG NGC4418 \citep[][also in the GOALS
sample]{spoon4418} and in the ULIRG IRAS00183-7111 \citep{tranice}. 
The application of CAFE
allows for the untangling of even weak to moderate ices from the
nearby 6.2-\micron~PAH feature and continuum, as well as for the strongest ice absorption features. To produce an accurate fit, 37 GOALS galaxies (15\%) require the inclusion of an ice absorption component at 6.0\micron~and the measured $\tau_{ice}$ values are given in Table \ref{pahtable}. Thirteen of these sources are ULIRGs for a detection rate of 56.5\% among ULIRGs, similar to the rate found for the ULIRGs in an ISO study of a sample of 103 IR galaxies \citep{spoonices}. The remaining 24 water ice detections were found in LIRGs for a detection rate of 10.7\% among LIRGs which is comparable to that observed for the lower luminosity starbursts \citep{brandlEW}. 

\begin{figure}[h!]
\begin{center}
\includegraphics[height=1.5in,width=3.5in,viewport=10 0 500 225,clip]{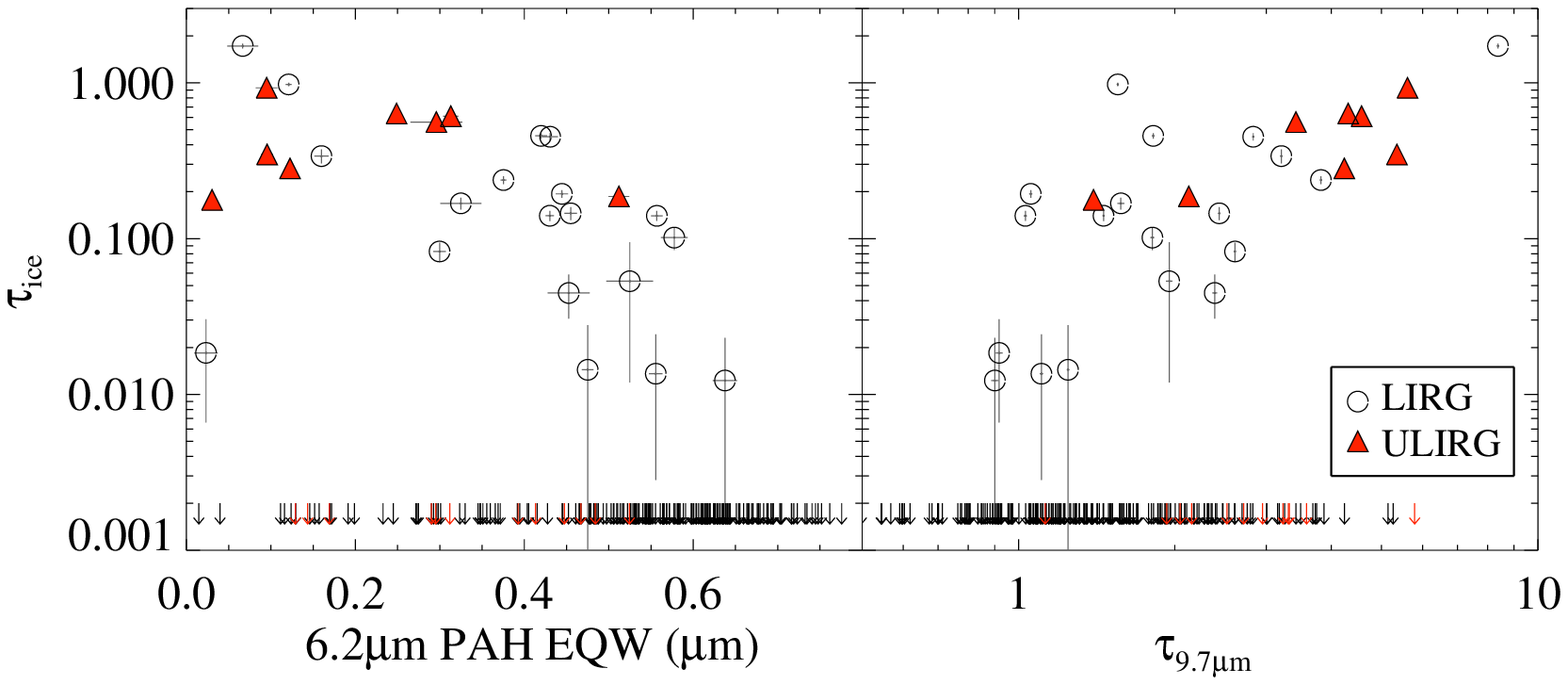}
\caption{Water ice optical depth at 6.0\micron~as a function of 6.2$\mu$m PAH EQW (left panel) and silicate optical depth at 9.7\micron~(right panel) for 227 GOALS LIRGs (open circles) and ULIRGs (red triangles). LIRGs (ULIRGs) with no ice detection are represented by black (red) arrows. Starburst galaxies with EQW$_{6.2\mu m} > $0.7$\mu$m do not show water ice absorption. Although there is a general trend for icier sources to also be dustier, even the dustiest sources can still have no ice features.
\label{icefig}}
\end{center}
\end{figure}

Galactic observations of water ice absorption features are usually attributed to icy mantles of frozen water molecules surrounding silicate dust grains \citep{galaxyice1, galaxyice2}. Such a scenario would suggest a tight relationship between $\tau_{ice}$ and $\tau_{9.7\mu m}$. As shown for the GOALS LIRGs (open circles) and ULIRGs (red triangles) in the right panel of Figure \ref{icefig}, among those galaxies that show ice absorption features, there is a general trend for icier sources to also be dustier. However, even the dustiest sources can still have no ice features.

Since the absorption feature at 6.0\micron~usually overlaps with the 6.2-\micron~PAH emission feature, the equivalent width of the PAH feature must be measured slightly differently in icy sources. As described in \cite{paperI}, for the GOALS sources, the ice absorption is assumed to affect the underlying continuum but not the PAH emission, and the EW$_{6.2\mu m}$ is calculated using the absorption-corrected continuum following the method described in \cite{spoon}. Therefore, a strong ice feature can lower the apparent continuum and thus artificially raise the \eqw, in some cases by 10-20\%. The optical depth of the water ice absorption feature $\tau_{ice}$ is plotted as a function of 6.2$\mu$m PAH EQW in the left panel of Figure \ref{icefig}. Of the 24 LIRGs (open circles) where ices are present, only four have EQW$_{6.2\mu m} \geq $0.54\micron~and no starburst galaxies with EQW$_{6.2\mu m} > $0.7\micron~show any water ice absorption at all. There is a general trend for the sources with the strongest ice absorption to have the smallest \eqw. Despite composing a much smaller fraction of the GOALS sample, LIRGs with low EQW$_{6.2\mu m} < $0.27\micron~(i.e. LIRGs likely dominated by an AGN) make up 1/3 (8 out of 24) of the LIRGs with water ice detections. However, the majority of the iciest sources are still ULIRGs (red triangles).

Together, both panels in Figure \ref{icefig} suggest that a special set of conditions are required for ices to survive. The correlation between $\tau_{9.7\mu m}$ and $\tau_{ice}$ for icy sources supports the idea that ices form as mantles around silicate dust grains. Silicates tend to build up in the nuclei of merging LIRG and ULIRG systems as the merger progresses \citep{paperI} but so may the contribution from an AGN which can dominate the MIR emission and heat the surrounding dust so that ices cannot form. Thus a galaxy must have enough silicate dust to support the existence of icy mantles, but the effects of its AGN must also be weak enough to keep the surrounding dust cool (70\micron/24\micron~$\lesssim$ 7) to allow for ices. These conditions are not typically met in starbursting LIRGs that are not heavily obscured. In ULIRGs, however, there is typically a significant amount of molecular hydrogen gas and dust that has been funneled into the galaxy core allowing for self-shielding by the silicate grains from the influence of any present AGN. Thus the detection rate of ices is much higher among ULIRGs \citep[as was observed for ISO spectra by][]{spoonices} and composite LIRGs (0.27\micron~$\leq EQW_{6.2\mu m} <$ 0.54\micron).

We also note that in an analysis of the dust features in the ULIRG IRAS08572+3915, \cite{dartois} present compelling evidence that part or all of the absorption in the 5.7-7\micron~range is not associated with water ice absorption but instead originates in carbonaceous materials associated with the absorption features seen at 6.85 and 7.25\micron. This type of absorption may dominate over that due to ices in sources like IRAS08572+3915 and ESO374-IG032, discussed in Section \ref{bfit}, given their warm SEDs.

\subsection{Comparison of PAHs with UV Properties}\label{secUV}

While PAH emission can be used as an effective proxy for ongoing star formation because it traces the PDRs and is relatively immune to the effects of extinction, a more direct tracer of young/moderate age stars (A to OB type) is the detection of UV emission from their photospheres. Empirical starburst reddening relations have been observed for normal, lower luminosity galaxies \citep{cortese}, for starbursts \citep[L$_{IR} <$10$^{11}$L$_{\odot}$; ][]{meurersbs}, and for UV-selected galaxies \citep{irxb}. \cite{charlotfall} suggest that the relation between IR excess (IRX=log(FIR/FUV)) and UV slope ($\beta$) is a sequence of effective optical depth for star forming late-type galaxies. As a consequence, one can estimate the IR luminosity of a source, albeit with considerable uncertainty, from rest frame UV observations even when little is known about the intrinsic dust attenuation, as is often the case for dusty galaxies at high redshift \citep[i.e.][]{reddy}. As part
of the larger GOALS legacy project, 135 LIRGs were observed by GALEX and compared to {\it{Spitzer}} imaging to determine whether or not such dusty galaxies also followed a similar starburst reddening relation. \cite{GALEX} found that LIRGs cover a wide range in IRX and $\beta$, with many sources falling between normal star-forming galaxies and ULIRGs, which typically have very high IRX and lie off the normal starburst IRX-$\beta$ relation \citep{meurersbs}. In these LIRGs, the UV emission is likely "decoupled" from the source of most of the IR light. In addition, LIRGs can scatter to high $\beta$ at relatively low IRX, and in these sources the red UV colors may indicate significant populations of older stars that dominate the GALEX bands.

\begin{figure}[htp]
\begin{center}
\includegraphics[height=2.6in,width=3.2in]{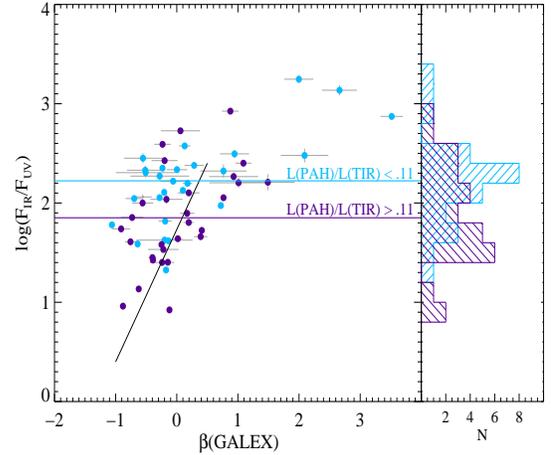}
\includegraphics[height=2.6in,width=3.2in]{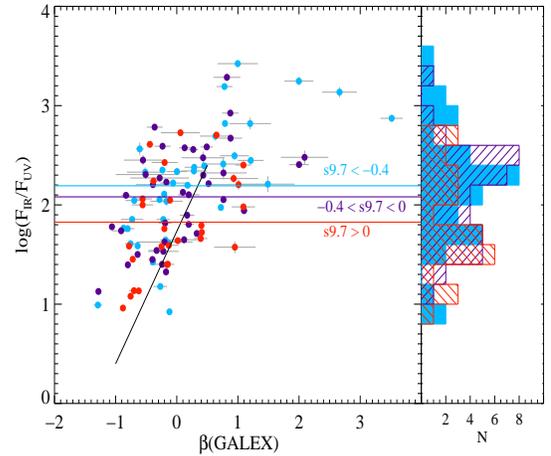}
\caption{Infrared excess (IRX; log(F$_{IR}$/F$_{UV}$)) versus UV
  spectral slope $\beta$ color coded by L(PAH)/L(IR) (left panel; 57 GOALS galaxies) and by silicate
  strength (right panel; 106 GOALS galaxies). Histogram insets show the range of IRX values
  represented by each color and the average for each colored bin is
  marked by a solid horizontal line. Sources with low L(PAH)/L(IR) and higher
  silicate optical depths tend to have higher IRX on average but no real correlation is observed. The black solid line marks the relation
  followed by starbursts \citep{meurersbs}.
\label{IRXbetaplot}}
\end{center}
\end{figure}

The IRX-$\beta$ relation determined by \cite{GALEX} is re-plotted in Figure \ref{IRXbetaplot} but here we color code the galaxies by L(PAH)/L(IR) (left panel) and by \sil~(right panel). The average IRX for each colored bin is marked by a horizontal line. On average, more obscured LIRGs with PAH emission that contributes $<$10\% to the total LIR tend to have a higher IRX than LIRGs with less silicate obscuration and more significant PAH emission. However, the mean values are all consistent within a standard deviation, and LIRGs at all silicate depths and L(PAH)/L(IR) are observed to cover the full range in IRX. 

\begin{figure}[ht]
\begin{center}
\includegraphics[height=2.25in,width=3in]{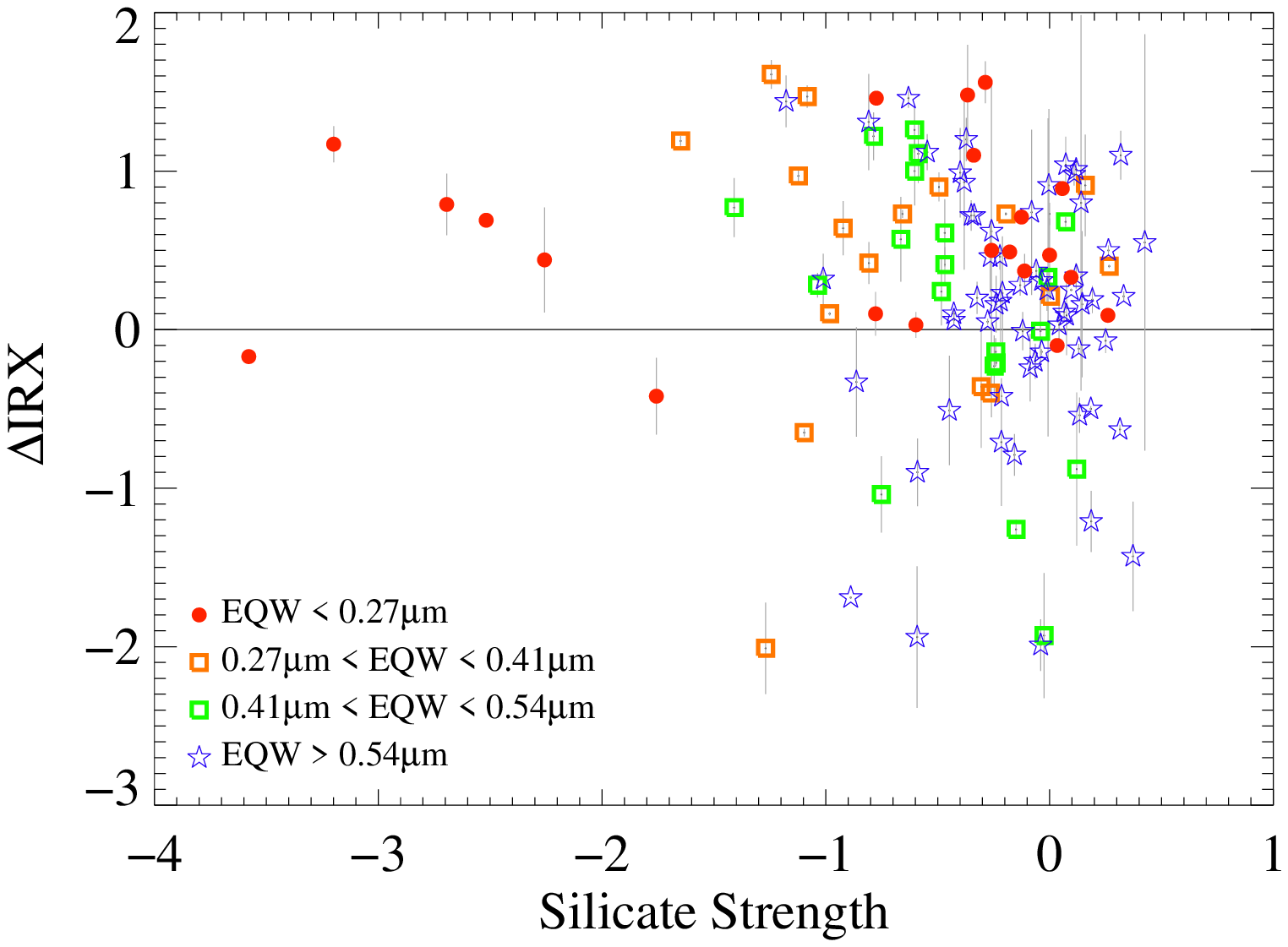}
\includegraphics[height=2.25in,width=3in]{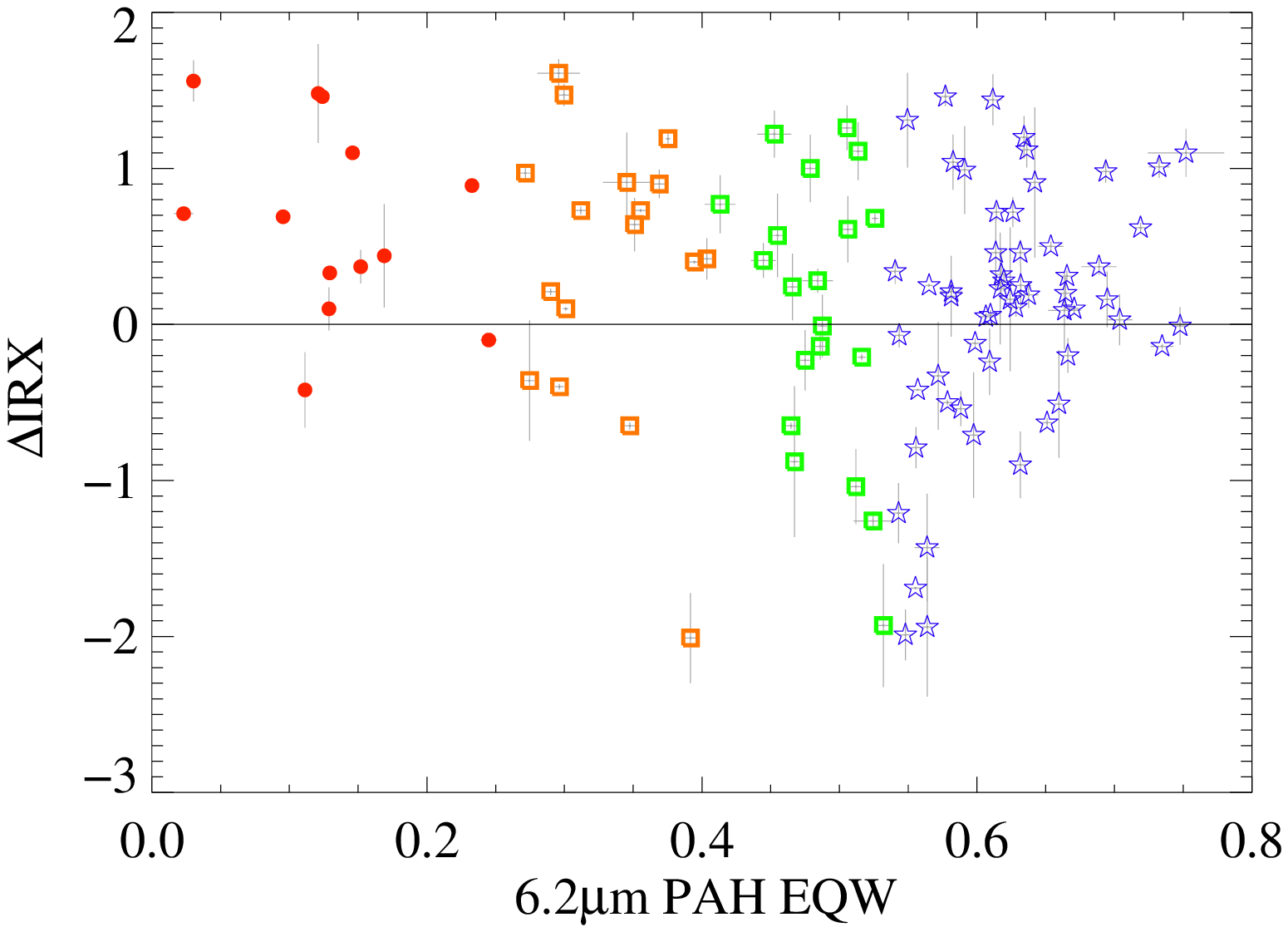}
\caption{The effects of obscuring silicate dust ($s_{9.7\mu m}$; left panel) and
  the presence of a weak AGN (as represented by EQW$_{6.2\mu m}$; right
  panel) on the decoupling of IR and UV emission ($\Delta$IRX) in
  106 GOALS (U)LIRGs. $\Delta$IRX represents the deviation of each galaxy from the
  IRX-$\beta$ relation derived for lower luminosity
  starbursts (solid line in Figure \ref{IRXbetaplot}). 
\label{dirx}}
\end{center}
\end{figure}

In Figure \ref{dirx}, we examine directly whether the amount of obscuring silicate dust or the presence of a weak AGN contributes to the observed decoupling of the IR and UV emission by comparing $\Delta$IRX to $s_{9.7\mu m}$ and EQW$_{6.2\mu m}$. Most (22 of 29) LIRGs with EQW$_{6.2\mu m} < $0.3\micron, including those that are highly obscured ($s_{9.7\mu m} < $-1.75), have positive $\Delta$IRX. \cite{GALEX} also saw a positive $\Delta$IRX for the GOALS LIRGs observed to have IRAC colors indicative of the presence of an AGN. However, as also shown in Figure \ref{dirx}, there are starbursting LIRGs (EQW$_{6.2\mu m} \geq $0.54) with no signs of an AGN and completely unobscured LIRGs ($s_{9.7\mu m} > 0$) that deviate just as much from the IRX-$\beta$ starburst relation as their lower EQW, more obscured counterparts. Thus high IRX is not exclusively associated with low \eqw.

\section{Discussion}\label{discussion}

\subsection{Warm Molecular Hydrogen in LIRGs}\label{miniH}

The rotational lines of warm molecular hydrogen (H$_2$) have three main
sources of excitation in luminous IR galaxies: UV photons from young stars, AGN via UV, soft, or hard X-ray photons, and
shocks \citep[see for example][]{higdon, ogle, pierre}. When far-UV photons emitted by massive stars
are absorbed by dust grains, the resulting de-excitation produces PAH emission, while the subsequent release of
photoelectrons heats the surrounding gas leading to the emission of H$_2$ lines. Shocks arising from colliding molecular clouds, molecular gas outflows, or supernova remnants, as well as X-ray emission from an AGN, can heat and ionize any surrounding H$_2$, but do not excite the PAH molecules found in PDRs. Thus linking warm H$_2$ and PAH emission can help determine the mechanism responsible for the H$_2$ lines observed in the MIR.

Several of the lowest energy pure-rotational transitions in the H$_2$ molecule
are observed in the MIR between 1-30\micron. After eliminating the
ten spectra for which firm detections are not possible (5 with off-center SL slit placements, 4 without
complete low resolution IRS data, and 1 which saturates the spectrograph, see Section \ref{recap}), 97.5\% of GOALS galaxies are detected in at least one of the four lowest transitions (H$_2$S(0) at 28.2\micron, H$_2$S(1) at 17.0\micron, H$_2$S(2) at 12.3\micron, and H$_2$S(3) at 9.7\micron). The H$_2$S(3) line which
falls directly in the middle of the 9.7-\micron~silicate absorption
feature has the highest detection rate and is observed in $\sim$90\% of our sources. The
H$_2$S(0) line has the lowest detection rate (in part due to the lower sensitivity of the LL1 module compared to SL) and is observed in only 6\% of the sample. Line fluxes (and the associated uncertainties) derived from the low resolution spectra for the seven lowest H$_2$ transitions (H$_2$S(0) to H$_2$S(7)) were given in Table \ref{H2table}. Upper limits are set at 3$\sigma$ of the residual after fits to the dust continuum and PAH emission features are subtracted. For the determination of fluxes and upper limits for sources without adequate fits from CAFE due to an overall lack of spectral features in the MIR (Section \ref{recap}), an unusually shallow spectral slope (Section \ref{bfit}), or crystalline silicates (Section \ref{crystals}), the continuum is determined via a spline fit to the spectrum covering 0.5\AA~on either side of the line. These sources are all marked with an asterisk in Table \ref{H2table}.

\subsubsection{Linking PAHs \& Warm H$_2$}
\begin{figure*}[htp]
\begin{center}
\includegraphics[height=3in,width=4in]{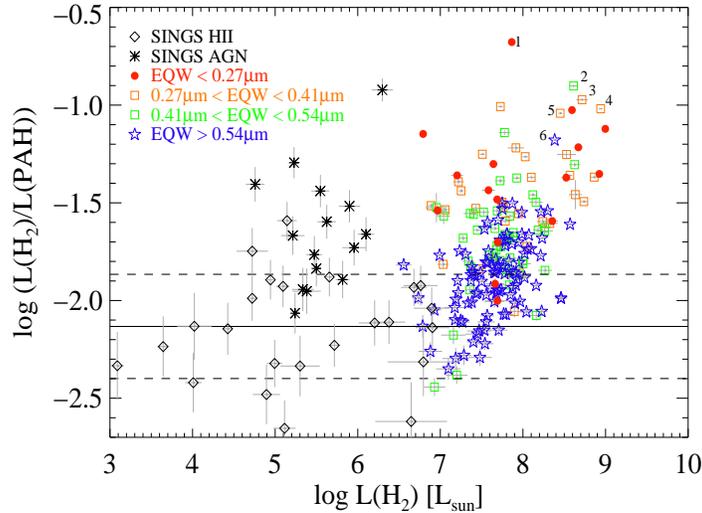}
\caption{PAHs and H$_2$: The ratio of L(H$_2$)/L(PAH) uncorrected for extinction versus
  L(H$_2$) for 196 GOALS galaxies (color-coded by EQW$_{6.2\mu m}$) and for lower luminosity SINGS galaxies (black diamonds and asterisks). L(PAH) is represented by the sum of the emission of the PAH features at 7.7 and 8.6\micron~(see footnote). L(H$_2$) is determined from the three
lowest transitions (H$_2$(S0-S2)). A clear trend of increasing luminosity ratio with increasing L(H$_2$)
is seen for the GOALS sample. The majority of the GOALS LIRGs have L(H$_2$)/L(PAH) ratios that fall above the solid (and dashed) lines showing the mean value ($\pm$ 1$\sigma$) for the SINGS HII nuclei. A few sources with notably high L(H$_2$)/L(PAH) are marked and discussed in Section \ref{extremesec} (i.e. 1$=$NGC1961; 2$=$IRASF16399-0937; 3$=$NGC3690\_E; 4$=$NGC6240; 5$=$IRAS01364-1042; 6$=$ESO507-G070).
\label{H2PAH}}
\end{center}
\end{figure*}

In low luminosity star-forming galaxies, the luminosity ratio L(H$_2$)/L(PAH) is roughly constant (albeit with a fairly large scatter) over several orders of magnitude in L(H$_2$) \citep[SINGS;][]{roussel} suggesting that both the observed H$_2$ and PAH emission mainly arise from PDRs. AGN, however, show L(H$_2$)/L(PAH) ratios higher by factors of 15-16 \citep{roussel} suggesting an enhancement in the H$_2$ emission, a decrease in PAH emission, or both compared to normal, star-forming galaxies. In Figure \ref{H2PAH}, we plot the ratio of H$_2$-to-PAH emission as a function of H$_2$ luminosity for the GOALS sample. Since we are examining both extinction and excitation effects, we do not correct either the H$_2$ emission lines or the PAH fluxes for extinction in Figure \ref{H2PAH}. To represent L(H$_2$) we use only the three
lowest transitions (H$_2$(S0-S2)) as the H$_2$S(3) line is most likely
to be affected by the silicate absorption at 9.7\micron~(although the distribution observed in Figure \ref{H2PAH} holds if the H$_2$S(3) emission is included). For the purposes of comparison to lower luminosity galaxies, we also performed an identical spectral decomposition using CAFE on the low resolution SINGS nuclear spectra \citep{jdsings, roussel} (represented by black asterisks and open diamonds). When quoting the L(H$_2$)/L(PAH) ratio, L(PAH) represents the luminosity of the 7.7-\micron~and 8.6-\micron~PAH complexes\footnote{As a result of using L(PAH) =
  L[7.7\micron] + L[8.6\micron] when computing the
  L(H$_2$)/L(PAH) ratio, the L(PAH) used in this section is not the
  same L(PAH) given in Section \ref{pahfrac}
in Figure \ref{lpahmerge}.}.

A clear trend of increasing H$_2$-to-PAH ratio with increasing L(H$_2$) is seen in Figure \ref{H2PAH} for the GOALS sample as a whole (filled circles, open squares, and stars, all color coded by EQW$_{6.2\mu m}$ as detailed in the legend). While there are a number of starburst-dominated GOALS sources with L(H$_2$)/L(PAH) values consistent with the nearly constant value found for lower luminosity star-forming SINGS galaxies (solid line), most of the GOALS sample lies well above this value. Combining the data from both the SINGS and GOALS samples, a threshold is reached near L(H$_2$)$\sim$ 10$^7$L$_{\odot}$ where the L(H$_2$)/L(PAH) ratio changes from a roughly constant (low) value (open diamonds and the solid line) to an increasing function of L(H$_2$). The fact that the rising L(H$_2$)/L(PAH) ratio is seen among the pure starburst LIRGs (i.e. those lacking enough hot dust in the MIR to decrease the \eqw~and thus indicate an AGN) is evidence that it is not the presence of an AGN that drives up the L(H$_2$)/L(PAH) ratio among LIRGs. In fact, of the 102 highest equivalent width sources (\eqw$>$0.54\micron; blue stars), only eight have any indication of an AGN at all (based on a detection of the [NeV] line; \cite{petric}). As discussed in more detail in Sections \ref{extremesec} \& \ref{outpdrs}, turbulence and shocks also present in star forming systems can result in enhanced L(H$_2$)/L(PAH) ratios \citep[see also ][]{cluvershocks,lesaffreshocks}.

Some process directly associated with the star formation must also result in enhanced L(H$_2$)/L(PAH) ratios in these LIRGs. The six GOALS sources with no detectable MIR H$_2$ emission in the low resolution spectra also have very low PAH equivalent width (i.e. \eqw$<$ 0.05 \micron) and are likely to harbor AGN.

\subsubsection{Enhanced H$_2$ Emitters}\label{extremesec}
\begin{figure}[htp]
\begin{center}
\includegraphics[height=2.5in,width=3.5in]{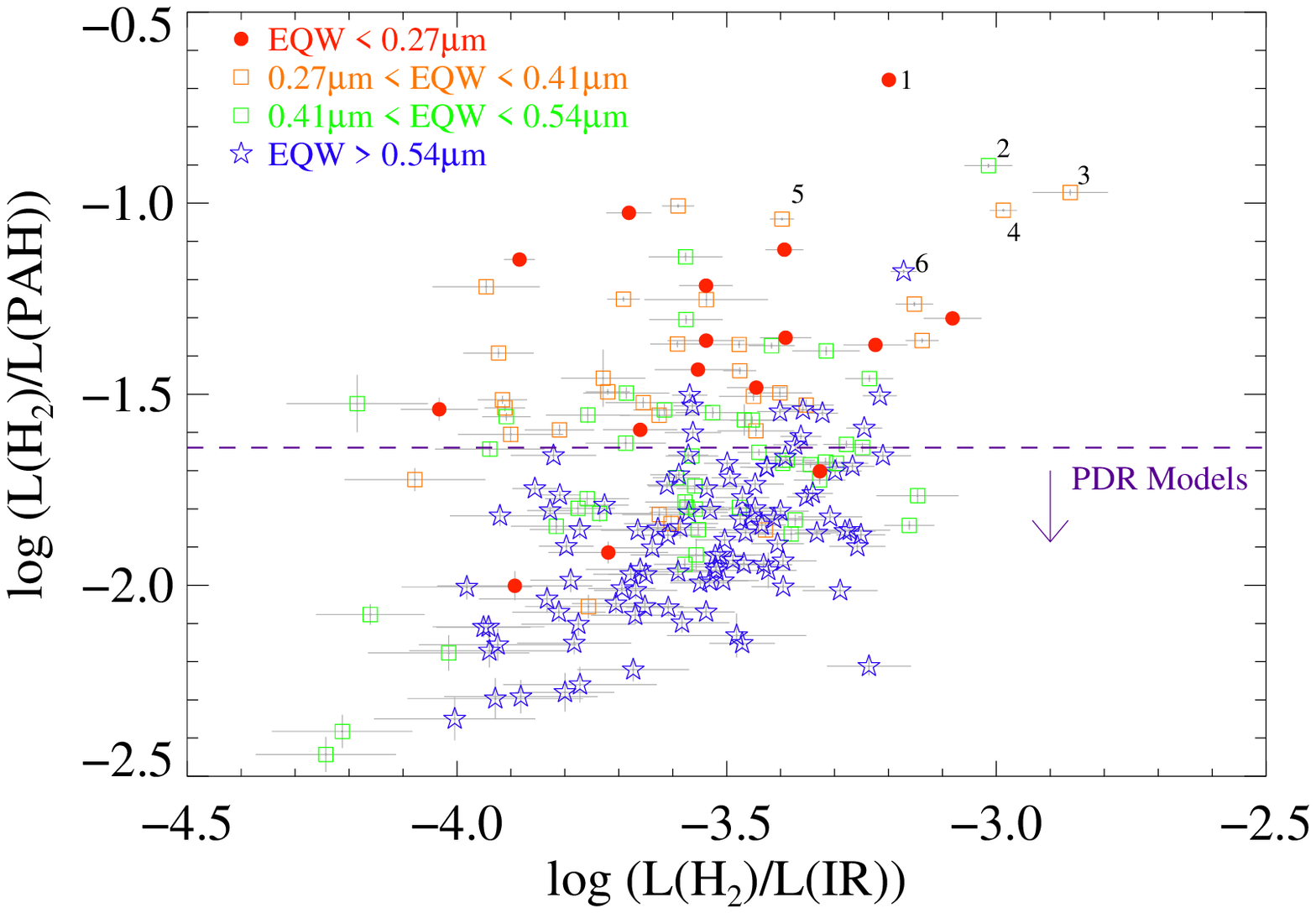}
\caption{Luminosity ratio L(H$_2$)/L(PAH) vs L(H$_2$)/L(IR) for 194 GOALS galaxies color-coded by \eqw. The dashed purple line indicates the upper limit on the L(H$_2$)/L(PAH) ratio that is consistent with PDR emission as derived from the Meudon PDR models \citep[assuming n$_H = $10$^4$cm$^{-3}$ and G$_{UV} = 10$;][]{lepetit} and presented in \cite{pierre}. The six galaxies with notably high H$_2$ emission labeled in Figure \ref{H2PAH} are marked by the same numbers here (i.e. 1$=$NGC1961; 2$=$IRASF16399-0937; 3$=$NGC3690\_E; 4$=$NGC6240; 5$=$IRAS01364-1042; 6$=$ESO507-G070).
\label{extreme}}
\end{center}
\end{figure}

\begin{figure*}[htp]
\begin{center}
\includegraphics[height=5in,width=3.2in]{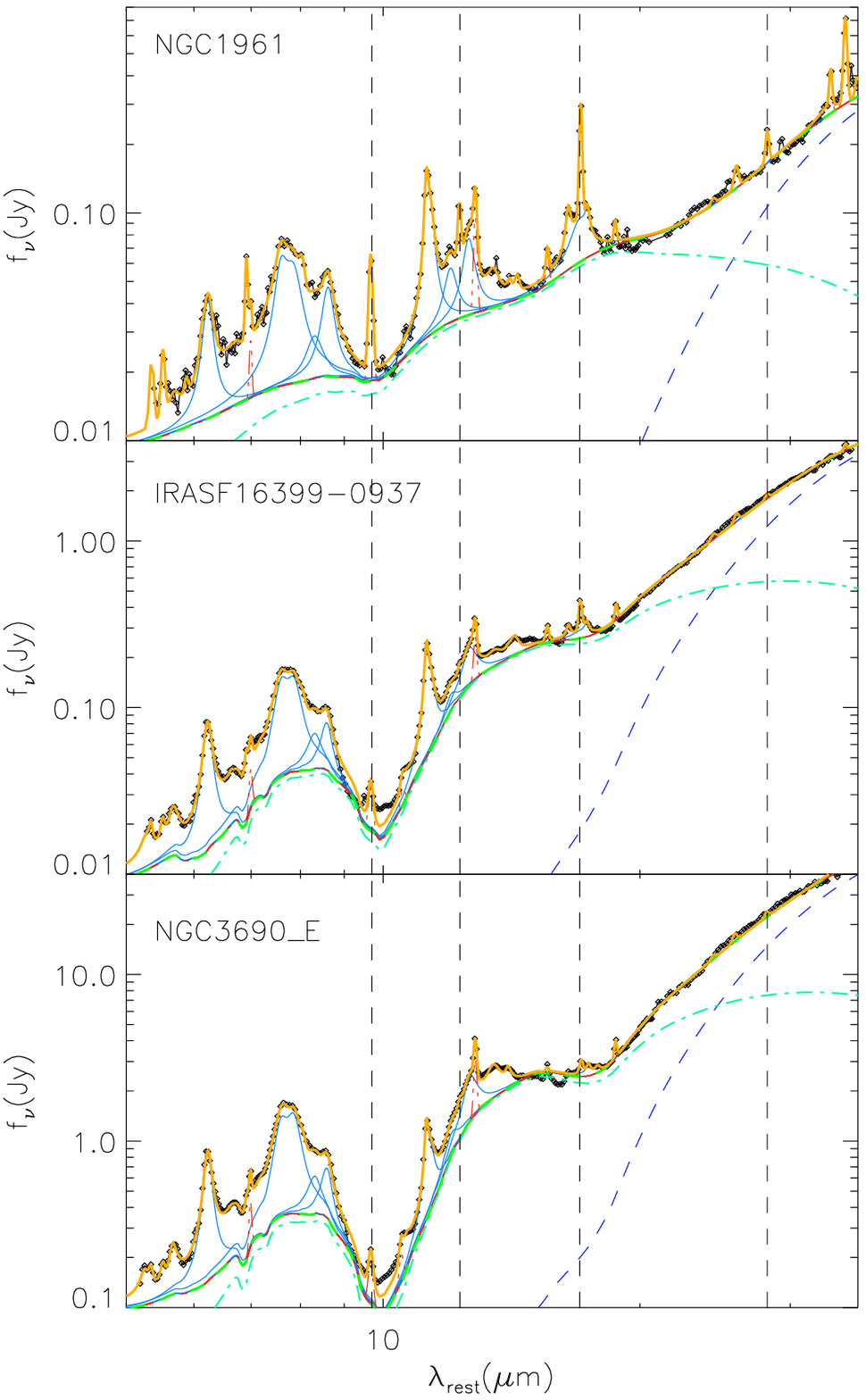}
\includegraphics[height=5in,width=3.2in]{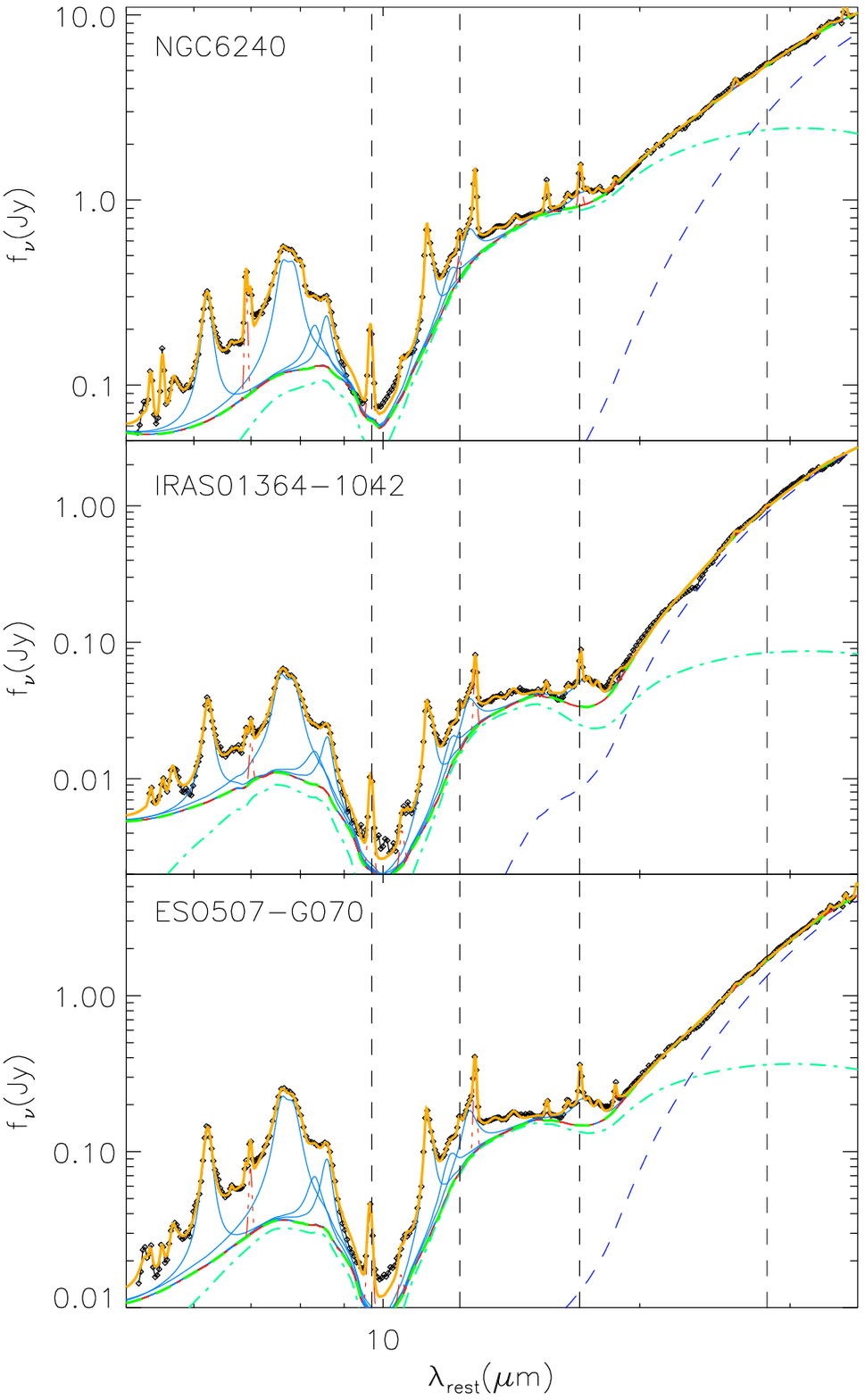}
\caption{Low Resolution IRS Spectra with CAFE spectral decomposition results for GOALS sources with excess H$_2$ emission.
  All 6 galaxies have high L(H$_2$)/L(PAH)) luminosity ratios (log(L(H$_2$)/L(PAH)) $>$ -1.3) and above average L(H$_2$)/L(IR) (log(L(H$_2$)/L(IR)) $> -3.5$).
  The overall fitted model is shown in yellow and model components are
  color-coded as in Figure \ref{repspecs}. The four lowest transitions (H$_2$S(0) at 28.2\micron, H$_2$S(1) at 17.0\micron, H$_2$S(2) at 12.3\micron, and H$_2$S(3) at 9.7\micron) are marked with vertical dashed lines. The locations of each galaxy are labeled in Figures \ref{H2PAH} and \ref{extreme} above.
\label{h2specs}}
\end{center}
\end{figure*}

To more clearly separate those GOALS galaxies that have high L(H$_2$)/L(PAH) luminosity ratios due to excess H$_2$ emission rather than weak PAHs, we plot the L(H$_2$)/L(PAH) ratio versus L(H$_2$)/L(IR) in Figure \ref{extreme}. Most of the LIRGs with the highest L(H$_2$)/L(PAH) ratios have \eqw$<$0.54\micron. In those LIRGs with high L(H$_2$)/L(PAH) but below average H$_2$ emission (i.e. log(L(H$_2$)/L(IR)) $< -3.5$, the median for the GOALS sample; upper left in Figure \ref{extreme}), supressed PAH emission likely drives up the L(H$_2$)/L(PAH) ratios. In $\sim$10\% of GOALS LIRGs, however, log(L(H$_2$)/L(PAH)) $>$ -1.5 $and$ log(L(H$_2$)/L(IR)) $> -3.5$, thus affirming their L(H$_2$)/L(PAH) luminosity ratios are high due to excess H$_2$ emission and not simply caused by weak PAH emission. Such enhanced H$_2$ emission likely requires the presence of X-rays or shocks produced by either an intense starburst or by AGN outflows. The MIR spectra for a sampling of these enhanced H$_2$ emitters are shown in Figure \ref{h2specs}. 

Also shown in Figure \ref{extreme} is an approximate translation of the upper limit set by PDR models in which both the warm H$_2$ and PAHs are directly heated by young stars (dashed purple line). We adopt the limiting PDR ratio presented in \cite{pierre} as derived from the Meudon models \citep[assuming n$_H = $10$^4$cm$^{-3}$ and G$_{UV} = 10$;][]{lepetit}. 
Since \cite{pierre} use different H$_2$ emission line and PAH features to represent the L(H$_2$)/L(PAH) ratio, we adjust their reported value for this upper limit by a multiplicative factor of 0.6, the average ratio between L(H$_2$)/L(PAH) for GOALS galaxies calculated using the H$_2$S(0-3) and 7.7\micron~PAH features \citep[as in][]{pierre} and using the H$_2$S(0-2) and both the 7.7 and 8\micron~PAH features (as presented in this work). Thus, while the GOALS LIRGs have high L(H$_2$)/L(PAH) ratios compared to lower luminosity starforming galaxies (as seen in Figure \ref{H2PAH}), the majority still lie below the maximum value set by PDR models, and so their higher L(H$_2$)/L(PAH) ratios may still be associated with PDRs. 

X-rays produced by AGN can also heat the surrounding H$_2$, and a subset of the GOALS sample has been observed in X-ray emission \citep{CHANDRAGOALS}. For LIRGs with enhanced H$_2$ emission (upper right in Figure \ref{extreme}), the average X-ray hardness ratio (i.e. X-ray color as defined by HR = (Hard \text{--} Soft)/(Hard + Soft)) and the average ratio of soft X-ray-to-IR luminosity are within 1$\sigma$ of the same values averaged for sources consistent with the limiting ratio set by PDR models (i.e. log( L(H$_2$)/L(PAH)) $< -1.8$). Thus X-ray heating is unlikely to be the only excitation mechanism producing the excess H$_2$ seen in Figures \ref{H2PAH}, \ref{extreme}, \& \ref{h2specs}. 

High velocity molecular outflows have recently been seen in some AGN-dominated ULIRGs like Mrk 231 with Herschel \citep{fischeroutflow, feruoutflow, sturmoutflow, spoonoutflow}, and nuclear outflows on sub-kpc scales have been observed via the line profiles of cold molecular gas in the nearby ULIRG system Arp220 \citep{sakamotooutflow, gonzalezoutflow} and in stacked local ULIRG spectra \citep{chungoutflow}. These detections are consistent with the production of shocks in the process of feedback on the interstellar medium. Based on a link between warm H$_2$ and [FeII] emission, \cite{zakamska2} cite shocks as the cause of the enhanced warm H$2$ emission in ULIRGs. Of the 16 GOALS LIRGs observed with the largest excess of warm H$_2$ emission (log(L(H$_2$)/L(PAH)) $>$ -1.3), none have observed velocity offsets ($\delta$v$>$200 km/s) as observed for the H$_2$S(0-3) emission lines in the high resolution IRS spectra. However, eight show at least one of the H$_2$ lines to be spectroscopically resolved (FWHM $>$ 600 km/s) and four have at least one nearly resolved H$_2$ emission line (525 km/s $<$ FWHM $<$ 600 km/s). So although there is no clear evidence for bulk motion from an outflow, shocks may still be stirring up the molecular gas in these systems and causing the excess H$_2$ emission. The remaining four galaxies with high L(H$_2$)/L(PAH) are likely AGN-dominated (EQW$_{6.2\mu m} <$ 0.27\micron) and thus are stronger candidates for X-ray excited H$_2$. However, the [NeV] emission line, usually an indicator of a strong AGN, is only detected in one of these four LIRGs, NGC6240 \citep{petric}.

Although the LINER galaxy NGC1961 does not have an exceptionally high L(H$_2$), it stands out in Figure \ref{H2PAH} as having the largest L(H$_2$)/L(PAH) ratio in our sample. The H$_2$ luminosity for NGC1961 is lower than that for NGC6240 \citep[a well-studied bright H$_2$ emitter, see][]{lee6240}, but the L(H$_2$)/L(PAH) ratio is nearly 3 times brighter in NGC1961 and is actually similar to those observed for the extreme H$_2$ emitters called MOHEGs \citep[molecular hydrogen emitting galaxies;][]{ogle}. Just as shocks are cited as the cause of the elevated warm H$_2$ emission in MOHEGs, shocks or a possible outflow likely play a significant role in elevating the L(H$_2$)/L(PAH) ratio in NGC1961. 

Optical integral field (IFU) observations of U/LIRGs lend further support for shocks as a mechanism driving elevated H$_2$ emission. IFU observations have revealed extended LINER-like emission in the absence of AGN activity \citep{ibero1,ibero2,rich}. This emission is correlated with an increase in velocity dispersion and in some instances with observed outflows, both of which indicate shocked gas is driving the extended LINER emission in U/LIRGs. Further, although shock emission cannot account for a substantial portion of the energy budget in merging U/LIRGs, shocks can strongly affect observed line emission; in the optical they may account for up to half of the observed line emission in some later stage mergers \citep{rich14}.

\subsubsection{Warm H$_2$ Emission Outside of PDRs}\label{outpdrs}

\begin{figure*}[htp]
\begin{center}
\includegraphics[height=3in,width=6in,viewport=10 0 500 230,clip]{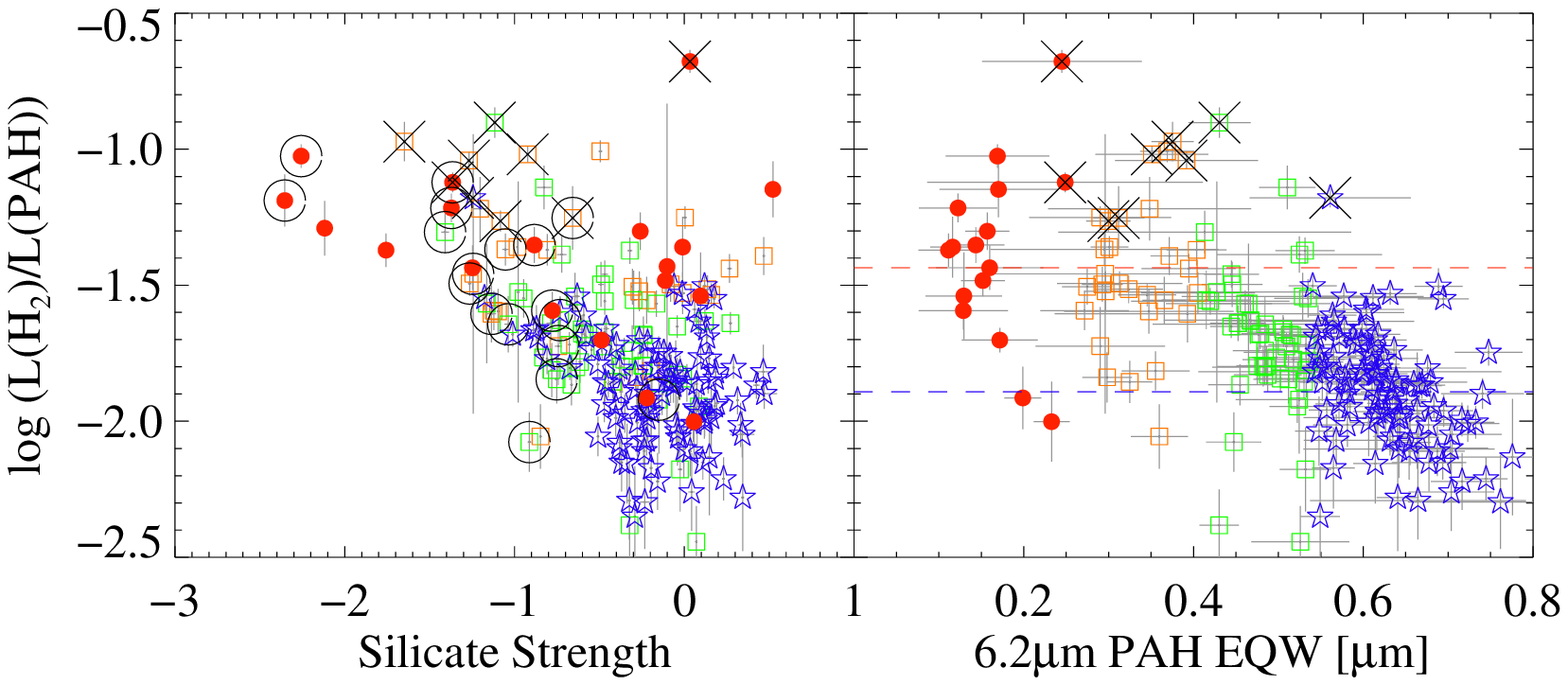}
\caption{Ratio of L(H$_2$)/L(PAH) uncorrected for extinction versus {\it{Left:}} silicate depth at 9.7\micron\ and {\it{Right:}} the equivalent width of the 6.2$\mu m$ PAH for 209 GOALS galaxies color-coded as in Figure \ref{extreme}. On average, the more heavily obscured GOALS ULIRGs (marked by open circles) have higher L(H$_2$)/L(PAH) ratios, a trend that is less clear among the LIRGs.
The mean ratio for the lowest equivalent width
  sources ($\langle$log (L(H$_2$)/L(PAH))$\rangle$ = -1.44; red dashed line) is $\sim$3 times higher
  than the mean ratio for highest equivalent width sources ($\langle$log
    ((LH$_2$)/L(PAH))$\rangle$ = -1.89; blue dashed line). Sources with excess H$_2$ emission (see Section \ref{extremesec}) are marked by X's.
\label{H2PAHEW}}
\end{center}
\end{figure*}

In a study of local ULIRGs, \cite{zakamska} suggest that a significant fraction of the warm H$_2$ emission is produced outside of PDRs, because the L(H$_2$)/L(PAH) ratio increases with silicate optical depth. The relation between the dust obscuration and the L(H$_2$)/L(PAH) ratio for the GOALS sample is examined directly in the left panel of Figure \ref{H2PAHEW}. At low levels of dust obscuration ($s_{9.7\mu m} > $-0.5) where almost all of the starburst-dominated LIRGs (EQW$_{6.2\mu m} \geq $0.54) are found, there is no correlation between the L(H$_2$)/L(PAH) ratio and silicate depth. As the level of obscuration increases, the distribution of sources begins to shift to higher luminosity ratios until reaching the most highly obscured sources ($s_{9.7\mu m} < $-1.75) which, given a reliable 7.7-\micron~PAH detection, are all found in our sample of extreme H$_2$ emitters (marked by X's). 

The substantial scatter in the left panel of Figure \ref{H2PAHEW} does not reveal the correlation among the GOALS LIRGs that was observed for ULIRGs \citep{zakamska}. However, there is a noticeable dearth of obscured sources with low L(H$_2$)/L(PAH), possibly due to a combination of the selective extinction and excitation. The GOALS ULIRGs (marked by open circles) also exhibit more of a correlation for increasing L(H$_2$)/L(PAH) with increasing silicate strength than is observed for the GOALS sample as a whole. Thus, the processes at work to excite warm H$_2$ emission outside of PDRs (i.e. shocks via starburst or AGN driven outflows or X-ray heating via an AGN) may be more common in ULIRG environments. 

Also shown in the right panel of Figure \ref{H2PAHEW} is a trend of increasing L(H$_2$)/L(PAH) with decreasing PAH equivalent width (right panel). The relation has a larger dispersion at low EQW$_{6.2\mu m}$ where there are fewer GOALS sources. The luminosity ratio is best described as decreasing with increasing EQW for EQW$_{6.2\mu m} \geq$ 0.41\micron~(e.g. $log(L(H_2)/L(PAH))= -1 (\pm0.1) - 1.4 (\pm0.2) \times $\eqw~for blue stars and green squares). The lower equivalent width sources (EQW$_{6.2\mu m} < $ 0.41\micron; orange squares and red circles) are roughly consistent with this trend but with an increased scatter. The mean ratios for the lowest (red circles) and the highest (blue stars) equivalent width bins are offset by a factor of 3 (red and blue dashed lines). 

While the underlying physics behind the increasing luminosity ratio observed in Figure \ref{H2PAH} is not entirely clear, taken together, Figures \ref{H2PAH}, \ref{extreme}, \& \ref{H2PAHEW} suggest it may reflect a few different physical processes. Some of the most heavily obscured LIRGs have elevated L(H$_2$)/L(PAH) ratios compared to the majority of the star-forming LIRGs (left panel of Figure \ref{H2PAHEW}), but not all. On average, lower equivalent width sources have higher L(H$_2$)/L(PAH) ratios (right panel of Figure \ref{H2PAHEW}) suggesting PAH emission may simply be lower in the sources with high L(H$_2$)/L(PAH) \emph{or} the hot dust may be rising while the PAH emission remains steady. However, the extreme emitters cover a range of equivalent widths, and we find no correlation between L(H$_2$)/L(PAH) and the compactness of the source (as traced by IR8 $=$L$_{IR}$/L$_{8\mu m}$). Thus, none of these processes (i.e. preferential extinction of PAHs over H$_2$, decreased PAH emission, or a rising hot dust continuum) are individually driving the enhanced ratio for all or even the majority of LIRGs.

\begin{figure}[htp]
\begin{center}
\includegraphics[height=2.4in,width=3.5in,viewport=20 0 500 345,clip]{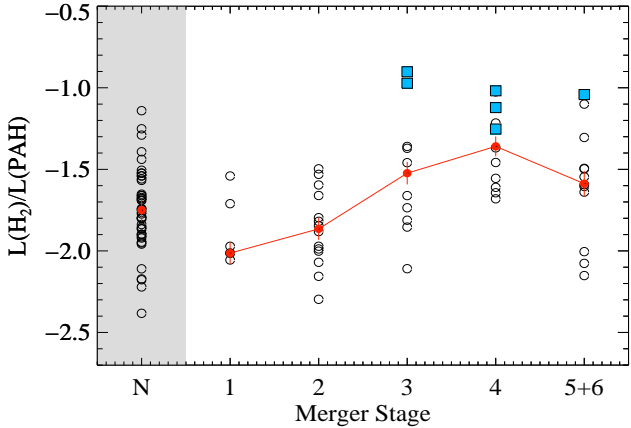}
\caption{Luminosity ratio L(H$_2$)/L(PAH) for 104 GOALS galaxies traced through merger stage. Mergers (Stages $1$-$6$) are represented by the 58 GOALS galaxies for which high resolution HST imaging is available \citep[see][]{haanHST} and for which reliable PAH and H$_2$ measurements were determined. The 46 nonmergers (Stage $N$) are classified using IRAC 3.6\micron~images and the literature \citep[see][for details.]{paperI}. Mean values for each merger stage clipped at
  3$\sigma$ are shown in red with their associated standard deviations. The extreme H$_2$ emitters (blue squares) are all mid- to late-stage mergers.
\label{H2merge}}
\end{center}
\end{figure}

Finally, Figure \ref{H2merge} shows that while enhanced L(H$_2$)/L(PAH) emission compared to low-luminosity SF galaxies \citep{roussel} is found in LIRGs of all merger stages, there is a trend of rising L(H$_2$)/L(PAH) on average (red squares) for late-stage mergers. Sources with excess H$_2$ emission from the GOALS sample (blue squares; discussed in Section \ref{extremesec}) are only found in mid- to late- merger stages. Later merger stages have also been linked to higher L(IR), higher dust temperatures, and increased starburst strength \citep[i.e.][]{paperI,inamihires}, and thus the higher L(H$_2$)/L(PAH) may be associated with more powerful starbursts.

\subsection{PAHs and MIR-Weak AGN}\label{agndisc}

Previous PAH studies have claimed that galaxies hosting an AGN separate themselves on PAH feature ratio plots, favoring lower L($7.7\mu m$ Complex)/L($11.3\mu m$ Complex) ratios \citep{5muses, SSGSS}. These studies credit the preferential destruction of smaller PAHs by the AGN for the lowered ratio, but have only a small number of data points (i.e. sources with an AGN that is weak enough that PAH emission is still detected). For a sample of LIRGs at slightly higher redshifts (0.02 $< z <$ 0.6), \cite{shipley} find that LIRGs with an AGN do not favor lower L($7.7\mu m$ Complex)/L($11.3\mu m$ Complex) ratios and blame aperture effects for damping out this effect. However, \cite{shipley} still find that half of their AGN sources dominate the extreme low end of the PAH ratio, and their low-EQW sources are not scattered with the same increased dispersion observed by the AGN-dominated sources in GOALS (see Figures \ref{ODfig5}, \ref{JDfig13}, \& \ref{JDfig14}). The clearest evidence that galaxies containing an AGN skew toward lower L($7.7\mu m$ Complex)/L($11.3\mu m$ Complex) ratios is observed for the normal, star-forming galaxies of the SINGS sample \citep{jdsings}. Because the sources with the lowest PAH feature ratios (i.e. L($7.7\mu m$ Complex)/L($11.3\mu m$ Complex) $<$ 2) were those that showed spectral signatures indicative of an AGN, \cite{jdsings} concluded this large spread was likely due to the preferential destruction of smaller dust grains caused by the presence of an AGN.

Within the GOALS sample, AGN contribute $<$15\% of the luminosity in the infrared \citep{petric} and 
in order to assess their influence on the PAHs, we consider here only those sources with a relatively ``weak'' AGN not strong enough to have overwhelmed all of the PAH emission (i.e. those sources with 0 $<$ EQW$_{6.2\mu m} <$ 0.27\micron). Despite covering the same range as the SINGS sample in L($11.3\mu m$ Complex)/L($17\mu m$ Complex), L($6.2\mu m$)/L($7.7\mu m$ Complex), and L([NeIII] 15.6$\mu m$)/L([NeII] 12.8$\mu m$), the GOALS LIRGs and ULIRGs do not produce a similarly large range in L($7.7\mu m$ Complex)/L($11.3\mu m$ Complex) (see Figure \ref{JDfig14}), nor do LIRGs with signatures of a weak AGN favor low L($7.7\mu m$ Complex)/L($11.3\mu m$ Complex) ratios. In the remainder of this section, we explore four potential causes for this differing behavior between the SINGs and GOALS galaxies: 1) the differing median distance for the two samples, 2) larger errors associated, almost by definition, with the PAH fluxes for lower-EQW GOALS sources, 3) different dust characteristics in LIRGs versus normal, lower luminosity star-forming galaxies, or 4) differing AGN behavior between the two samples.

The staring mode observations presented here are all aimed at representing the nuclear region of each GOALS galaxy. However, depending on the size and distance of each galaxy, the portion of the galaxy contained within the IRS slit changes. The nuclear spectra for the SINGS sample were derived from mapping mode observations and thus the sizes of the extraction apertures were tailored for each galaxy based on distance to only focus on the nucleus \citep{jdsings}. The median distance of the SINGS galaxy sample is also only 10 Mpc \citep{dalesings}, compared to the GOALS median distance of 100 Mpc. With a fixed IRS slit size and larger median distance, the nuclear GOALS spectra will in some cases be contaminated by emission outside the nucleus. If the distribution of the 7.7-\micron~PAH were more extended beyond the nucleus than the 11.3-\micron~PAH, the lack of GOALS sources with low L($7.7\mu m$ Complex)/L($11.3\mu m$ Complex) could be explained by extra-nuclear emission falling within the slit. However, the opposite was shown to be true for the GOALS LIRGs: the 11.3-$\mu m$ PAH appears more extended than the other PAH features \citep{tanio2}. There are also several GOALS LIRGs that likely have a contribution from a weak AGN (0 $<$ \eqw\ $<$ 0.27\micron) but do not have a low L($7.7\mu m$ Complex)/L($11.3\mu m$ Complex) PAH ratio despite being nearby (D $<$ 60 Mpc). Thus distance cannot entirely explain the differences in the PAH feature ratio observed between the GOALS and SINGS samples.

As shown in Figures \ref{ODfig5}, \ref{JDfig13}, \& \ref{JDfig14},
LIRGs with low PAH equivalent widths are not offset in their PAH
feature ratios but
instead show a larger dispersion in PAH values than their starbursting
counterparts. Low EQW$_{6.2\mu m}$ indicates either low PAH emission overall or high continuum,
and lower detected PAH fluxes may decrease the signal-to-noise ratio for PAH
detections in low-EQW LIRGs. While the sources with weak PAH emission have larger uncertainties in their PAH ratios, a Kolmogorov-Smirnov (K-S) test confirms the low
and high EQW LIRGs are distinctly different populations at a
confidence level of $>$96\%. Even if the errors on the PAH feature ratios
are doubled and every low-EQW LIRG is forced inward by this amount toward the center
of the distribution for the high-EQW LIRGs, the K-S test still
confirms the two distributions as distinct at a 90\% confidence
level. Even if higher uncertainties do contribute in a small way to
the higher dispersion among low-EQW LIRGs, it is highly unlikely that errors might be so large that any GOALS galaxies would 
actually have L($7.7\mu m$ Complex)/L($11.3\mu m$
Complex) $<$ 2 to match the ratios observed for the SINGS AGN.

Since PAH feature ratios ultimately tell us about the dust and ISM properties in galaxies (see theoretical tracks in Figure \ref{ODfig5}),
the differing ratios between SINGS and GOALS may represent physical
differences in the dust grains in normal, star-forming galaxies versus
LIRGs and ULIRGs. The rougly constant L($7.7\mu m$ Complex)/L($11.3\mu m$
Complex) ratio observed for GOALS could be caused by competing effects: the
presence of an AGN may destroy smaller PAH molecules causing a decrease in the 7.7-\micron~PAH emission, while the AGN also ionizes those PAHs, causing the relative emission from the neutral 11.3-\micron~PAH to decrease. These two effects may be equally important in the GOALS LIRGs, having the net outcome a constant PAH ratio, while the destruction of small PAH grains may dominate in the lower luminosity SINGS galaxies.

However, if competing processes are at work to produce Figure \ref{JDfig14}, we would expect the effects of the destruction of the smaller PAHs to be more pronounced in the L($6.2\mu m$)/L($7.7\mu
m$ Complex) ratio in Figure \ref{ODfig5} or in the L($11.3\mu m$ Complex)/L($17\mu m$ Complex) ratio in Figure \ref{JDfig13} which are not expected to depend on the ionization state of the grains. Thus the differing L($7.7\mu m$ Complex)/L($11.3\mu m$ Complex)
ratios may instead be due to the relative power of the AGNs
themselves and/or their viewing geometry. 

\begin{figure*}[ht]
\begin{center}
\includegraphics[height=3in,width=6in,viewport=10 0 500 230,clip]{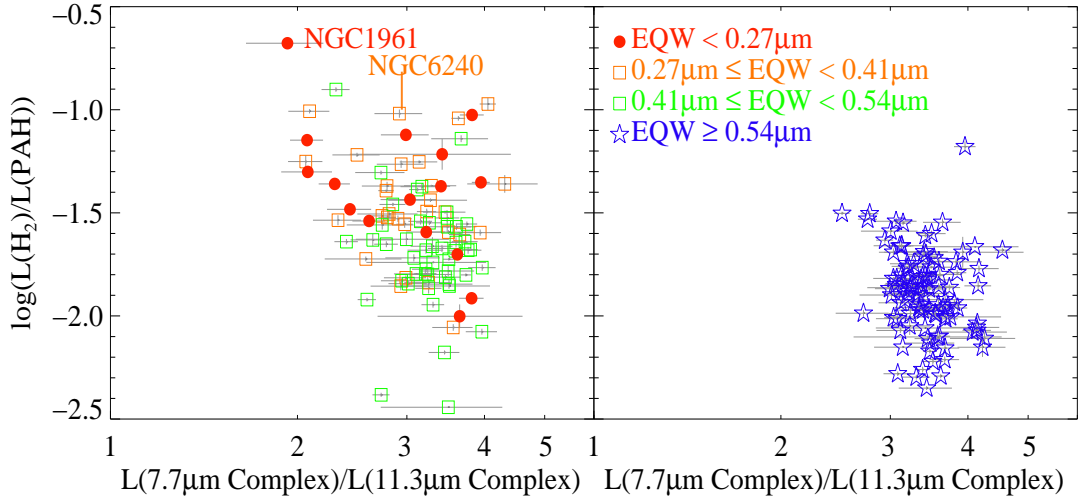}
\caption{The L($7.7\mu m$ Complex)/L($11.3\mu m$ Complex) PAH ratio
 versus the L(H$_2$)/L(PAH) ratio for 203 GOALS galaxies color coded by \eqw. Those LIRGs with the lowest PAH band ratios are all observed to have high  L(H$_2$)/L(PAH) ratios suggesting that the mechanism responsible for elevating their H$_2$ emission may also be linked to the destruction of the 7.7\micron\ PAHs.
\label{compactvPAHs}}
\end{center}
\end{figure*}

Significant 7.7\micron~PAH destruction like that found for SINGS is not observed for the GOALS LIRGs and ULIRGs. However, as shown in Figure \ref{compactvPAHs}, the sources that come closest to showing such an effect (i.e. those with the lowest L($7.7\mu m$ Complex)/L($11.3\mu m$ Complex) ratio), are also observed to have high L(H$_2$)/L(PAH). Thus, the mechanism responsible for elevating the L(H$_2$)/L(PAH) ratio may also be destroying the 7.7\micron~PAHs (although less effectively than the small grain destruction via AGN observed for SINGS galaxies). All of the sources with high L(H$_2$)/L(PAH) do not also have low L($7.7\mu m$ Complex)/L($11.3\mu m$ Complex) ratios, but that is as expected. As discussed in Section \ref{miniH}, there are likely a variety of causes for the elevated L(H2)/L(PAH) ratios, and some may be more effective at also destroying 7.7\micron~PAHs than others. One source with both exceptionally high L(H2)/L(PAH) \emph{and} low L($7.7\mu m$ Complex)/L($11.3\mu m$ Complex) is NGC1961 (labeled in Figure \ref{compactvPAHs}) which, as discussed in Section \ref{extremesec}, is a strong candidate for shocks playing a significant role in the resulting emission in the MIR.

\section{Conclusions}\label{conc}
We presented the detailed multi-component spectral decomposition of the low resolution IRS spectra for 244 galaxy nuclei in the GOALS sample of 180 LIRGs and 22 ULIRGs. The GOALS galaxies cover a range of spectral types, silicate strengths, and merger stages, and represent a complete subset of the IRAS Revised Bright Galaxy Sample. We have investigated the MIR properties derived from the spectral fits and concluded the following:

\noindent 1) Despite the wide range of MIR spectral properties observed for the GOALS U/LIRGs, as long as some PAH emission is detected, little variation is seen in the PAH feature ratios after correcting for extinction. The ranges of the L($6.2\mu m$)/L($7.7\mu m$ Complex) and L($11.3\mu m$ Complex)/L($17\mu m$ Complex) PAH luminosity ratios for the GOALS LIRGs are similar to those observed for the nearby, normal star-forming galaxies from SINGS \citep{jdsings}, the 24-\micron-selected galaxies of 5MUSES \citep{5muses}, the UV-selected nearby galaxies in SSGSS \citep{SSGSS}, and the higher redshift LIRGs of \cite{shipley}. One exception is the L($7.7\mu m$)/L($11.3\mu m$) ratio which reaches much lower values for the SINGS galaxies than for any other sample.

\noindent 2) The contribution from PAH emission to the total IR luminosity (L(PAH)/L(IR)) in LIRGs varies from 0-23\%. LIRG systems containing pairs of galaxies that appear to be prior to their first encounter show higher median L(PAH)/L(IR) ratios than those observed for other LIRGs. Local LIRGs have a constant (high) average \eqw\ over nearly two orders of magnitude in $\nu$L$_{\nu}$ similar to high redshift SMGs and star forming galaxies at higher $\nu$L$_{\nu}$ but unlike local ULIRGs which show a trend for decreasing \eqw\ with increasing 24-\micron\ luminosity. This is consistent with the larger sizes and gas fractions observed for SMGs compared to local ULIRGs \citep[i.e.][]{daddiGasFrac, genzelGasFrac, elbaz}. 

\noindent 3) No change in the grain size or ionization distribution is observed with IR8, and so the destruction of smaller dust grains is not the cause of the higher IR8 observed in more compact starburst systems. Instead, since the overall flux contribution from the PAHs is decreasing for the higher \eqw\ sources, there instead is likely less PDR emission relative to the emission in the IR. 

\noindent 4) Absorption features similar to those attributed to crystalline silicates \citep{spoonULIRGs} are observed at 23\micron\ only once deep levels of silicate absorption are reached ($s_{9.7\mu m} < $1.24) in $\sim$6\% of the GOALS (U)LIRGs. Absorption due to water ices at 6.0\micron~is observed in 13 ULIRGs and 24 LIRGs for detection rates of 56.5\% among ULIRGs and 10.7\% among LIRGs.

\noindent 5) Although (U)LIRGs with low L(PAH)/L(IR) or deep silicate absorption show higher IRX on average, the spread in these MIR parameters is too large to indicate any correlation with $\Delta$IRX. No other MIR properties were found to correlate with the decoupling of the IR and UV fields, $\Delta$IRX, including EQW$_{6.2\mu m}$ and $s_{9.7\mu m}$.

\noindent 6) While there are a number of starburst-dominated GOALS LIRGs with L(H$_2$)/L(PAH) values consistent with the nearly constant (low) ratio found for lower luminosity star-forming galaxies \citep{roussel}, most of the GOALS galaxies lie well above this value in an increasing function of L(H$_2$). The fact that the rising L(H$_2$)/L(PAH) ratio is seen among the pure starburst LIRGs is evidence that it is not the presence of an AGN that drives up the L(H$_2$)/L(PAH) ratio among LIRGs.  Turbulence and shocks also present in star forming systems likely result in the observed enhanced L(H$_2$)/L(PAH) ratios.

\noindent 7) A subset of GOALS (U)LIRGs covering a range of EQW$_{6.2\mu m}$ show enhanced H$_2$ emission in excess over that suggested by PDR models which only allow for young stars to excite the H$_2$. These galaxies can be identified by their high L(H$_2$)/L(PAH) ratio (log(L(H$_2$)/L(PAH)) $>$ -1.5). One quarter of these galaxies are dominated by AGN (i.e. EQW$_{6.2\mu m} < $0.27\micron) and thus are strong candidates for X-ray heating of the warm H$_2$. However, half of the extreme H$_2$ emitters show resolved H$_2$ lines in the high resolution IRS spectra which indicate that the shocks may be stirring up the molecular gas, even in the absence of evidence for larger scale coherent outflows.

\noindent 8) A correlation between increasing silicate strength and L(H$_2$)/L(PAH), like that observed for the \cite{zakamska} sample of ULIRGs, is not observed for the GOALS LIRGs indicating that warm H$_2$ emitted outside of PDRs does not dominate their MIR H$_2$ emission. The GOALS ULIRGs are closer to exhibiting such a trend, suggesting that processes at work to excite warm H$_2$ emission outside of PDRs may be more common in ULIRG environments.

\noindent 9) While high-EQW LIRGs are very consistent in their MIR properties, low-EQW LIRGs usually cover the full spread of any given MIR parameter. Starbursting LIRGs (EQW$_{6.2\mu m} \geq$ 0.54\micron) which make up the majority of the GOALS sample all have very similar values for s$_{9.7\mu m}$ and F$_{\nu}$[30\micron]/F$_{\nu}$[15\micron] \citep{paperI}, for neon ratios, $\tau_{ice}$, and all 3 PAH feature ratios presented here. The large range of values observed for such a variety of galaxy characteristics may suggest that
before the emission from an AGN grows enough to dominate in the MIR, a weak AGN leads to varied dust conditions.

The {\it{Spitzer}} Space Telescope is operated by the Jet Propulsion
Laboratory, California Institute of Technology, under NASA contract
1407. This research has made use of the NASA/IPAC Extragalactic
Database (NED) which is operated by the Jet Propulsion Laboratory,
California Institute of Technology, under contract with the National
Aeronautics and Space Administration. This research has made use of
the NASA/IPAC Infrared Science Archive, which is operated by the Jet
Propulsion Laboratory, California Institute of Technology, under
contract with the National Aeronautics Space Administration. This work was supported in part by National Science Foundation Grant No. PHYS-1066293 and the hospitality of the Aspen Center for Physics. VC would like to acknowledge partial support from the EU FP7 Grant PIRSES-GA-2012-316788. The authors would
like to thank M. Cluver for many helpful discussions and the anonymous referee for her/his help in improving the paper. 

\bibliography{LIRGreferences}{}
\newpage
\begin{sidewaystable}
\caption{Fitted Mid-IR Spectral Parameters for the GOALS Sample\label{pahtable}}
\resizebox{\linewidth}{!}{
\tabcolsep=3pt
\begin{tabular}{l c r r r r r c c c}
\hline
\hline
Source Name$^{*,**}$ & EQW$_{6.2}$($\sigma$) & 6.2$\mu m$ PAH($\sigma$) & 7.7$\mu m$ PAH($\sigma$) & 11.3$\mu m$ PAH($\sigma$) & 17$\mu m$ PAH($\sigma$)  & L(PAH)($\sigma$) & $\tau_{ice}(\sigma)$ & $\tau_{9.7\mu m}(\sigma)$ & $\eta(\sigma)$\\
 & [$\mu m$]$^{i}$ & & & & & [10$^9$L$_{\odot}$] & & & \\
\hline
NGC0023 & 0.58 (0.01) &  6.118 (0.136) & 24.779 (0.790) &  7.883 (0.080) &  4.473 (0.206) &  8.52 (0.56) & 0.00 & 0.81 (0.01) & 0.10 (0.05) \\
NGC0034 & 0.45 (0.02) &  9.214 (0.177) & 38.081 (1.162) & 10.234 (0.303) &  5.996 (0.436) & 21.30 (1.37) & 0.04 (0.01) & 2.39 (0.01) & -0.07 (0.05) \\
Arp256 & 0.72 (0.01) &  4.407 (0.104) & 14.505 (0.404) &  4.099 (0.070) &  3.027 (0.250) & 17.09 (0.90) & 0.00 & 1.20 (0.02) &  ---~~~~~~~  \\
ESO350-IG038 & 0.15 (0.00) &  1.448 (0.071) &  6.189 (0.495) &  1.703 (0.310) & $<$ 0.534 (0.346) & $<$ 3.82 (0.53) & 0.00 & 1.26 (0.01) & 0.39 (0.03) \\
NGC0232\_W & 0.55 (0.01) &  5.674 (0.118) & 22.873 (0.468) &  6.701 (0.368) &  4.652 (0.167) & 16.92 (0.84) & 0.00 & 1.63 (0.01) & -0.01 (0.05) \\
NGC0232\_E & 0.16 (0.00) &  0.966 (0.044) &  3.337 (0.143) &  1.606 (0.064) &  0.785 (0.104) &  2.79 (0.32) & 0.00 & 1.19 (0.02) & -0.02 (0.05) \\
\hline\end{tabular}}
\end{sidewaystable}
\begin{sidewaystable}
\resizebox{\linewidth}{!}{
\tabcolsep=3pt
\begin{tabular}{l c r r r r r c c c}
\hline
\hline
Source Name$^{*,**}$ & EQW$_{6.2}$($\sigma$) & 6.2$\mu m$ PAH($\sigma$) & 7.7$\mu m$ PAH($\sigma$) & 11.3$\mu m$ PAH($\sigma$) & 17$\mu m$ PAH($\sigma$)  & L(PAH)($\sigma$) & $\tau_{ice}(\sigma)$ & $\tau_{9.7\mu m}(\sigma)$ & $\eta(\sigma)$\\
 & [$\mu m$]$^{i}$ & & & & & [10$^9$L$_{\odot}$] & & & \\
\hline
NGC0992 & 0.72 (0.01) &  5.728 (0.130) & 22.742 (0.808) &  7.490 (0.172) &  3.695 (0.169) &  6.61 (0.48) & 0.00 & 0.87 (0.02) & 0.18 (0.04) \\
UGC02238 & 0.63 (0.01) &  6.922 (0.140) & 29.345 (0.699) &  8.820 (0.163) &  4.665 (0.164) & 20.55 (0.85) & 0.00 & 2.06 (0.02) & 0.11 (0.04) \\
IRASF02437+2122 & 0.16 (0.01)$^{i}$ &  1.297 (0.045) &  4.999 (0.276) &  1.647 (0.061) &  0.821 (0.074) &  4.45 (0.47) & 0.34 (0.03) & 3.21 (0.02) & -1.64 (0.13) \\
UGC02369 & 0.57 (0.01) &  2.909 (0.071) &  9.712 (0.377) &  2.338 (0.044) &  2.166 (0.207) & 15.35 (1.10) & 0.00 & 0.89 (0.01) & 0.03 (0.05) \\
UGC02608\_N & 0.20 (0.00) &  2.736 (0.086) & 13.970 (0.386) &  3.663 (0.098) &  3.757 (0.248) & 11.29 (0.92) & 0.00 & 1.45 (0.02) & 0.01 (0.05) \\
UGC02608\_S & 0.25 (0.14) & --- ~~~~~~~~ & --- ~~~~~~~~ & --- ~~~~~~~~ & --- ~~~~~~~~ & --- ~~~~~~ & --- & --- & 0.24 (0.04) \\
\hline\end{tabular}}
\end{sidewaystable}
\begin{sidewaystable}
\resizebox{\linewidth}{!}{
\tabcolsep=3pt
\begin{tabular}{l c r r r r r c c c}
\hline
\hline
Source Name$^{*,**}$ & EQW$_{6.2}$($\sigma$) & 6.2$\mu m$ PAH($\sigma$) & 7.7$\mu m$ PAH($\sigma$) & 11.3$\mu m$ PAH($\sigma$) & 17$\mu m$ PAH($\sigma$)  & L(PAH)($\sigma$) & $\tau_{ice}(\sigma)$ & $\tau_{9.7\mu m}(\sigma)$ & $\eta(\sigma)$\\
 & [$\mu m$]$^{i}$ & & & & & [10$^9$L$_{\odot}$] & & & \\
\hline
IRAS05223+1908$^*$ & $<$0.01 & & & & & & & & -0.02 (0.05) \\
MCG+08-11-002 & 0.56 (0.01) &  8.642 (0.141) & 37.418 (0.543) &  9.119 (0.202) &  7.203 (0.167) & 19.92 (0.68) & 0.00 & 3.73 (0.02) & -0.01 (0.05) \\
NGC1961 & 0.24 (0.01) &  0.806 (0.025) &  3.230 (0.138) &  1.676 (0.038) &  0.940 (0.088) &  0.99 (0.08) & 0.00 & 1.05 (0.02) & 0.87 (0.01) \\
UGC03351 & 0.53 (0.01) &  5.187 (0.112) & 23.361 (0.422) &  7.589 (0.148) &  3.729 (0.109) &  7.74 (0.35) & 0.00 & 2.32 (0.02) & 0.20 (0.04) \\
IRAS05442+1732 & 0.66 (0.01) &  7.168 (0.141) & 24.746 (0.623) &  6.079 (0.380) &  5.200 (0.243) & 13.15 (0.73) & 0.00 & 1.58 (0.01) & 0.01 (0.05) \\
IRASF06076-2139$^{**}$ & 0.33 (0.02) &  1.066 (0.048) &  3.724 (0.309) &  1.645 (0.248) & $<$ 0.323 (0.074) &  $<$ 9.78 (1.51) & 0.00 & 3.18 (0.02) & ---~~~~~~~ \\
\hline\end{tabular}}
\end{sidewaystable}
\begin{sidewaystable}
\resizebox{\linewidth}{!}{
\tabcolsep=3pt
\begin{tabular}{l c r r r r r c c c}
\hline
\hline
Source Name$^{*,**}$ & EQW$_{6.2}$($\sigma$) & 6.2$\mu m$ PAH($\sigma$) & 7.7$\mu m$ PAH($\sigma$) & 11.3$\mu m$ PAH($\sigma$) & 17$\mu m$ PAH($\sigma$)  & L(PAH)($\sigma$) & $\tau_{ice}(\sigma)$ & $\tau_{9.7\mu m}(\sigma)$ & $\eta(\sigma)$\\
 & [$\mu m$]$^{i}$ & & & & & [10$^9$L$_{\odot}$] & & & \\
\hline
IRASF08572+3915$^*$ & $<$0.03 & & & & & & & & 0.00 (0.05) \\
IRAS09022-3615 & 0.14 (0.00) &  2.963 (0.102) & 11.019 (0.909) &  2.791 (0.062) &  2.126 (0.116) & 63.87 (7.97) & 0.00 & 2.71 (0.01) & -0.01 (0.05) \\
IRASF09111-1007\_W & 0.47 (0.02) &  1.348 (0.057) &  4.755 (0.602) &  1.588 (0.036) &  1.006 (0.053) & 25.73 (3.46) & 0.00 & 2.15 (0.03) & -0.06 (0.05) \\
IRASF09111-1007\_E & 0.52 (0.03) &  0.945 (0.057) &  3.040 (0.554) &  1.175 (0.137) &  0.588 (0.075) & 16.08 (5.96) & 0.00 & 1.13 (0.04) & -0.04 (0.05) \\
UGC04881\_W & 0.61 (0.01) &  1.315 (0.030) &  4.911 (0.207) &  1.435 (0.020) &  0.896 (0.062) & 12.22 (0.90) & 0.00 & 1.51 (0.02) & ---~~~~~~~ \\
UGC04881\_E & 0.40 (0.01) &  0.840 (0.023) &  3.328 (0.149) &  1.194 (0.039) &  0.719 (0.047) &  9.12 (0.99) & 0.00 & 2.00 (0.02) & ---~~~~~~~ \\
\hline\end{tabular}}
\end{sidewaystable}
\begin{sidewaystable}
\resizebox{\linewidth}{!}{
\tabcolsep=3pt
\begin{tabular}{l c r r r r r c c c}
\hline
\hline
Source Name$^{*,**}$ & EQW$_{6.2}$($\sigma$) & 6.2$\mu m$ PAH($\sigma$) & 7.7$\mu m$ PAH($\sigma$) & 11.3$\mu m$ PAH($\sigma$) & 17$\mu m$ PAH($\sigma$)  & L(PAH)($\sigma$) & $\tau_{ice}(\sigma)$ & $\tau_{9.7\mu m}(\sigma)$ & $\eta(\sigma)$\\
 & [$\mu m$]$^{i}$ & & & & & [10$^9$L$_{\odot}$] & & & \\
\hline
ESO267-G030\_E & 0.50 (0.01) &  3.924 (0.105) & 15.785 (0.442) &  5.234 (0.067) &  2.774 (0.139) &  8.19 (0.58) & 0.00 & 1.00 (0.01) & 0.03 (0.05) \\
IRAS12116-5615 & 0.36 (0.01) &  5.589 (0.123) & 23.100 (0.419) &  6.485 (0.363) &  4.426 (0.237) & 26.18 (1.34) & 0.00 & 2.61 (0.02) & -0.01 (0.05) \\
IRASF12224-0624 & 0.07 (0.02)$^{i}$ &  0.393 (0.030) &  2.516 (0.301) &  0.578 (0.035) &  0.427 (0.040) &  2.70 (0.49) & 1.72 (0.06) & 8.38 (0.09) & -0.02 (0.05) \\
NGC4418$^*$ & $<$0.02 & & & & & & & & -0.09 (0.05) \\
UGC08058$^*$ & 0.01 (0.00) & & & & & & & & 0.02 (0.05) \\
NGC4922 & 0.16 (0.00) & --- ~~~~~~~~ & --- ~~~~~~~~ & --- ~~~~~~~~ & --- ~~~~~~~~ & --- ~~~~~~~~ & --- & --- & ---~~~~~~~ \\
\hline\end{tabular}}
\end{sidewaystable}
\begin{sidewaystable}
\resizebox{\linewidth}{!}{
\tabcolsep=3pt
\begin{tabular}{l c r r r r r c c c}
\hline
\hline
Source Name$^{*,**}$ & EQW$_{6.2}$($\sigma$) & 6.2$\mu m$ PAH($\sigma$) & 7.7$\mu m$ PAH($\sigma$) & 11.3$\mu m$ PAH($\sigma$) & 17$\mu m$ PAH($\sigma$)  & L(PAH)($\sigma$) & $\tau_{ice}(\sigma)$ & $\tau_{9.7\mu m}(\sigma)$ & $\eta(\sigma)$\\
 & [$\mu m$]$^{i}$ & & & & & [10$^9$L$_{\odot}$] & & & \\
\hline
Arp84\_S & 0.17 (0.02) &  0.181 (0.012) &  0.753 (0.112) &  0.364 (0.012) &  0.187 (0.018) &  0.18 (0.09) & 0.00 & 0.35 (0.04) & 1.10 (0.06) \\
CGCG247-020 & 0.56 (0.01) &  2.547 (0.057) &  9.820 (0.227) &  2.828 (0.077) &  2.083 (0.128) & 10.23 (0.46) & 0.00 & 1.17 (0.01) & -0.02 (0.05) \\
NGC5653$^{**}$ & 0.54 (0.01) &  4.371 (0.102) & 19.341 (0.595) &  6.455 (0.066) & $<$ 3.491 (0.163) &  $<$ 4.08 (0.22) & 0.00 & 0.60 (0.01) & ---~~~~~~~ \\
IRASF14348-1447 & 0.25 (0.01)$^{i}$ &  1.277 (0.024) &  4.812 (0.065) &  1.610 (0.036) &  1.316 (0.078) & 66.34 (3.72) & 0.63 (0.03) & 4.31 (0.03) & 0.04 (0.05) \\
IRASF14378-3651 & 0.39 (0.03) &  1.082 (0.053) &  3.612 (0.185) &  0.990 (0.048) &  0.816 (0.050) & 30.98 (3.90) & 0.00 & 2.95 (0.03) & 0.08 (0.05) \\
NGC5734\_N & 0.47 (0.01) &  3.247 (0.082) & 13.797 (0.361) &  5.219 (0.055) &  2.700 (0.142) &  4.08 (0.18) & 0.00 & 0.81 (0.01) & 0.18 (0.04) \\
\hline\end{tabular}}
\end{sidewaystable}
\begin{sidewaystable}
\resizebox{\linewidth}{!}{
\tabcolsep=3pt
\begin{tabular}{l c r r r r r c c c}
\hline
\hline
Source Name$^{*,**}$ & EQW$_{6.2}$($\sigma$) & 6.2$\mu m$ PAH($\sigma$) & 7.7$\mu m$ PAH($\sigma$) & 11.3$\mu m$ PAH($\sigma$) & 17$\mu m$ PAH($\sigma$)  & L(PAH)($\sigma$) & $\tau_{ice}(\sigma)$ & $\tau_{9.7\mu m}(\sigma)$ & $\eta(\sigma)$\\
 & [$\mu m$]$^{i}$ & & & & & [10$^9$L$_{\odot}$] & & & \\
\hline
IRASF17138-1017 & 0.68 (0.01) &  8.761 (0.174) & 34.707 (1.058) & 11.049 (0.266) & $<$ 3.896 (2.600) & $<$ 15.04 (2.11) & 0.00 & 1.29 (0.01) & -0.04 (0.05) \\
IRASF17207-0014 & 0.31 (0.01)$^{i}$ &  6.428 (0.102) & 24.176 (1.054) &  7.487 (0.540) &  3.783 (0.091) & 71.13 (5.02) & 0.61 (0.03) & 4.58 (0.04) & -0.01 (0.05) \\
ESO138-G027 & 0.52 (0.01) &  2.356 (0.065) &  8.627 (0.278) &  2.679 (0.046) &  1.512 (0.136) &  5.78 (0.45) & 0.00 & 1.17 (0.01) & 0.03 (0.05) \\
UGC11041 & 0.58 (0.01) &  5.892 (0.126) & 24.553 (0.761) &  7.604 (0.129) &  4.098 (0.170) &  9.39 (0.51) & 0.00 & 1.07 (0.01) & 0.06 (0.05) \\
CGCG141-034 & 0.48 (0.01) &  2.884 (0.072) & 12.862 (0.488) &  3.445 (0.075) &  2.505 (0.081) &  7.44 (0.49) & 0.00 & 2.47 (0.02) & -0.01 (0.05) \\
IRAS17578-0400\_W & 0.62 (0.02) &  1.001 (0.032) &  3.374 (0.380) &  1.242 (0.016) &  0.592 (0.036) &  1.08 (0.19) & 0.00 & 0.62 (0.03) & 0.05 (0.05) \\
\hline\end{tabular}}
\end{sidewaystable}
\begin{sidewaystable}
\resizebox{\linewidth}{!}{
\tabcolsep=3pt
\begin{tabular}{l c r r r r r c c c}
\hline
\hline
Source Name$^{*,**}$ & EQW$_{6.2}$($\sigma$) & 6.2$\mu m$ PAH($\sigma$) & 7.7$\mu m$ PAH($\sigma$) & 11.3$\mu m$ PAH($\sigma$) & 17$\mu m$ PAH($\sigma$)  & L(PAH)($\sigma$) & $\tau_{ice}(\sigma)$ & $\tau_{9.7\mu m}(\sigma)$ & $\eta(\sigma)$\\
 & [$\mu m$]$^{i}$ & & & & & [10$^9$L$_{\odot}$] & & & \\
\hline
CGCG448-020\_E$^{**}$ & 0.27 (0.01) &  1.572 (0.043) &  7.058 (0.401) &  1.240 (0.200) & $<$ 0.828 (0.236) & $<$ 12.45 (1.65) & 0.00 & 2.31 (0.01) & ---~~~~~~~ \\
IRAS20551-4250 & 0.10 (0.01)$^{i}$ &  2.695 (0.143) &  9.947 (0.582) &  3.245 (0.162) & & $<$ 38.32 (3.86) & 0.35 (0.02) & 5.35 (0.03) & 0.05 (0.05) \\
ESO286-G035 & 0.69 (0.01) &  6.045 (0.126) & 25.372 (0.901) &  6.588 (0.483) &  4.363 (0.127) & 10.55 (0.80) & 0.00 & 1.64 (0.02) & 0.06 (0.05) \\
IRAS21101+5810 & 0.55 (0.02) &  1.621 (0.053) &  5.811 (0.561) &  1.859 (0.039) &  1.001 (0.077) & 14.20 (1.97) & 0.00 & 2.36 (0.02) & 0.01 (0.05) \\
ESO343-IG013\_S$^{**}$ & 0.61 (0.01) &  2.966 (0.061) & 12.508 (0.336) &  3.239 (0.065) & $<$ 1.838 (0.066) &  $<$ 6.38 (0.34) & 0.00 & 1.70 (0.01) & ---~~~~~~~ \\
ESO343-IG013\_N$^{**}$ & 0.47 (0.01) &  1.250 (0.032) &  3.856 (0.155) &  1.067 (0.031) & $<$ 0.483 (0.090) &  $<$ 2.35 (0.20) & 0.00 & 1.52 (0.01) & ---~~~~~~~ \\
\hline\end{tabular}}
\end{sidewaystable}
\begin{sidewaystable}
\resizebox{\linewidth}{!}{
\tabcolsep=3pt
\begin{tabular}{l c r r r r r c c c}
\hline
\hline
Source Name$^{*,**}$ & EQW$_{6.2}$($\sigma$) & 6.2$\mu m$ PAH($\sigma$) & 7.7$\mu m$ PAH($\sigma$) & 11.3$\mu m$ PAH($\sigma$) & 17$\mu m$ PAH($\sigma$)  & L(PAH)($\sigma$) & $\tau_{ice}(\sigma)$ & $\tau_{9.7\mu m}(\sigma)$ & $\eta(\sigma)$\\
 & [$\mu m$]$^{i}$ & & & & & [10$^9$L$_{\odot}$] & & & \\
\hline
NGC7771\_W & 0.39 (0.01) &  0.997 (0.036) &  3.955 (0.402) &  1.208 (0.029) &  0.838 (0.069) &  1.22 (0.14) & 0.00 & 0.45 (0.02) & 0.31 (0.03) \\
NGC7771\_S & 0.36 (0.01) &  1.790 (0.049) &  6.188 (0.430) &  2.070 (0.032) &  1.190 (0.103) &  1.95 (0.19) & 0.00 & 1.05 (0.02) & 0.03 (0.05) \\
NGC7771\_N & 0.52 (0.01) &  7.584 (0.162) & 30.037 (1.011) &  8.380 (0.158) &  5.288 (0.266) &  8.79 (0.52) & 0.00 & 1.42 (0.01) & 0.12 (0.04) \\
MRK0331 & 0.63 (0.01) & 10.017 (0.205) & 37.361 (0.790) & 11.169 (0.121) &  7.129 (0.292) & 19.09 (0.77) & 0.00 & 1.59 (0.01) & 0.01 (0.05) \\
\hline\end{tabular}}
~~Fitted MIR Spectral Parameters of the GOALS Sample. Column (1): Source Name, Column (2): the equivalent width of the 6.2\micron~PAH feature in \micron~as calculated using a spline fit, Columns (3)-(6): the fluxes for the 6.2\micron, 7.7\micron~Complex, 11.3\micron~Complex, and 17\micron~Complex PAH features in units of 10$^{-12}$ ergs ~s$^{-1}$ cm$^{-2}$, Column (7): total PAH luminosity summed from all fitted PAH features in units of 10$^9$L$_{\odot}$, Columns (8)-(9): optical depths of the 6.0\micron~ice and 9.7\micron~silicate features, and Column (10): $\eta = log (F_{tot}^{IRAC}[8\mu m] / F_{slit}^{IRS}[8\mu m])$, a measure of the total-to-slit flux ratio at 8\micron. A dash indicates that no IRS data were available, and nondetections are left blank. Uncertainties are given in ( ).\\
$*$ Spectrum could not be adequately fit by spectral decomposition due to the presence of crystalline silicates, an unusually shallow spectral slope, or a lack of features in the MIR (see text).\\
$**$ Multiple nuclei fall within the larger LL slit. When spectral lines are well-fit, their parameters are given, but they likely represent contributions from both galaxies and so are marked as upper limits.\\
$^i$ 6.2$\mu$m PAH equivalent width was calculated assuming an icy source.\\
\end{sidewaystable}
\clearpage

\begin{sidewaystable}
\caption{H$_2$ Line Fluxes for the GOALS Sample \label{H2table}}
\resizebox{\linewidth}{!}{
\tabcolsep=3pt
\begin{tabular}{l r r r r r r r r}
\hline
\hline
Source Name$^{*,**}$ & H$_2$S(0) ($\sigma$) & H$_2$S(1) ($\sigma$) & H$_2$S(2) ($\sigma$) & H$_2$S(3) ($\sigma$) & H$_2$S(4) ($\sigma$) & H$_2$S(5) ($\sigma$) & H$_2$S(6) ($\sigma$) & H$_2$S(7) ($\sigma$)\\
\hline
NGC0023 & $<$0.30 &  2.469 (0.238) &  1.143 (0.286) &  2.045 (0.238) & $<$1.77 &  0.354 (0.140) & $<$3.01 & 0.223 (0.083) \\
NGC0034 & $<$1.08 &  4.763 (0.786) &  2.148 (0.446) &  4.065 (0.251) & $<$2.68 & $<$5.97 & $<$5.07 &  0.612 (0.215) \\
Arp256 & $<$0.01 & $<$0.68 & $<$0.44 &  0.773 (0.134) & $<$0.90 & $<$4.65 & $<$2.36 & $<$1.49 \\
ESO350-IG038 & $<$0.16 & $<$0.38 & $<$0.59 &  0.645 (0.224) & $<$1.31 & $<$1.27 & $<$1.14 & ---\\
NGC0232\_W & $<$0.09 &  2.223 (0.266) &  0.945 (0.275) &  1.716 (0.150) &  1.990 (0.573) &  0.478 (0.149) & $<$3.03 & $<$2.24 \\
NGC0232\_E & $<$0.01 &  1.249 (0.164) &  0.306 (0.099) &  0.566 (0.090) & $<$0.19 &  0.281 (0.093) & $<$0.35 & $<$0.65 \\
MCG+12-02-001 & $<$0.24 &  2.361 (0.564) & $<$1.29 &  2.369 (0.292) & $<$2.26 & $<$9.12 & $<$5.16 & $<$3.21 \\
\hline\end{tabular}}
\end{sidewaystable}
\begin{sidewaystable}
\resizebox{\linewidth}{!}{
\tabcolsep=3pt
\begin{tabular}{l r r r r r r r r}
\hline
\hline
Source Name$^{*,**}$ & H$_2$S(0) ($\sigma$) & H$_2$S(1) ($\sigma$) & H$_2$S(2) ($\sigma$) & H$_2$S(3) ($\sigma$) & H$_2$S(4) ($\sigma$) & H$_2$S(5) ($\sigma$) & H$_2$S(6) ($\sigma$) & H$_2$S(7) ($\sigma$)\\
\hline
NGC0992 & $<$0.36 &  1.519 (0.239) &  0.773 (0.226) &  0.637 (0.208) & $<$1.26 & $<$3.48 & $<$2.72 & $<$1.76 \\
UGC02238 & $<$0.02 &  1.980 (0.256) &  1.206 (0.246) &  1.094 (0.196) & $<$1.85 & $<$4.35 & $<$3.60 & $<$1.92 \\
IRASF02437+2122 & $<$0.17 &  0.839 (0.196) &  0.419 (0.122) &  0.431 (0.141) &  1.122 (0.269) & $<$1.31 & $<$0.52 & $<$0.70 \\
UGC02369 & $<$0.03 &  0.854 (0.244) &  0.486 (0.150) &  1.305 (0.136) & $<$0.41 &  0.414 (0.108) & $<$2.36 & $<$1.22 \\
UGC02608\_N & $<$0.07 &  1.478 (0.378) & $<$0.58 &  0.691 (0.160) &  1.418 (0.488) & $<$4.22 & $<$1.60 &  0.575 (0.177) \\
UGC02608\_S & --- & --- & --- & --- & ---  & ---  & ---  & ---  \\
NGC1275$^*$ & $<$0.12 & $<$1.42 & $<$0.36 & $<$1.00 & $<$0.46 & 5.635 (0.708) & $<$0.71 & 2.670 (1.098) \\
\hline\end{tabular}}
\end{sidewaystable}
\begin{sidewaystable}
\resizebox{\linewidth}{!}{
\tabcolsep=3pt
\begin{tabular}{l r r r r r r r r}
\hline
\hline
Source Name$^{*,**}$ & H$_2$S(0) ($\sigma$) & H$_2$S(1) ($\sigma$) & H$_2$S(2) ($\sigma$) & H$_2$S(3) ($\sigma$) & H$_2$S(4) ($\sigma$) & H$_2$S(5) ($\sigma$) & H$_2$S(6) ($\sigma$) & H$_2$S(7) ($\sigma$)\\
\hline
IRAS05223+1908$^*$ & $<$0.10 & $<$0.09 & $<$0.22 & $<$0.14 & $<$0.39 & $<$0.50 & $<$1.03 & $<$0.73 \\
MCG+08-11-002 & $<$0.59 &  3.739 (0.357) &  1.664 (0.309) &  2.396 (0.217) &  4.335 (0.948) & $<$6.60 & $<$3.68 & $<$3.78 \\
NGC1961 &  0.854 (0.062) &  4.339 (0.116) &  1.556 (0.070) &  2.774 (0.085) &  1.001 (0.204) &  1.537 (0.067) &  0.205 (0.047) &  0.989 (0.051) \\
UGC03351 & $<$0.25 &  2.137 (0.244) &  1.457 (0.268) &  1.153 (0.181) &  2.835 (0.650) & $<$3.44 & $<$2.07 & $<$1.69\\
IRAS05442+1732 & $<$0.13 &  1.829 (0.345) & $<$0.65 &  1.451 (0.169) & $<$1.09 & $<$5.10 & $<$2.72 & $<$2.18\\
IRASF06076-2139$^{**}$ & $<$0.17 & $<$0.64 & 0.522 (0.147) & 1.261 (0.197) & $<$0.37 & $<$1.06 & $<$0.50 & $<$0.51\\
UGC03410\_W &  0.160 (0.031) &  0.526 (0.044) &  0.195 (0.049) &  0.274 (0.073) & $<$0.37 & $<$0.79 & $<$0.77 & $<$0.48\\
\hline\end{tabular}}
\end{sidewaystable}
\begin{sidewaystable}
\resizebox{\linewidth}{!}{
\tabcolsep=3pt
\begin{tabular}{l r r r r r r r r}
\hline
\hline
Source Name$^{*,**}$ & H$_2$S(0) ($\sigma$) & H$_2$S(1) ($\sigma$) & H$_2$S(2) ($\sigma$) & H$_2$S(3) ($\sigma$) & H$_2$S(4) ($\sigma$) & H$_2$S(5) ($\sigma$) & H$_2$S(6) ($\sigma$) & H$_2$S(7) ($\sigma$)\\
\hline
IRASF08572+3915$^*$ & $<$0.10 & $<$0.18 & $<$0.06 & $<$0.04 & $<$0.18 & $<$3.21 & $<$1.56 & $<$0.72 \\
IRAS09022-3615 & $<$0.38 &  2.449 (0.304) &  1.220 (0.263) &  2.886 (0.263) & $<$1.54 &  1.133 (0.327) & $<$1.06 & $<$1.61 \\
IRASF09111-1007\_W & $<$0.06 &  0.630 (0.130) &  0.341 (0.104) &  0.400 (0.142) & $<$0.68 & $<$0.76 & $<$0.64 & $<$0.74  \\
IRASF09111-1007\_E & $<$0.11 &  0.364 (0.102) & $<$0.24 &  0.355 (0.100) & $<$0.44 & $<$0.81 & $<$0.45 & $<$0.43 \\
UGC04881\_W & $<$0.09 &  0.637 (0.083) &  0.253 (0.057) &  0.350 (0.042) & $<$0.14 &  0.288 (0.060) & $<$0.63 & $<$0.48 \\
UGC04881\_E & $<$0.24 &  0.950 (0.117) &  0.328 (0.078) &  0.365 (0.050) & $<$0.13 &  0.303 (0.062) & $<$0.28 &  0.161 (0.056) \\
UGC05101 & $<$0.18 &  1.252 (0.270) &  1.066 (0.247) & $<$0.31 &  2.539 (0.536) &  0.816 (0.246) & $<$2.67 &  3.856 (0.343) \\
\hline\end{tabular}}
\end{sidewaystable}
\begin{sidewaystable}
\resizebox{\linewidth}{!}{
\tabcolsep=3pt
\begin{tabular}{l r r r r r r r r}
\hline
\hline
Source Name$^{*,**}$ & H$_2$S(0) ($\sigma$) & H$_2$S(1) ($\sigma$) & H$_2$S(2) ($\sigma$) & H$_2$S(3) ($\sigma$) & H$_2$S(4) ($\sigma$) & H$_2$S(5) ($\sigma$) & H$_2$S(6) ($\sigma$) & H$_2$S(7) ($\sigma$)\\
\hline
ESO267-G030\_E & $<$0.13 &  2.231 (0.238) & $<$0.63 &  0.639 (0.160) & $<$0.57 & $<$2.74 & $<$2.43 & $<$1.18 \\
IRAS12116-5615 & $<$0.05 &  1.540 (0.416) & $<$0.56 &  1.299 (0.201) &  1.981 (0.728) &  0.708 (0.240) & $<$2.39 & $<$2.06 \\
IRASF12224-0624 & $<$0.05 &  0.435 (0.078) & $<$0.06 &  0.689 (0.107) &  1.468 (0.138) & $<$0.18 & $<$0.06 & $<$0.20 \\
NGC4418$^*$ & $<$0.62 & $<$1.02 & $<$0.27 & 4.091 (0.410) & $<$0.64 & $<$8.58 & $<$0.97 & $<$2.57 \\
UGC08058$^*$ & $<$0.28 & $<$3.42 & $<$0.54 & $<$0.40 & $<$0.97 & $<$2.56 & $<$4.33 & $<$1.09 \\
NGC4922 & --- & --- & --- & --- & ---  & ---  & ---  & ---  \\
CGCG043-099 & $<$0.08 &  1.238 (0.159) &  0.477 (0.115) &  1.015 (0.083) & $<$0.20 &  0.600 (0.102) & $<$1.46 & $<$1.28 \\
\hline\end{tabular}}
\end{sidewaystable}
\begin{sidewaystable}
\resizebox{\linewidth}{!}{
\tabcolsep=3pt
\begin{tabular}{l r r r r r r r r}
\hline
\hline
Source Name$^{*,**}$ & H$_2$S(0) ($\sigma$) & H$_2$S(1) ($\sigma$) & H$_2$S(2) ($\sigma$) & H$_2$S(3) ($\sigma$) & H$_2$S(4) ($\sigma$) & H$_2$S(5) ($\sigma$) & H$_2$S(6) ($\sigma$) & H$_2$S(7) ($\sigma$)\\
\hline
Arp84\_S &  0.140 (0.024) &  0.354 (0.026) &  0.081 (0.014) &  0.118 (0.036) & $<$0.09 &  0.130 (0.036) & $<$0.20 & $<$0.17 \\
CGCG247-020 & $<$0.18 &  0.630 (0.215) & $<$0.38 &  0.688 (0.073) & $<$0.52 &  0.191 (0.075) & $<$1.47 & $<$1.12 \\
NGC5653$^{**}$ & $<$0.40 & $<$ 1.74 & 0.983 (0.238) & 1.099 (0.144) & $<$1.48 & $<$2.93 & $<$2.50 & $<$1.14 \\
IRASF14348-1447 & $<$0.53 &  1.487 (0.149) &  0.641 (0.086) &  1.635 (0.094) &  1.411 (0.174) &  0.846 (0.225) & $<$0.53 &  0.915 (0.087) \\
IRASF14378-3651 & $<$0.43 &  0.343 (0.106) &  0.343 (0.112) &  0.388 (0.149) & $<$0.31 & $<$0.42 & $<$0.39 & $<$0.70 \\
NGC5734\_N & $<$0.56 &  2.495 (0.250) &  0.768 (0.200) &  0.768 (0.161) &  1.389 (0.385) &  0.365 (0.132) & $<$1.45 &  0.320 (0.085) \\
NGC5734\_S & $<$0.25 &  1.228 (0.145) &  0.432 (0.157) &  0.671 (0.154) &  0.924 (0.316) & $<$1.72 & $<$1.29 & $<$0.78 \\
\hline\end{tabular}}
\end{sidewaystable}
\begin{sidewaystable}
\resizebox{\linewidth}{!}{
\tabcolsep=3pt
\begin{tabular}{l r r r r r r r r}
\hline
\hline
Source Name$^{*,**}$ & H$_2$S(0) ($\sigma$) & H$_2$S(1) ($\sigma$) & H$_2$S(2) ($\sigma$) & H$_2$S(3) ($\sigma$) & H$_2$S(4) ($\sigma$) & H$_2$S(5) ($\sigma$) & H$_2$S(6) ($\sigma$) & H$_2$S(7) ($\sigma$)\\
\hline
IRASF17138-1017 & $<$0.85 & $<$3.84 &  2.313 (0.563) &  2.348 (0.229) & $<$2.15 &  0.451 (0.160) & $<$4.13 & $<$0.16 \\
IRASF17207-0014 & $<$0.47 &  3.081 (0.313) &  1.391 (0.280) &  4.139 (0.361) & $<$1.65 &  2.923 (0.420) & $<$1.81 & $<$1.99 \\
ESO138-G027 & $<$0.14 &  1.481 (0.265) & $<$0.45 &  0.912 (0.128) & $<$0.47 & $<$2.67 & $<$1.66 &  0.208 (0.066) \\
UGC11041 & $<$0.08 &  2.271 (0.218) &  0.941 (0.205) &  1.362 (0.201) & $<$1.41 & $<$4.21 & $<$3.15 & $<$1.83 \\
CGCG141-034 & $<$0.16 &  1.109 (0.162) &  0.489 (0.146) &  0.825 (0.171) & $<$0.76 & $<$3.16 & $<$1.24 & $<$1.56 \\
IRAS17578-0400\_W & $<$0.07 &  0.230 (0.044) &  0.140 (0.046) & $<$0.21 & $<$0.33 & $<$0.66 & $<$0.61 & $<$0.38 \\
IRAS17578-0400\_N & $<$0.65 &  1.571 (0.174) &  0.949 (0.172) &  1.214 (0.215) & $<$0.84 & $<$2.78 & $<$2.76 &  0.402 (0.088) \\
\hline\end{tabular}}
\end{sidewaystable}
\begin{sidewaystable}
\resizebox{\linewidth}{!}{
\tabcolsep=3pt
\begin{tabular}{l r r r r r r r r}
\hline
\hline
Source Name$^{*,**}$ & H$_2$S(0) ($\sigma$) & H$_2$S(1) ($\sigma$) & H$_2$S(2) ($\sigma$) & H$_2$S(3) ($\sigma$) & H$_2$S(4) ($\sigma$) & H$_2$S(5) ($\sigma$) & H$_2$S(6) ($\sigma$) & H$_2$S(7) ($\sigma$)\\
\hline
CGCG448-020\_E$^{**}$ & $<$0.49 & $<$ 2.29 &  0.564 (0.204) & 1.829 (0.126) & $<$0.66 & 0.433 (0.101) & $<$0.62 & $<$0.72 \\
IRAS20551-4250 & $<$0.71 &  3.859 (0.527) & $<$0.93 &  2.684 (0.403) &  4.214 (1.173) &  1.837 (0.657) & $<$0.78 &  1.652 (0.373) \\
ESO286-G035 & $<$0.30 &  1.677 (0.231) &  0.753 (0.209) &  1.209 (0.171) & $<$0.87 & $<$4.78 & $<$3.24 & $<$1.89 \\
IRAS21101+5810 & $<$0.11 &  0.707 (0.145) &  0.323 (0.119) &  0.925 (0.158) & $<$0.20 & $<$1.61 & $<$0.94 & $<$0.65 \\
ESO343-IG013\_S$^{**}$ & $<$0.01 & $<$ 0.96 & 0.549 (0.102) &  0.664 (0.064) & $<$0.98 & $<$2.07 & $<$1.51 & $<$0.91 \\
ESO343-IG013\_N$^{**}$ & $<$0.42 & $<$ 1.00 & $<$0.26 &  0.517 (0.078) & $<$0.29 & $<$1.37 & $<$0.74 & $<$0.37 \\
NGC7130 & $<$0.32 &  2.698 (0.473) &  0.823 (0.311) &  0.638 (0.182) &  1.431 (0.471) &  0.510 (0.175) & $<$1.37 &  0.424 (0.155) \\
\hline\end{tabular}}
\end{sidewaystable}
\begin{sidewaystable}
\resizebox{\linewidth}{!}{
\tabcolsep=3pt
\begin{tabular}{l r r r r r r r r}
\hline
\hline
Source Name$^{*,**}$ & H$_2$S(0) ($\sigma$) & H$_2$S(1) ($\sigma$) & H$_2$S(2) ($\sigma$) & H$_2$S(3) ($\sigma$) & H$_2$S(4) ($\sigma$) & H$_2$S(5) ($\sigma$) & H$_2$S(6) ($\sigma$) & H$_2$S(7) ($\sigma$)\\
\hline
NGC7771\_W & $<$0.29 &  1.174 (0.084) &  0.354 (0.066) &  0.873 (0.109) & $<$0.21 & $<$0.50 & $<$0.59 & $<$0.52 \\
NGC7771\_S & $<$0.04 &  0.926 (0.147) & $<$0.34 & $<$0.30 & $<$0.52 & $<$1.09 & $<$0.61 & $<$0.80 \\
NGC7771\_N & $<$0.62 &  4.177 (0.328) &  1.410 (0.303) &  1.915 (0.236) & $<$1.58 & $<$7.00 & $<$3.38 & $<$2.57 \\
MRK0331 & $<$0.38 &  2.623 (0.457) &  0.967 (0.349) &  2.377 (0.257) &  3.160 (0.919) & $<$7.58 & $<$5.42 & $<$3.97 \\
\hline
\end{tabular}}
H$_2$ Line Fluxes the GOALS Sample. Column (1): Source Name, Columns (2)-(7): the fluxes for the H$_2$S(0), H$_2$S(1), H$_2$S(2), H$_2$S(3), H$_2$S(4), H$_2$S(5), H$_2$S(6), and H$_2$S(7) line features in units of 10$^{-13}$ ergs ~s$^{-1}$ cm$^{-2}$. A dash indicates that no IRS data were available, and 3-$\sigma$ upper limits are given for nondetections. Uncertainties are given in ( ).\\
$*$ Spectrum could not be adequately fit by spectral decomposition due to the presence of crystalline silicates, an unusually shallow spectral slope, or a lack of features in the MIR (see text).\\
$**$ Multiple nuclei fall within the larger LL slit. When spectral lines are well-fit, their parameters are given, but they likely represent contributions from both galaxies and so are marked as upper limits.\\
\end{sidewaystable}

\end{document}